%% file: 7921.tex
\documentclass[a4paper, oldversion] 
{aa}

\usepackage{natbib}
\usepackage[english]{babel}
\usepackage{latexsym}
\usepackage{epsfig}
\usepackage{amsmath}
\usepackage{txfonts}
\usepackage{rotating}
\usepackage[usenames]{color}

\input{Befehle}

\renewcommand{\kB}{k}

\newcommand{\change}[1]{{#1}}

\begin{document}

\title{Two-photon transitions in hydrogen\\ and cosmological recombination}
\titlerunning{Two-photon transitions in hydrogen and cosmological recombination}

\author{J. Chluba\inst{1} \and R.A. Sunyaev\inst{1,2}}
\authorrunning{Chluba \and Sunyaev}

\institute{Max-Planck-Institut f\"ur Astrophysik, Karl-Schwarzschild-Str. 1,
85741 Garching bei M\"unchen, Germany 
\and 
Space Research Institute, Russian Academy of Sciences, Profsoyuznaya 84/32, 117997 Moscow, Russia}

\offprints{J. Chluba, \\ \email{jchluba@mpa-garching.mpg.de}}
\date{Received 22 May 2007 / Accepted 26 November 2007}
\abstract
{
  We study the two-photon process for the transitions $n{\rm s}\rightarrow
    1{\rm s}$ and $n{\rm d}\rightarrow 1{\rm s}$ in hydrogen up to large $n$.
  For $n\leq 20$ we provide simple analytic fitting formulae to describe the
  non-resonant part of the two-photon {emission} profiles.  Combining these
  with the analytic form of the cascade-term yields a simple and accurate
  description of the full two-photon decay spectrum, which {only involves} a
  sum over a few intermediate states.
  We demonstrate that the cascade term naturally leads to a nearly Lorentzian
  shape of the two-photon profiles in the vicinity of the resonances.
  However, due to quantum-electrodynamical corrections, the two-photon
  emission spectra deviate significantly from the Lorentzian shape in the very
  distant wings of the resonances. We investigate up to which distance the
  two-photon profiles are close to a Lorentzian and discuss the role of the
  interference term.
  We then analyze how the deviation of the two-photon profiles from the
  Lorentzian shape affects the dynamics of cosmological hydrogen
  recombination.
{\change{Since in this context the escape of photons from the Lyman-$\alpha$
resonance plays a crucial role, we concentrate on the two-photon corrections in
the vicinity of the Lyman-$\alpha$ line.}}
  Our computations show that the \change{changes in} the ionization history
    due to the additional two-photon process from high shell ($n>2$) likely do
    not reach the percent-level. For conservative assumptions we find a
    correction $\Delta N_{\rm e}/N_{\rm e}\sim-0.4\%$ at redshift $z\sim
    1160$.
    This is numerically similar to the result of another recent study;
      however, the physics leading to this conclusion is rather different. In
      particular, our calculations of the effective two-photon decay rates
      yield significantly different values, {\change{where the}} destructive
      interference of the resonant and non-resonant terms plays a crucial role
      in this context. We also show that the bulk of the corrections to the
      ionization history is only due to the 3s and 3d-states and that the
      higher states do not contribute significantly.
}


\maketitle

\section{Introduction}
During the epoch of cosmological hydrogen recombination (typical redshifts
$800\lesssim z \lesssim 1600$), any direct recombination of electrons to the
ground state of hydrogen is immediately followed by the ionization of a
neighboring neutral atom due to re-absorption of the newly released
Lyman-continuum photon.
In addition, because of the enormous difference in the $2{\rm
  p}\leftrightarrow 1{\rm s}$ dipole transition rate and the Hubble expansion
{rate}, photons emitted close to the center of the Lyman-$\alpha$
line scatter $\sim 10^8-10^9$ times before they can finally escape further
interaction with the medium and thereby permit a successful {settling
  of electrons in the hydrogen ground state}.
It is due to these very peculiar circumstances that the $2{\rm
  s}\leftrightarrow 1{\rm s}$-two-photon decay process, which is $\sim 10^8$
orders of magnitude slower than the Lyman-$\alpha$ resonance transition, is
able to substantially control the dynamics of cosmological hydrogen
recombination \citep{Zeldovich68, Peebles68},
allowing about 57\% of all hydrogen atoms in the Universe to recombine at
redshift $z\lesssim 1400$ \citep{Chluba2006b}.

The tremendous success in observations of the cosmic microwave background
temperature and polarization anisotropies \citep{Hinshaw2006, Page2006} has
recently motived several works on high precision computations of the
cosmological hydrogen \citep{Dubrovich2005, Chluba2006, Kholu2006,
  Novosyadlyj2006, Jose2006, Chluba2007, Wong2006b, Chluba2007b} and helium
\citep{SwitzerHirata2007I, SwitzerHirata2007II, SwitzerHirata2007III,
  Kholu2007} recombination history.

One interesting additional physical process, which had been neglected in
earlier computations \citep{SeagerRecfast1999, Seager2000}, is connected to
the two-photon transitions {from} high $n$s and $n$d-states to the
ground state of hydrogen and was first proposed by \citet{Dubrovich2005}.
In their computations a simple scaling for the total two-photon decay rate of
the s and d-states in hydrogen was given and, including these additional
channels leading to the 1s-level, corrections to the ionization history were
found that exceed the percent-level.
These modifications would have a strong impact on the determination of the key
cosmological parameters \citep{Lewis2006} and therefore require careful
consideration.

More recently, theoretical values for the {\it non-resonant}\footnote{{We base
  our definition of this terminus} on the energies of the involved
  intermediate states (see Sect.~\ref{sec:theory} for details).}  two-photon
  decay rates of the 3s and 3d level based on the work of \citet{Cresser1986}
  were utilized to improve the computations of \citet{Dubrovich2005}, showing
  that the effect of two-photon transitions from higher levels on the
  recombination history is likely to be less than $\sim 0.4\%$
  \citep{Wong2006b}.
However, to our knowledge, neither extensive calculations of the two-photon
decay rates for the transitions $n{\rm s}\rightarrow 1{\rm s}$ and $n{\rm
  d}\rightarrow 1{\rm s}$ from high levels exist nor detailed reports of
direct measurements can be found in the literature, so that \citet{Wong2006b}
also had to extrapolate existing values towards higher levels, largely relying
on the previous estimates by \citet{Dubrovich2005} and the values given in
\citet{Cresser1986}.
Here we argue that, due to quantum-interference, it is difficult to separate
the contributions of the {\it pure} two-photon process from the {\it resonant}
single photon processes.
Therefore to answer the question how much two-photon processes are affecting
the recombination history requires a more rigorous treatment in connection
with the radiative transfer and escape of photons \citep{Varshalovich1968,
Grachev1991, Rybicki1994, Chluba2007c} from the main resonances (especially
the Lyman-$\alpha$ line).
Also \citet{SwitzerHirata2007II} re-analyzed the importance of the two-photon
process in the context of cosmological helium recombination and showed that,
for high values of $n$, the rate estimates by \citet{Dubrovich2005} are rather
rough and that in particular the linear scaling with $n$ fails.
Here we provide some conservative lower limits on the possible impact of the
two-photon transitions on the hydrogen recombination history.
We show that the non-resonant contribution to the two-photon decay rate indeed
scales $\propto n$ (see Sect.~\ref{sec:nr_large_n}). However, due to
destructive interference between the resonant and non-resonant terms, the
effective two-photon decay rate is much lower and \change{actually} decreases
with $n$.

If one considers an isolated neutral hydrogen atom\footnote{We restrict
  ourselves to the non-relativistic formulation of the hydrogen atom and
  assume that it is at rest in the lab frame.} with the electron in some
  excited state $(n, l)$, then because of the finite lifetime of the level,
  the electron will reach the ground-state after some short time (typically
  $\sim 10^{-8}\,$s), in general releasing more than one photon.
Astrophysicists usually describe this {\it multi-photon cascade}, \change{also
known as Seaton-cascade, as a sequence of independent, single-step, one-photon
processes \citep[e.g. see][]{Seaton1959}, where every resonance has a pure Lorentzian
shape.}
This approximation should be especially good in the presence of many
perturbing particles (free electrons and ions), such as in {stellar
atmospheres}, which destroy the {\it coherence} of
processes\footnote{\change{More rigorously, collisions more likely destroy the
coherence for photons emitted very close to the line center, since in the
distant wings the typical characteristic time of processes involving the
corresponding quantum-states is much shorter
\citep[e.g.][]{KarshenboimPrep}.}}
involving more than one intermediate transition.
However, in extremely low density environments, like the expanding Universe
during cosmological hydrogen recombination, hardly any perturbing particle is
within the Weisskopf-radius \citep{Weisskopf1932, Sobel1995}, so that the
coherence of two-photon and possibly multi-photon transitions is maintained at
least for the lower shells.
Here we consider the simplest extension to the classical treatment of the
multi-photon cascade and focus only \change{on} the two-photon process.

Beginning with the paper of \citet{Mayer1931}, several textbooks of quantum
electrodynamics \citep{AkhiezerQED, LandauQED} discuss the two-photon emission
process.
Using the formulation of quantum-electrodynamics, one naturally obtains a
nearly Lorentzian shape of the line profiles in the vicinity of the
resonances, which also allows us to check for tiny deviations in the real
two-photon emission from the spectrum obtained using the simplest classical
cascade treatment.
As we discuss below, \change{at least} in the decay of high s and d-states
quantum-electrodynamical corrections lead to additional broad {\it continuum
emission} and strong deviations of the profiles from the natural Lorentzian
shape in the very distant wings of the resonant lines.
In this paper we investigate up to what distance the wings of the two-photon
emission spectrum in the vicinity of the Lyman-$\alpha$ line continue to have
a Lorentzian shape.
These deviations in the red wings are the reason for the corrections
to the hydrogen recombination history due to the two-photon transitions from
high s and d-states.
\change{Similarly, these modifications of the Lyman-$\alpha$ line profile
could also be important during the initial stages of reionization in the
low-$z$ Universe.
One should mention that the developed picture is valid for the primordial
chemical composition of the Universe, which is characterized by a complete
absence of heavy elements, e.g. dust and low ionization energy species that
would influence the escape of Lyman-$\alpha$ photons in planetary nebulae and
$\ion{H}{ii}$ regions in present-day galaxies.}

For hydrogen, several publications on the theoretical value of the total $\rm
2s\rightarrow 1s$ two-photon decay rate can be found \citep{Breit1940,
  Kipper1950, Spitzer1951, Klarsfeld1969, Johnson1972, Goldman1981, Drake1986,
  Goldman1989} with recent computations performed by \citet{Labzowsky2005}
yielding $A_{2{\rm s}1{\rm s}}=8.2206\,{\rm s}^{-1}$.
In these calculations one has to consider all the possible intermediate states
(bound and continuum) with angular momentum quantum number $l=1$, i.e.
p-states.
Within the non-relativistic treatment of the hydrogen atom for the metastable
2s-level, no p-state with energy lower than the 2s-state exists; hence the
two-photon process only involves transitions via {\it virtual} intermediate
states, \change{without any resonant contributions}.
Therefore the total two-photon decay rate of the 2s-level is very low and the
2s-state has \change{an} extremely long lifetime ($\sim 0.12\,$sec).
We show that the formulae obtained by \citet{Cresser1986} are not applicable
in this case (see Sect. \ref{sec:2s}).

Also, some calculations for the 3s and 3d two-photon transitions to the ground
state have been carried out \citep{Quattropani1982, Tung1984, Florescu1984},
but here a problem arises in connection with the contribution from the
intermediate 2p-state, which has an energy {\it below} the initial level.
\change{The corresponding} term is dominating the total two-photon decay
probability for the 3s and 3d two-photon process and is connected with the
{\it resonant transition} via an energetically lower level. It can be
interpreted as a {\it cascade} involving the {\it quasi-simultaneous emission
of two-photons}.
As in the case of the 2s-level, a broad {\it continuum emission} also appears
due to transitions via virtual intermediate states, with energies above the
initial level. This continuum is not connected to any resonances and therefore
has a much lower amplitude.
In addition to the cascade-term and this non-resonant term, an {\it
  interference}-term also appears for which a clear interpretation is
  difficult within the classical formulation.
Similarly, in the two-photon decay process of higher $n$s and $n$d-states to
the ground state, $(2n-4)$-resonances appear, yielding complex structures in
the distribution of emitted photons. 
\change{Some additional examples of $2\gamma$ emission spectra can also be
found in \citet{Quattropani1982} and \citet{Tung1984}.}

As mentioned above, astrophysicists usually interpret the two-photon cascade
as a $1+1$-single photon process. Since even in the full two-photon
formulation, the cascade-term dominates the total two-photon decay rate (hence
defining the lifetime of the initial $n$s and $n$d-states), in a vacuum the
total two-photon decay rate should be very close to the $1+1$-single photon
rate of the considered level.
In the $1+1$-photon picture, the spontaneous two-photon decay rate is simply
given by the sum of all spontaneous one-photon decay rates from the initial
state, since after the detection of one photon, say a Balmer-$\alpha$ photon
in the 3s$\rightarrow$1s transition, in vacuum the presence of a
Lyman-$\alpha$ photon is certain and therefore should not affect the total 3s
decay probability.
For the 3s and 3d-states \change{\citet{Florescu1984}} computed the total
spontaneous two-photon decay rate and indeed found $A^{2\gamma}_{\rm
3s\rightarrow 1s}\sim A^{1\gamma}_{\rm 3s\rightarrow 2p}\approx
\pot{6.317}{6}\,\rm s^{-1}$ and $A^{2\gamma}_{\rm 3d\rightarrow 1s}\sim
A^{1\gamma}_{\rm 3d\rightarrow 2p}\approx \pot{6.469}{7}\,\rm s^{-1}$.
This also suggests that, very close to the resonances, the $1+1$-photon
description provides a viable approximation, in which the line profile is very
close to a Lorentzian.
However, as we show below, quantum-electrodynamical corrections {(e.g. virtual
intermediate states, interference, correlations of the photons in energy)}
lead to differences in the two-photon profiles in comparison with the
$1+1$-single photon profile, which are significant especially in the distant
wings, far from the resonances.
In particular the interference term plays a crucial role in this context and
cannot be neglected.

\section{Two-photon transitions of the $n\rm s$ and $n\rm d$-states}
\label{sec:theory}
Considering only cases when the initial states is either an $n$s or $n$d-level
and the final state corresponds to a s-level, one can simplify the general
formula for the two-photon transition probability as given by
\citet{Mayer1931} considerably.
First, the average over the directions and polarizations of the emitted
photons can be carried out immediately, since within the non-relativistic
formulation, one can separate the {\it radial} and {\it angular} parts of the
wave function. 
For $\rm s\rightarrow s$-transition, this leads to a global factor of $1/27$,
while this average yields $2/135$ for $\rm d\rightarrow s$-transitions
\citep[see][]{Tung1984}.
Afterwards, the probability for the decay $n_il_i\rightarrow 1{\rm s}$ (where
$l_i=0$ or $l_i=2$) with the emission of two photons can be written in terms
of the integrals $\left<R_{n'l'}|r| R_{nl}\right>=\left<R_{n'l'}|r|
  R_{nl}\right>^\ast=\int_0^\infty R_{n'l'} r^3 R_{nl}\id r$ over the
normalized radial functions, $R_{nl}(r)$, for which explicit expressions can
be found in the literature \citep[e.g. \S 52 in][]{LandauQED}. Then the
probability of emitting one photon at frequency $\nu$ and another at $\nu'$ in
the transition $(n_i l_i)\rightarrow \rm 1s$ is given by 
\bsub
\label{eq:general_nsnd}
\beal
\id W_{n_i l_i\rightarrow 1{\rm s}}&=C_{l_i}\,\nu^3\nu'^3
\left|M\right|^2\id \nu
\\[1mm]
M&=\sum_{n=2}
\left<R_{1{\rm s}}|\,r\,| R_{n{\rm p}}\right>\left<R_{n{\rm p}}|\, r \,| R_{n_i l_i}\right>
g_n(\nu)
\\
\label{eq:g_n}
g_n(\nu)&=\frac{1}{E_n-E_{n_i}+h\nu}+\frac{1}{E_n-E_{n_i}+h\nu'}.
\end{align}
\esub
Here $C_l=9\alpha^6 c R_{\rm H}/2^{10}\times a_l/\nu_{\rm 2s1s}^{5}\approx
4.3663\,\text{s}^{-1}\times a_l/\nu_{\rm 2s1s}^{5}$, with $a_0=1$ and
$a_2=2/5$, and where $\alpha$ is the fine structure constant, $c$ the speed of
light, $\nu_{\rm 2s1s}$ the 2s-1s transition frequency, and $R_{\rm H}$ the
Rydberg constant for hydrogen.
Due to energy conservation, the frequencies of the emitted photons are related
by $\nu+\nu'=~\nu_{i1{\rm s}}$, where here $\nu_{i1{\rm s}}$ is the transition
frequency between the initial level $n_i l_i$ and final 1s-state. The energy
of level $n$ is given by $E_n=-E_{\rm 1s}/n^2$, with the 1s-ionization energy
of the hydrogen atom $E_{\rm 1s}\approx 13.6\,$eV.
The sum in Eq.~\eqref{eq:general_nsnd} has to be extended by an integral over
the continuum states, but for convenience we omit these in our notation.

For $n_i>2$ and $n<n_i$, it is clear from Eq.~\eqref{eq:general_nsnd} that at
$h\nu=E_{n_i}-E_n$ and $h\nu=E_n-E_{1{\rm s}}$, i.e. corresponding to the
resonance frequencies to {\it energetically lower} levels, one of the
denominators inside the sum vanishes, leading to a divergence of the
expression.
As we discuss below (Sect. \ref{sec:cas_int}), including the {\it lifetime} of
  the intermediate states provides a possibility of removing these
  singularities \citep{Low1952, Labzowsky2004};
however, a consistent consideration of this problem requires a more
sophisticated treatment beyond the scope of this paper.
Physically, transitions to intermediate states with energies $E_n\geq E_{n_i}$
are {\it virtual}\footnote{\change{Strictly speaking, photons appearing in the
distant wings of the resonance are also connected with virtual states
\citep[e.g.][]{KarshenboimPrep}.}}.
We split up the sum over all the intermediate states like\footnote{Within the
  non-relativistic treatment of the hydrogen atom, this is equivalent to
  separating the cases $n<n_i$ and $n\geq n_i$. Since the energy of the
  continuum states is always greater than for the bound states, the former are
  associated with the case $n\geq n_i$.}
$\sum=\sum_{\rm v}+\sum_{\rm r}$, where $\sum_{\rm v}=\sum_{E_n\geq E_{n_i}}$
and $\sum_{\rm r}=\sum_{E_n<E_{n_i}}$ denote the sum over virtual and real
intermediate states, respectively.  Then we can write
\beal
\label{eq:M_contributions}
|M|^2\propto |{\sum}_{\rm r}|^2 +|{\sum}_{\rm v}|^2 
+{\sum}_{\rm r}^\ast\,{\sum}_{\rm v}+{\sum}_{\rm r}\,{\sum}_{\rm v}^\ast.
\end{align}
Henceforth, we refer to the contribution to the transition matrix element from
$|{\sum}_{\rm r}|^2$ as the {\it cascade} part, due to $|{\sum}_{\rm v}|^2$ as
{\it non-resonant} part, and the rest as the {\it interference}\footnote{In
principle one should be more accurate by calling this contribution
resonant/non-resonant-interference term, since also some level of interference
is already included inside the resonant and non-resonant-term, which
\change{is} absent in the $1+1$-photon picture.  However, we generally do not
make this distinction.} part.
This distinction is {\it ad hoc} and {\it not unique}, but only motivated by
our separation of the infinite sum, which will turn out to be fairly
convenient in terms of evaluation of the spectrum and total two-photon decay
rate.

The {\it non-resonant} contribution to the two-photon decay probability, $\id
W^{\rm nr}$, is then given by
\bsub
\beal
\label{eq:general_nsnd_nr}
\id W^{\rm nr}_{n_i l_i\rightarrow 1{\rm s}}&\!=C_{l_i}\,\nu^3\nu'^3
\left|M_{\rm nr}\right|^2\id \nu
\\[1mm]
M_{\rm nr}&\!=\!\sum_{n=n_i}^\infty
\left<R_{1{\rm s}}|\,r\,| R_{n{\rm p}}\right>\left<R_{n{\rm p}}|\, r \,| R_{n_i l_i}\right>
g_n(\nu).
\end{align}
\esub
For the cascade and interference terms, one instead has
\bsub
\beal
\label{eq:general_nsnd_cas}
\id W^{\rm cas}_{n_i l_i\rightarrow 1{\rm s}}&=C_{l_i}\,\nu^3\nu'^3
\left|M_{\rm cas}\right|^2\id \nu
\\[1mm]
M_{\rm cas}&=\!\sum_{n=2}^{n_i-1}
\left<R_{1{\rm s}}|\,r\,| R_{n{\rm p}}\right>\left<R_{n{\rm p}}|\, r \,| R_{n_i l_i}\right>
g_n(\nu)
\end{align}
\esub
and
\beal
\label{eq:general_nsnd_int}
\id W^{\rm int}_{n_i l_i\rightarrow 1{\rm s}}=C_{l_i}\,\nu^3\nu'^3
\left[
M_{\rm cas}^\ast \,M_{\rm nr}+M_{\rm cas} \,M_{\rm nr}^\ast
\right]\id \nu,
\end{align}
respectively.

\subsection{Total two-photon decay rate}
\label{sec:all_theory}
In order to obtain the {\it total} two-photon decay rate in vacuum one now has
to integrate Eq.~\eqref{eq:general_nsnd} over all possible frequencies $\nu$.
The corresponding integral can be cast into the form
\beal
\label{eq:A_nsnd}
A^{2\gamma}_{n_i l_i\rightarrow 1{\rm s}}
&=\frac{1}{2}\int_0^{\nu_{i1{\rm s}}} \!\!\!\id W_{n_i l_i\rightarrow 1{\rm s}} 
=\frac{1}{2}\int_0^1 \!\phi^{2\gamma}_{n_i l_i\rightarrow 1{\rm s}}(y)\id y,
\end{align}
with $y=\nu/\nu_{i1{\rm s}}$ and where the factor of 1/2 avoids
double-counting of photons.
In Eq. \eqref{eq:A_nsnd} we have introduced the two-photon decay profile
function, 
\beal
\label{eq:phi_2g}
\phi^{2\gamma}_{n_i l_i\rightarrow 1{\rm s}}(y)
=
\phi^{\rm nr}_{n_i l_i\rightarrow 1{\rm s}}(y)
+\phi^{\rm cas}_{n_i l_i\rightarrow 1{\rm s}}(y)
+\phi^{\rm int}_{n_i l_i\rightarrow 1{\rm s}}(y),
\end{align}
which is the sum of the two-photon profiles due to the non-resonant, cascade,
and interference terms, respectively. Below we discuss the contribution of
each term separately.

\subsubsection{Interpretation of the two-photon emission profile}
\label{sec:interpretation}
Physically the two-photon emission profile or spectrum $\phi^{2\gamma}_{n_i
  l_i\rightarrow 1{\rm s}}$ defines the number of photons that are released
per second in the frequency interval between $\nu$ and $\nu+\id\nu$. If one
integrates over the whole spectrum, this therefore yields the total number of
photons emitted per second due to the two-photon transition.  The two-photon
profile includes both photons at the same time, so that the total two-photon
transition rate per initial s or d-state has to be divided by a factor of 2
(see Eq.  \ref{eq:A_nsnd}).

Due to energy conservation it is clear that, when detecting a photon that was
produced in a particular two-photon transition from some initial $n$s or
$n$d-state to the ground state, at a frequency $\nu$, the other photon has
frequency $\nu'=\nu_{n\rm 1s}-\nu$. Therefore also the probability to release
a photon at $\nu$ should be equal to the probability for the emission of a
photon at $\nu_{n\rm 1s}-\nu$, a property that is reflected in the symmetry of
the two-photon profiles around $y=1/2$ (see Sect. \ref{sec:two_em} for more
explicit examples).

\subsubsection{Total two-photon decay rate due to the non-resonant contribution}
\label{sec:nr_theory}
In Eq. \eqref{eq:phi_2g} we have introduced the non-resonant two-photon decay
profile function, $\phi^{\rm nr}_{n_i l_i\rightarrow 1{\rm s}}(y)$, which can
be written as
\bsub
\label{eq:A_phi_nr}
\beal
\label{eq:A_phi_nr_a}
\phi^{\rm nr}_{n_i l_i\rightarrow 1{\rm s}}(y)&\!=G_{n_i l_i}\,y^3 (1-y)^3\left|\tilde{M}_{\rm nr}\right|^2 
\\
\label{eq:A_phi_nr_b}
\tilde{M}_{\rm nr}&\!=\!
\sum_{n=n_i}^\infty
\left<R_{1{\rm s}}|\,r\,| R_{n{\rm p}}\right>\left<R_{n{\rm p}}|\, r \,| R_{n_i l_i}\right>
f_n(y)
\\
f_n(y)&\!=
\frac{1}{y+y^{\rm nr,+}_n}-\frac{1}{y-y^{\rm nr,-}_n}
\end{align}
with $G_{n_i l_i}=\nu_{i1{\rm s}}^5 C_{l_i}$. Here we defined the frequencies
\beal
\label{eq:y_nr}
y^{\rm nr,+}_n&=\frac{n^2-n_i^{2}}{n^2(n_i^{2}-1)}
\\
y^{\rm nr,-}_n&=1+y^{\rm nr,+}_n=\frac{n_i^2(n^{2}-1)}{n^2(n_i^{2}-1)},
\end{align}
\esub
which for $n\geq n_i$ vary within the ranges $0\leq y^{\rm nr,+}_n\leq
1/(n^2_i-1)$ and $1\leq y^{\rm nr,+}_n\leq n^2_i/(n^2_i-1)$. For $n=n_i$ one
finds $y^{\rm nr,+}_n=0$ and $y^{\rm nr,-}_n=1$.
Since the sum in $\tilde{M}_{\rm nr}$ only involves intermediate states with
$n\geq n_i$, the denominators of $f_n$ never vanish within the interval $0< y
< 1$, and for $n=n_i$ the factors $y^3$ and $(1-y)^3$ ensure that
$\tilde{M}_{\rm nr}$ approaches zero within the limits $y\rightarrow 0$ and
$y\rightarrow 1$.
In addition $\tilde{M}_{\rm nr}$ is real and symmetric around $y=1/2$.

To compute the total rate one now only has to replace $\phi^{2\gamma}_{n_i
  l_i\rightarrow 1{\rm s}}(y)$ in Eq. \eqref{eq:A_nsnd}, by the expression
\eqref{eq:A_phi_nr}.
To evaluate the sum and integrals over the radial functions, we used {\sc
  Mathematica}.  Normally we restrict ourselves to the first 200 terms in the
  sum, but computations with up to 4000 terms were also performed for the 2s,
  3s, and 3d rates.
Within the assumptions the results for the other levels should be correct to
better than $\sim 1\%$. To make cross checks easier, we give the expression
for the necessary radial integrals $\left<R_{n'l'}|\,r\,| R_{nl}\right>$ up to
$n_i=5$ in Appendix~\ref{app:A}.

\subsubsection{Total two-photon decay rate due to the cascade and interference terms}
\label{sec:cas_int}
For the cascade and interference terms special care has to be taken close to
and at intermediate distances from the resonance frequencies
$h\nu=E_{n_i}-E_n$ and $h\nu=E_n-E_{\rm 1s}$.
As mentioned above, a consistent treatment of this problem requires more
sophisticated methods, including the amplitudes of several additional
processes \citep[e.g.][]{KarshenboimPrep}, than are within the scope of this
paper.
One simple approximate solution to this problem can be \change{given} when
taking the {\it lifetime} of the intermediate states into account as a small
imaginary contribution to their energy\footnote{We neglect the small
correction to the real part of the energy caused by the Lamb-shift.}.
Including this shift into the equations for the 2p-transition leads to the
classical expression of the Lorentzian within the formulation of Quantum
Electrodynamics and can be attributed to the first order radiative corrections
of the one-photon process \citep{Low1952, Labzowsky2004}.
\citet{Florescu1984} used this approach to compute the total 3s and 3d
two-photon decay rates and simply replaced the energy, $E_{\rm 2p}$, of the
2p-state by, $E_{\rm 2p}-i h \Gamma_{\rm 2p}/2$, where $\Gamma_{\rm 2p}$ is
the \change{width of the 2p-state due to spontaneous transitions}.
\change{Except for the 2s-state, summing all the one-photon decay rates
\citep[e.g. these values can be computed using the routines
of][]{StoreyHum1991} should yield a very good approximation for the total
lifetime of any given initial level in the hydrogen atom}.
\change{For estimates, we therefore follow this approximate procedure
and replace the energies of all intermediate p-states by
$E_{n\rm p}\rightarrow E_{n\rm p}-i h \Gamma^{1\gamma}_{n\rm p}/2$.
Here $\Gamma^{1\gamma}_{n\rm p}\equiv A^{1\gamma}_{n{\rm p}}=\sum_{n'l'}
  A^{1\gamma}_{n{\rm p}\rightarrow n'l'}$ is the total $1\gamma$-width of the
  intermediate $n$p-state due to spontaneous transitions}.
\change{Note that for $n>2$ the p-states can decay via channels, which do not
directly lead to the 1s-state, thereby leading to the possible emission of
more than two photons.}

\change{The two-photon contribution to the total lifetime of the s and
  d-levels for transitions to the ground state should be close to the value
  following from the sum of the rates for all possible $1\gamma$-transitions
  to lower-lying intermediate p-states multiplied by the probability that the
  electron will `afterwards' go directly to the 1s-level\footnote{\change{The
  authors wish to thank E.E.~Kholupenko for pointing out some inaccuracies
  related to this aspect. However, the modification did not affect the results
  of this paper.}}}:
\beal
\label{eq:Gamma_n}
\Gamma_{n{\rm s / d}}^{2\gamma}\approx\sum_{n'=2}^{n-1} A^{1\gamma}_{n{\rm s/d}\rightarrow
  n'\rm p}\times \frac{A^{1\gamma}_{n'{\rm p}\rightarrow
  \rm 1s}}{A^{1\gamma}_{n'{\rm p}}},
\end{align}
where $A^{1\gamma}_{n{\rm s/d}\rightarrow n'l'}$ is the spontaneous one-photon
Einstein coefficients of the {considered transition from the initial}
s/d-state\footnote{Here $A^{1\gamma}_{n{\rm s/d}\rightarrow n'l'}$ refers to
either the $A^{1\gamma}_{n{\rm s}\rightarrow n'l'}$ or the $A^{1\gamma}_{n{\rm
d}\rightarrow n'l'}$ decay rate. We use this notation more often below.}.
{Although for the s-levels, this already accounts for all the possible
one-photon decay channels, d-states with $n>4$ can also decay via
intermediate f-states.
However, in this case again more than two-photons have to be released in order
to reach the 1s-level and therefore do not contribute anything here.
The addition \eqref{eq:Gamma_n}} to the energies is not important for the
non-resonant term, but it becomes crucial very close to the resonances for
both the cascade and interference term. However, {here we are mainly
interested in the behavior at large distances from the poles ($|\Delta\nu| \gg
\Gamma^{1\gamma}_{n\rm p}$), so that this replacement has no \change{direct
effect on the discussed cosmological application (see
Sect.~\ref{sec:recombination})}}.

\change{Here one may ask, {\it why the lifetime of the initial level is not
    included}?
  Physically this is motivated by the idea that, following the interpretation
  of \citet{Weisskopf1930}, we consider one particular initial `energy
  sub-level' and do not specify the process that populated it.
  Therefore the final profile should be independent of the shape of the
  distribution of energy-sub-levels around the mean energy of the initial
  state.
  One can also consider this as equivalent to neglecting any possible
  reshuffling of the electron by perturbing particles while it is in the
  initial state.
  However, in the computation presented below we do not approach the
  resonances so close that these differences would play any role.  
}

With the notation of Sect. \ref{sec:nr_theory}, we now introduce the cascade
and interference two-photon emission profiles by
\bsub
\label{eq:A_phi_cas}
\beal
\label{eq:A_phi_cas_a}
\phi^{\rm cas}_{n_i l_i\rightarrow 1{\rm s}}(y)&\!=\!G_{n_i l_i}\,y^3 (1-y)^3\!\left|\tilde{M}_{\rm cas}\right|^2 
\\
\phi^{\rm int}_{n_i l_i\rightarrow 1{\rm s}}(y)&\!=\!G_{n_i l_i}\,y^3 (1-y)^3 \tilde{M}_{\rm nr}\!\left[
\tilde{M}_{\rm cas}^\ast+\tilde{M}_{\rm cas}
\right],
\end{align}
where the cascade matrix element is given by
\beal
\label{eq:A_phi_cas_b}
\tilde{M}_{\rm cas}&=
\sum_{n=2}^{n_i-1}
\left<R_{1{\rm s}}|\,r\,| R_{n{\rm p}}\right>\left<R_{n{\rm p}}|\, r \,| R_{n_i l_i}\right>
h_n(y)
\\
h_n(y)&=
\frac{1}{y-y^{\rm cas,+}_n-i\delta_n}-\frac{1}{y-y^{\rm cas,-}_n-i\delta_n}.
\end{align}
\esub
Here $\delta_n=\Gamma_{n{\rm p}}/4\pi\nu_{i1{\rm s}}$ accounts for the
energy-shifts due to the finite lifetime of the intermediate $n$p-state.
We have also introduced the resonance frequencies
\bsub
\label{eq:y_cas}
\beal
y^{\rm cas,+}_n&\equiv-y^{\rm nr,+}_n
\\
y^{\rm cas,-}_n&\equiv y^{\rm nr,-}_n.
\end{align}
\esub
{Since for the cascade and interference term $n<n_i$, these now have values
strictly within the range $0<y<1$.}

\begin{figure}
\centering 
\includegraphics[width=0.96\columnwidth]
{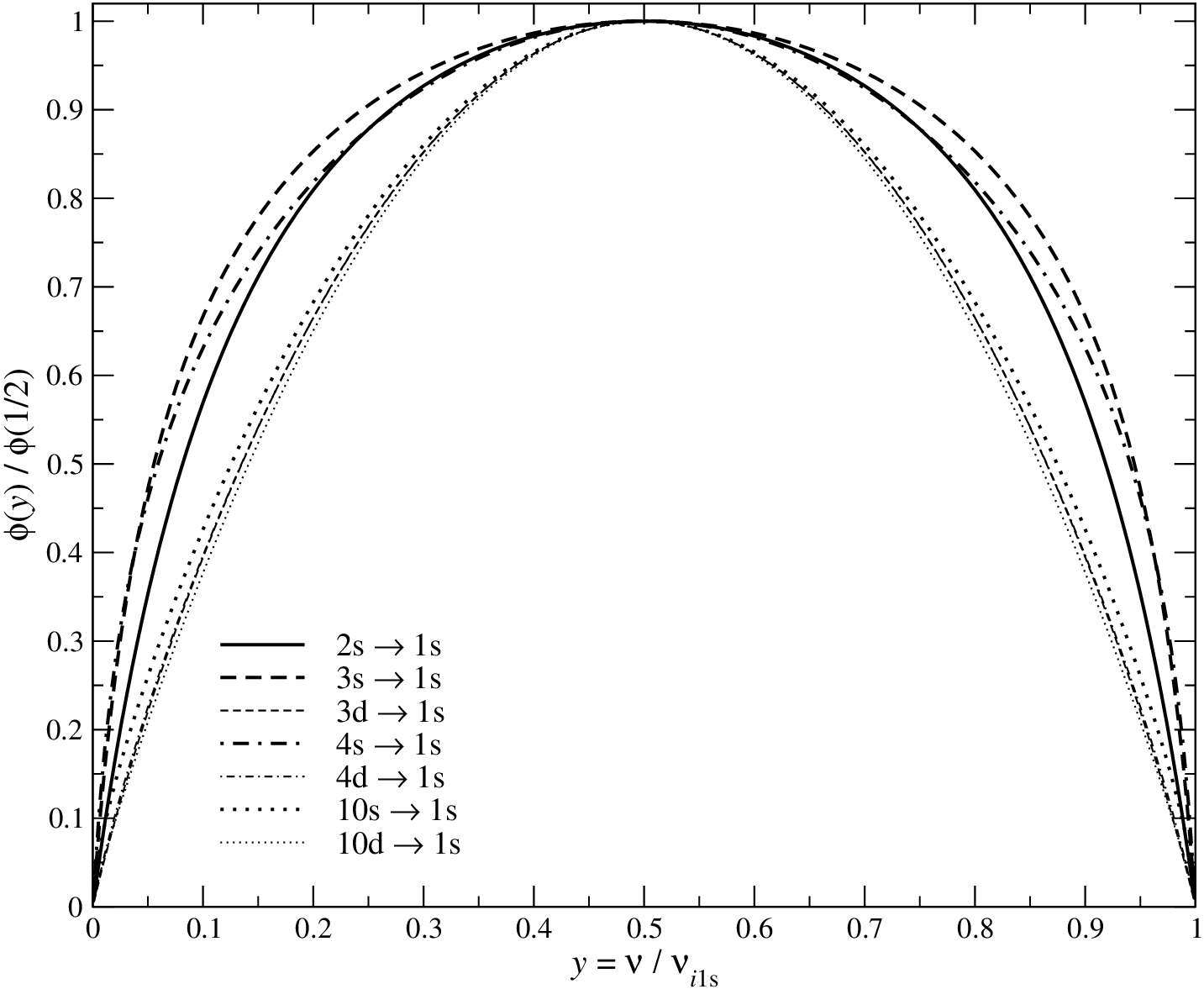}
\caption
{Non-resonant two-photon emission spectra, {Eq. \eqref{eq:A_phi_nr}}, for
  several transitions. All curves are normalized to unity at $y=1/2$. The
  values of $\phi^{\rm nr}(1/2)$ can be found in Appendix~\ref{app:fits}.}
\label{fig:three}
\end{figure}
%
{Defining the function}
\beal
\label{eq:L}
L(a, b)&=\frac{a}{a^2+b^2},
\end{align}
which actually has the generic shape of a Lorentzian, then 
$f_n(y)=f^{\rm r}_n(y)+i f^{\rm i}_n(y)$,
where the real and imaginary part of $f_n$ are defined by
\bsub
\beal
\label{eq:f_re_im}
f^{\rm r}_n(y)&=L(y-y^{\rm cas,+}_n, \delta_n)-L(y-y^{\rm cas,-}_n,\delta_n)
\\
f^{\rm i}_n(y)&=L(\delta_n, y-y^{\rm cas,+}_n)-L(\delta_n, y-y^{\rm cas,-}_n),
\end{align}
\esub
respectively. 
Introducing $\kappa_{n}=\left<R_{1{\rm s}}|\,r\,| R_{n{\rm
      p}}\right>\left<R_{n{\rm p}}|\, r \,| R_{n_i l_i}\right>$ one can
rewrite $|\tilde{M}_{\rm cas}|^2$ and $|\tilde{M}_{\rm int}|^2=\tilde{M}_{\rm
  nr}[\tilde{M}_{\rm cas}^\ast+\tilde{M}_{\rm cas}]$ as
\bsub
\label{eq:M2_cas_inf}
\beal
\label{eq:M2_cas_inf_a}
\left|\tilde{M}_{\rm cas}\right|^2 
&=
\sum_{n=2}^{n_i-1}
\kappa_{n}^2\,|f_n(y)|^2
+2\sum_{n=2}^{n_i-1}\sum_{m=2}^{n-1}
\kappa_{n}\kappa_{m}\left[f^{\rm r}_n\,f^{\rm r}_m+f^{\rm i}_n\,f^{\rm i}_m\right]
\\
\left|\tilde{M}_{\rm int}\right|^2 
&=2\tilde{M}_{\rm nr}
\times
\sum_{n=2}^{n_i-1}
\kappa_{n}\,f^{\rm r}_n(y).
\end{align}
\esub
To compute the total rate one now only has to replace $\phi^{2\gamma}_{n_i
l_i\rightarrow 1{\rm s}}(y)$ in Eq. \eqref{eq:A_nsnd}, by the corresponding
expressions \eqref{eq:A_phi_cas}.

\section{Two-photon emission spectra}
\label{sec:two_em}

\subsection{Non-resonant two-photon emission spectra}
In Fig.~\ref{fig:three} we present the profile functions for the non-resonant
contribution to the two-photon decay spectrum. All the profiles have a maximum
at $y=1/2$.
For the $n{\rm s}\rightarrow 1{\rm s}$-emission profiles the difference in the
shape of the curves is quite big, while for initial d-states in general the
profile does not vary as much. However, in both cases the amplitude at $y=1/2$
changes strongly, increasing towards larger $n$ (see Appendix~\ref{app:fits}).

Due to our separation of the {\it infinite} sum over the intermediate
substates, the sums in the cascade and interference terms become {\it
  finite}.  This allows us to evaluate $M_{\rm nr}$ numerically and use
convenient fitting formulae for their representation.
Realizing that $M_{\rm nr}$ is symmetric around $y=1/2$ and that it scales
like $\sim 1/y$ and $\sim 1/(1-y)$ at the boundaries, we approximated
$y(1-y)\,M_{\rm nr}$.
In Appendix~\ref{app:fits} the obtained formulae for all $n{\rm s}\rightarrow
  1{\rm s}$ and $n{\rm d}\rightarrow 1{\rm s}$ transitions up to $n=20$ are
  given.
For the non-resonant term within the range $10^{-3}\leq y\leq 0.999$, these
approximations should be accurate to better than 0.1\%.
Since these are fast and simple to evaluate they should be useful for analytic
estimates and numerical applications.
Note that all non-resonant matrix-elements are negative.

\begin{figure*}
\centering 
\includegraphics[width=0.96\columnwidth]
{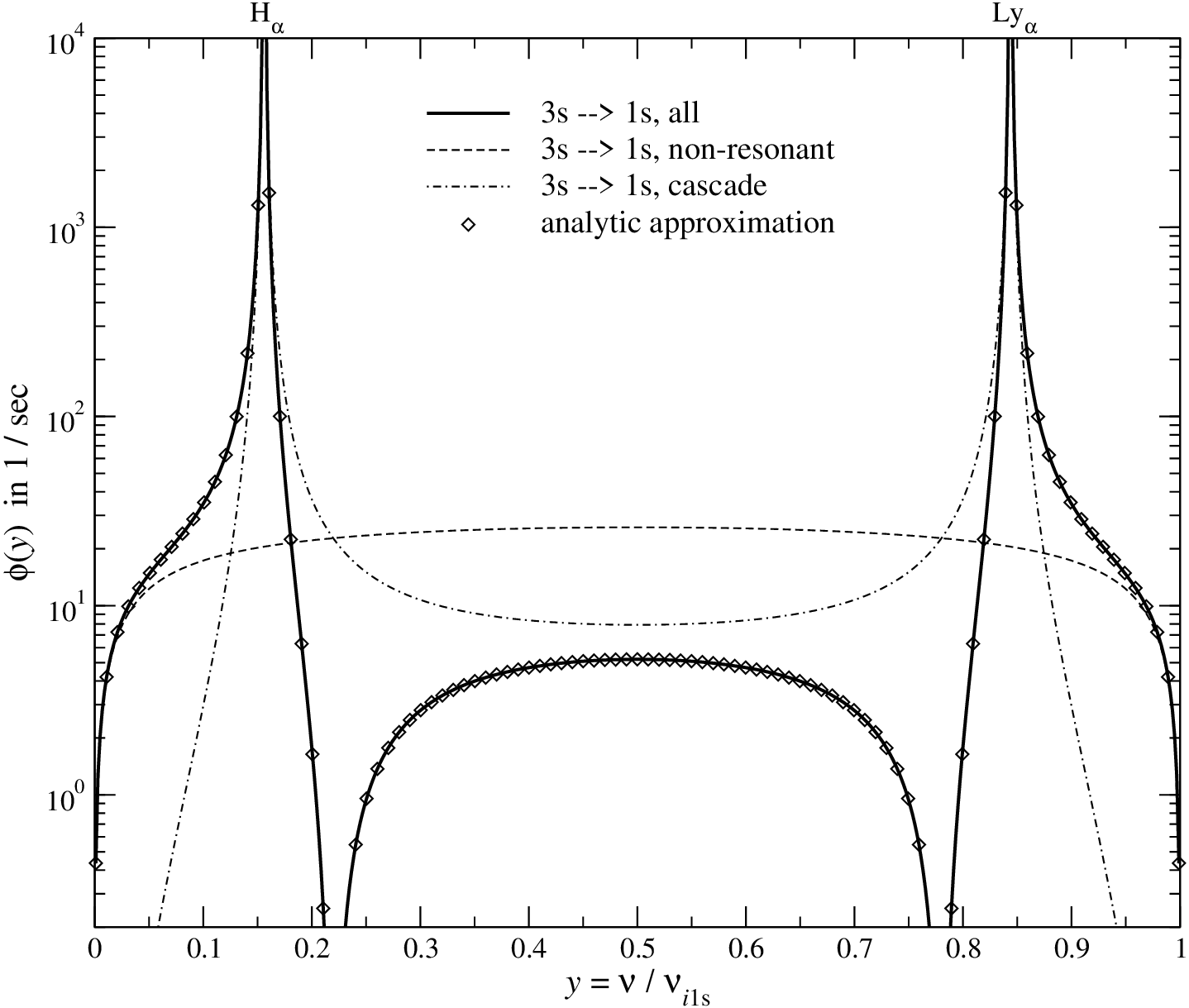}
\hspace{1mm}
\includegraphics[width=0.96\columnwidth]
{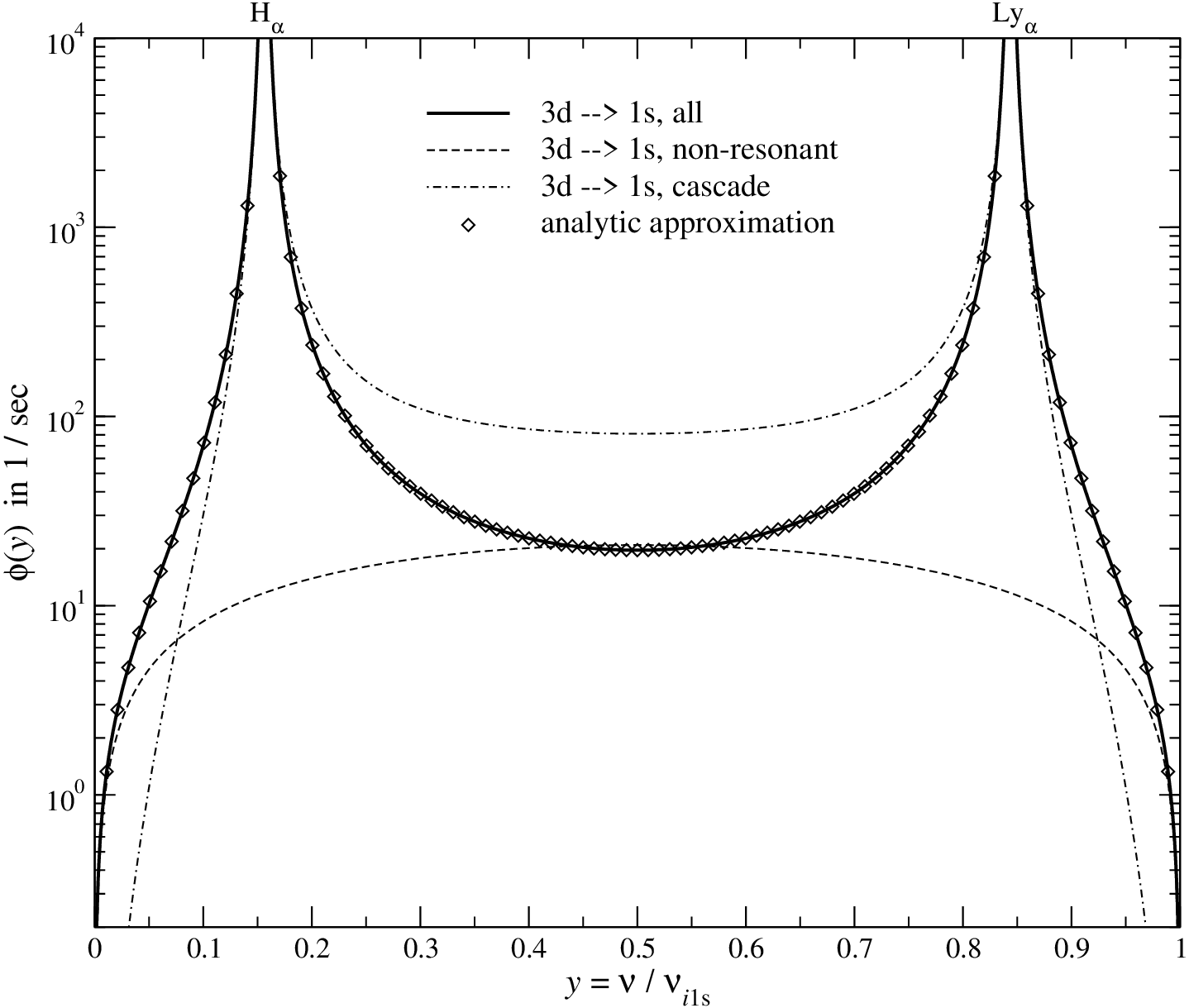}
\caption
{Two-photon emission spectra for the $3{\rm s}\rightarrow 1{\rm s}$ and $3{\rm
    d}\rightarrow 1{\rm s}$ transitions. The non-resonant, cascade, and
    combined spectra are shown as labeled. Also we give the analytic
    approximation based on the equations discussed in Sect. \ref{sec:cas_int},
    together with the fitting formula for the 2s and 3d non-resonant matrix
    element (see Appendix~\ref{app:fits}). The resonances correspond to the
    Balmer-$\alpha$ and Lyman-$\alpha$ transitions.}
\label{fig:3s3d_spec}
\end{figure*}
\subsection{The 3s and 3d two-photon emission spectrum}
In Fig. \ref{fig:3s3d_spec} we give the {full} two-photon emission spectra for
the $3{\rm s}~\rightarrow~1{\rm s}$ and $3{\rm d}~\rightarrow~1{\rm s}$
transition. In both cases, the \change{non-resonance} term increases the wings
of the profiles close to $y\sim 0$ and $y\sim 1$. However, the interference
between the non-resonant and cascade part is destructive in the central region
($y\sim 1/2$) and significantly reduces the amplitude of the total two-photon
emission.
It even leads to full cancellation at $y\sim 0.22$ and $y\sim 0.78$ for the
3s-level, whereas for the 3d-level the photon production does not vanish in
the region between the resonances \change{\citep[see also][]{Tung1984}}.

For both the $3{\rm s}~\rightarrow~1{\rm s}$ and $3{\rm d}~\rightarrow~1{\rm
 s}$ transitions, only one term in the cascade appears, which is related to the
 transition via the intermediate 2p-state. The matrix elements for these are
 $\kappa^{\rm 3s}_2\approx 1.211$ and $\kappa^{\rm 3d}_2\approx 6.126$, and
 according to Eq.  \eqref{eq:y_cas} the resonance frequencies are at $y^{\rm
 cas,+}_2=5/32$ (Balmer-$\alpha$ transition) and $y^{\rm cas,-}_2=27/32$
 (Lyman-$\alpha$ transition).
 With the Eqn. given in Sect. \ref{sec:cas_int} and using the fitting formulae
 according to Appendix~\ref{app:fits} one can analytically approximate the
 full two-photon emission spectrum. As Fig.  \ref{fig:3s3d_spec} shows the
 agreement is excellent at all considered frequencies.

\begin{figure*}
\centering 
\includegraphics[width=0.96\columnwidth]
{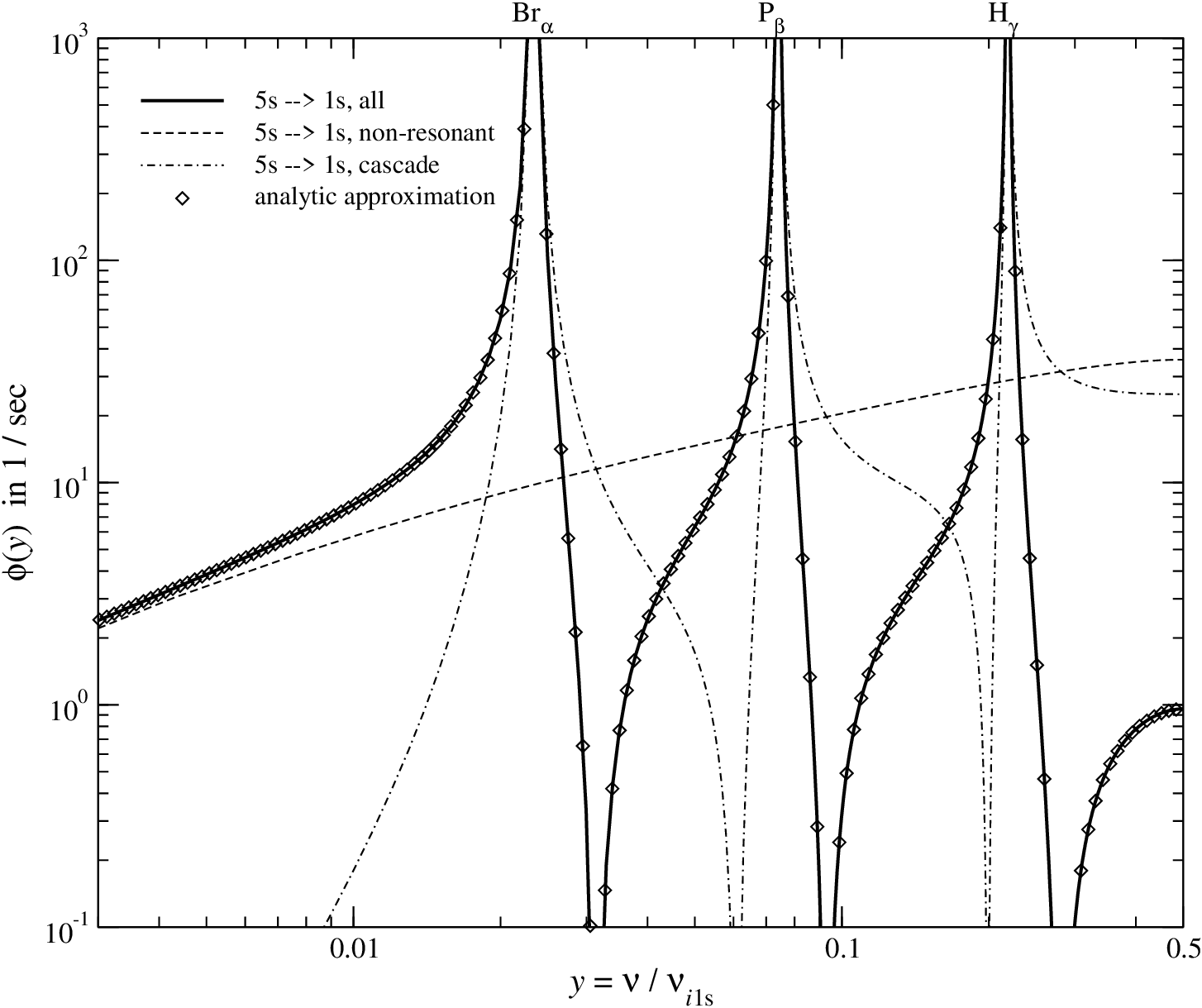}
\hspace{1mm}
\includegraphics[width=0.97\columnwidth]
{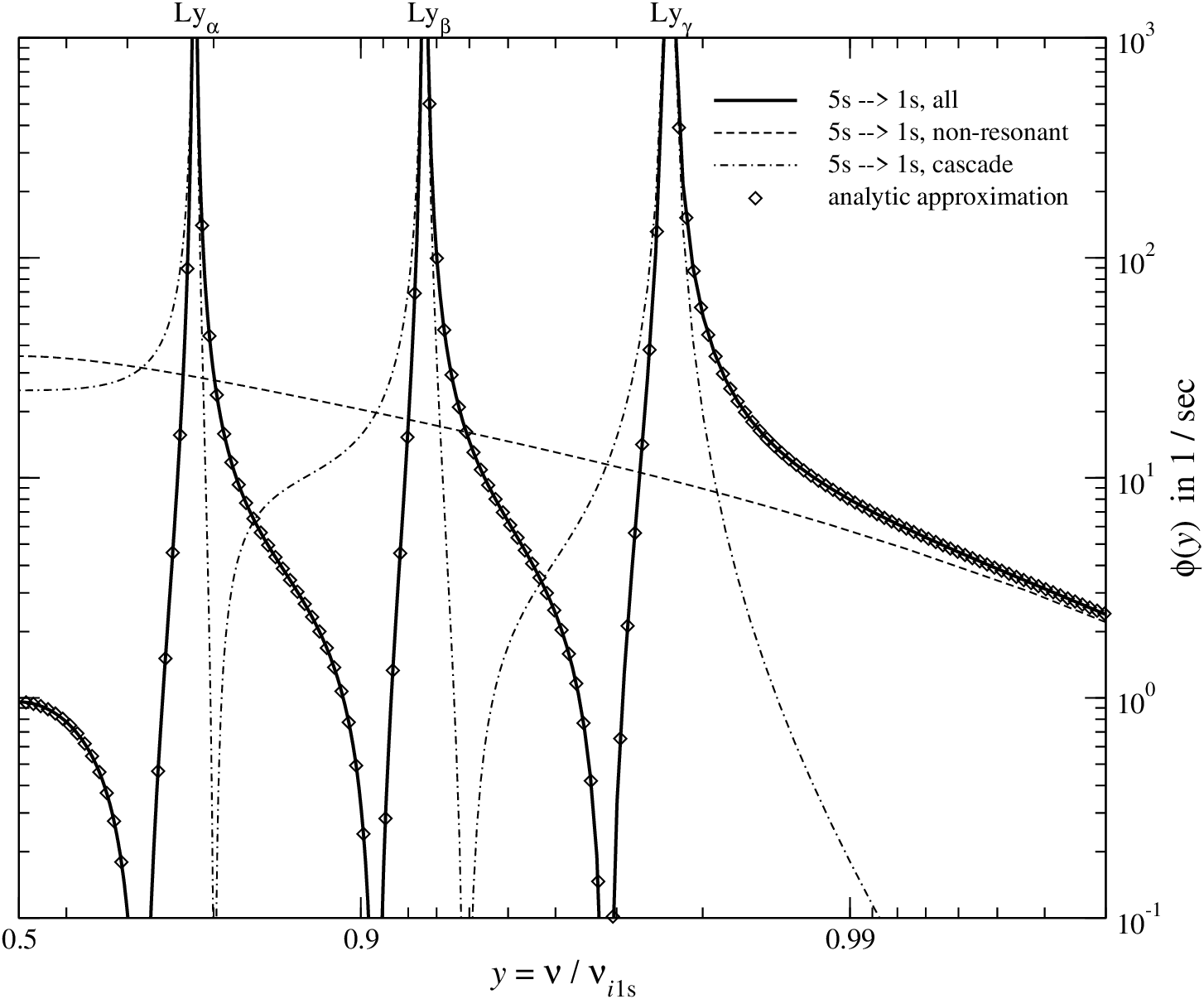}
\\[1mm]
\includegraphics[width=0.96\columnwidth]
{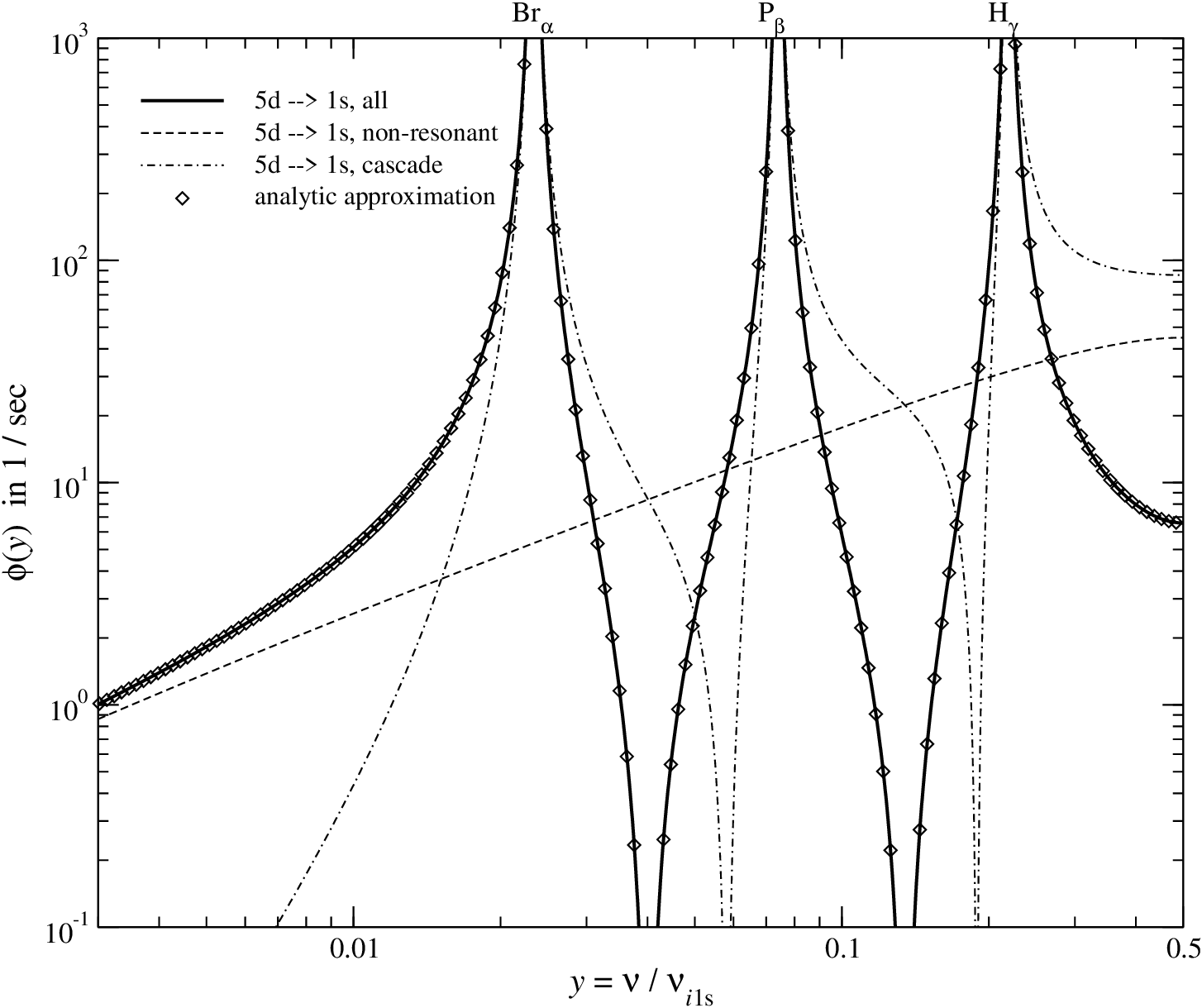}
\hspace{1mm}
\includegraphics[width=0.97\columnwidth]
{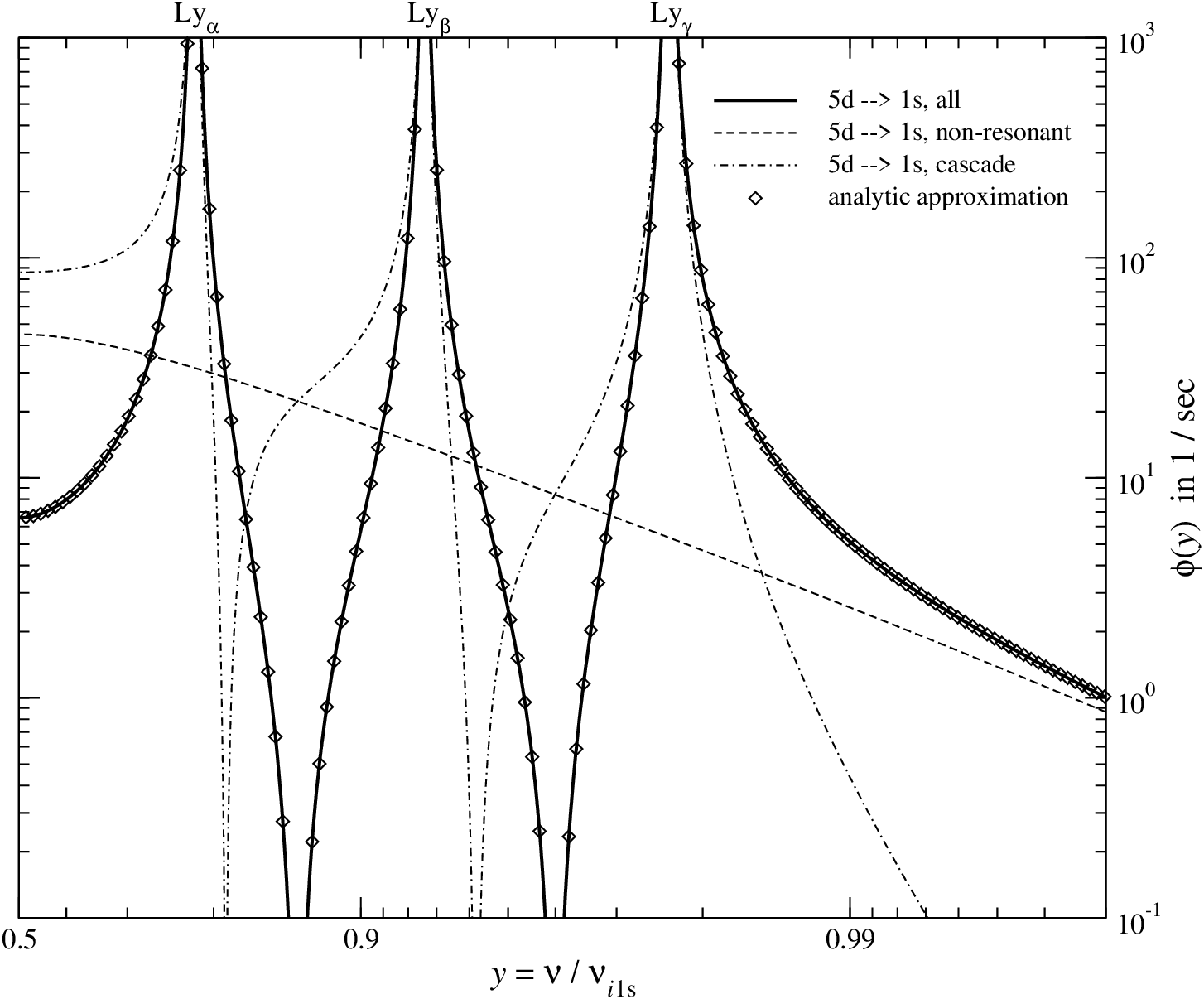}
\caption
{Two-photon emission spectra for the $5{\rm s}\rightarrow 1{\rm s}$ and $5{\rm
    d}\rightarrow 1{\rm s}$ transitions. The non-resonant, cascade, and
    combined spectra are shown as labeled. Also we give the analytic
    approximation based on the equations discussed in Sect. \ref{sec:cas_int},
    together with the fitting formula for the 5s and 5d non-resonant matrix
    element (see Appendix~\ref{app:fits}). The resonances
    correspond to the Brackett-$\alpha$, Paschen-$\beta$, and Balmer-$\gamma$
    transitions at $y<1/2$, and Lyman-$\alpha$, Lyman-$\beta$, and
    Lyman-$\gamma$ at $y>1/2$.}
\label{fig:5s5d_spec}
\end{figure*}
\subsection{The $n$s and $n$d two-photon emission spectrum}
As an example, in Fig. \ref{fig:5s5d_spec} we present the {full} two-photon
emission spectra for the $5{\rm s}~\rightarrow~1{\rm s}$ and $5{\rm
d}~\rightarrow~1{\rm s}$ transition.
Again one can see that the interference term strongly affects the shape of the
spectrum in the wings of the resonances. In particular, destructive
interference close to $y~=~1/2$ strongly reduces the total amplitude of the
emission.
For the 5s-level, interference leads to full cancellation of the photon
production ($y\sim 0.28$ and $y\sim 0.72$) in the region between the innermost
resonances, whereas the photon production does not vanish
within this range for the 5d-level.
This difference is characteristic of the shape of the s and d-two-photon
spectra, also for higher values of $n$.
It is also clear, that using the Eqn. given in Sect. \ref{sec:cas_int},
together with the fitting formulae according to Appendix~\ref{app:fits}, one
can analytically approximate the full two-photon emission spectrum with very
high accuracy in the full range of considered frequencies.

For initial states with a higher value of $n_i$, more resonances (in total
$2\,n_i-4$) appear, but otherwise the spectra look very similar and do not add
any deeper physical aspects.
We checked the analytic approximations for the full two-photon emission
spectrum of several $n{\rm s}~\rightarrow~1{\rm s}$ and $n{\rm
  d}~\rightarrow~1{\rm s}$ two-photon transition up to $n=20$ and always found
excellent agreement with the results from our full numerical treatment.

\section{Total two-photon decay rates}
\label{sec:tot_rate}

\subsection{The two-photon rate for the $2\rm s$-state}
\label{sec:2s}
It is clear that, within the non-relativistic formulation for the
2s$\rightarrow$1s-two-photon transition of hydrogen-like ions, {\it only}
non-resonant contributions to the total lifetime exist.
When we use Eq. \eqref{eq:A_nsnd} together with Eq. \eqref{eq:A_phi_nr} and
  include the first 4000 terms in the infinite sum, we obtain the value $A_{\rm
  2s1s}=8.2293 {\rm s}^{-1}$, which fully agrees with the result of earlier
  computations \citep{Breit1940, Kipper1950, Spitzer1951, Klarsfeld1969,
  Johnson1972, Goldman1981, Drake1986, Goldman1989, Labzowsky2005}.
With the approximate formula for the non-resonant two-photon emission spectrum
for the 2s-state as given in Appendix~\ref{app:fits}, we obtain $A_{\rm
  2s1s}=8.2297 {\rm s}^{-1}$, which shows the high accuracy of the
approximation.

Equation \eqref{eq:A_nsnd}, together with Eq. \eqref{eq:A_phi_nr}, is very
similar to Eq.~(14) in the work of \citet{Cresser1986}. The expression of
\citet{Cresser1986} was obtained using general arguments about the total
lifetime of the considered level, and according to their work it should be
applicable to all s and d-states of hydrogen, yielding the two-photon
correction to the lifetime.
However, if applied to the hydrogen 2s-level one finds $A_{\rm 2s1s}^{\rm
  Cr}=1.4607 {\rm s}^{-1}$ instead of $A_{\rm 2s1s}=8.2293 {\rm s}^{-1}$.
The difference stems from the fact that here, like in the publications
mentioned above, we included the term with $n\equiv n_i$ in the sum
\change{Eq.~\eqref{eq:A_phi_nr}}. This shows that {the largest} contribution
to the total 2s-two-photon decay rate (in this case equivalent to the
non-resonant contribution) actually comes from the transition via the
intermediate $2{\rm p}$-state, i.e.  the matrix element $\left<R_{1{\rm s}}|\,
r\,| R_{2{\rm p}}\right>\left<R_{2{\rm p}}|\, r\,| R_{2{\rm s}}\right>\approx
-6.704$, and cannot be neglected.
This suggests that the arguments by \citet{Cresser1986} are incomplete, or at
least not generally applicable.

Although to our knowledge only rough direct measurements of the two-photon
decay rate exist for the hydrogen 2s-state \citep{Krueger1975, Cesar1996},
one can find experimental confirmations \citep{Prior1972, Kocher1972,
Hinds1978} of the theoretical value for the two-photon decay rate of the
hydrogen-like helium ion ($\Gamma_{\rm th}\approx 8.229 Z^6 {\rm s}^{-1}$),
which do reach percent-level accuracy.
Also measurement for hydrogen-like Ar, F, and O exist \citep{Marrus1972,
  Cocke1974, Gould1983}, but with lower accuracy.
These experimental confirmations further support the idea that, in theoretical
computations of the total two-photon decay rate and in particular the
correction to the one-photon lifetime, it is not enough to consider only
intermediate states with energies $E_n> E_i$.

For hydrogen-like ions, care should be taken when computing the 2s-two photon
decay rate within the relativistic treatment. In this case the 2p$_{1/2}$
level due to the {\it Lamb-shift} and {\it fine-structure splitting}
energetically lies below the 2s$_{1/2}$ level. Increasing $Z$ will make this
shift even bigger, but as the measurements for He and Ar show, this
intermediate state cannot contribute beyond the percent-level to the total
lifetime of the corresponding 2s$_{1/2}$-state.
This is also expected because the lifetime of the 2s-state should not be
strongly altered by the slow 2s$_{1/2}\rightarrow 2{\rm p}_{1/2}$ transition
($\sim 1.6 \times 10^{-9}\,{\rm s^{-1}}$). In addition, the poles due to this
intermediate state lie very close to $\nu\rightarrow 0$ and $\nu\rightarrow
\nu_{i\rm1s}$ and are therefore suppressed by the factors of $\nu^3\nu'^3$ in
Eq.~\eqref{eq:general_nsnd}.

\subsection{The two-photon rates for the $3\rm s$ and $3\rm d$-states}

\subsubsection{The non-resonant contribution}
Using the formula given by \citet{Cresser1986}, i.e. {explicitly}
neglecting the transition via the intermediate 3p-state, we can reproduce
their values for the non-resonant contribution to the two-photon decay rates
of the 3s$\rightarrow$1s and 3d$\rightarrow$1s transitions.
Later \citet{Florescu1988} computed these values again within the framework of
\citet{Cresser1986} but to higher accuracy. We are also able to reproduce
these results ($A^{\rm nr, Cr}_{3{\rm s}\rightarrow 1{\rm s}}~=~8.22581\,{\rm
s}^{-1}$ and $A^{\rm nr, Cr}_{3{\rm d}\rightarrow 1{\rm s}}~=~0.131814\,{\rm
s}^{-1}$) up to {\it all} given figures.
Although the discussion in the previous sections has already shown that these
values probably have no direct relation to the total corrections in the
lifetime of the level due to the two-photon process, we computed them to check
our own computational procedure.

However, returning to our definition of the non-resonant two-photon decay
rate, the transition via the intermediate 3p-state has to be included. We then
obtain $A^{\rm nr}_{3{\rm s}\rightarrow 1{\rm s}}=10.556 (10.558)\,{\rm
s}^{-1}$ and $A^{\rm nr}_{3{\rm d}\rightarrow 1{\rm s}}=7.1474(7.1475)\,{\rm
s}^{-1}$, where the values in parenthesis were computed by integrating our
analytic approximation. In particular, for the 3d-level, this increases the
non-resonant contribution to the total two-photon decay rate by a factor of
$\sim 54$.
If in the sum \eqref{eq:A_phi_nr_b} we only consider the term
$\left<R_{1{\rm s}}|\,r\,| R_{3{\rm p}}\right>\left<R_{3{\rm p}}|\, r \,|
  R_{3{\rm d}}\right>\,f_3\approx-5.199 \,f_3$,
with the function $f_3=y^{-1}+(1-y)^{-1}$ and the integral $\int_0^1 y^3
(1-y)^3 f^2_3\id y=1/6$, then one obtains $A^{\rm nr}_{3{\rm d}\rightarrow 1{\rm
s}}\approx 9.199\,{\rm s}^{-1}$. 
This shows that indeed the main contribution to the non-resonant part of
3d-two-photon decay rate comes from the transition via the intermediate 3p-state.

\subsubsection{The cascade and interference terms}
\label{sec:3s3d_int}
With the formulae given in Sect. \ref{sec:theory}, it should be possible
to compute the total two-photon decay rate of the 3s and 3d-states.
Including the lifetime of the intermediate 2p-state as discussed in
Sect.~\ref{sec:cas_int}, we also computed the total lifetime of the 3s and
3d-states, and, in agreement with \citet{Florescu1984}, \change{obtained
values that were very close to the one expected from} the one-photon lifetime.
But as mentioned in Sect. \ref{sec:cas_int}, within the simple approximation
used to regularize the cascade and interference terms, it is not possible to
compute the total correction to the one-photon lifetime, consistent in the
considered order of the fine-structure constant $\alpha$.
In addition, as we will see in Sect. \ref{sec:astro_app}, this is not
necessary for our \change{cosmological application}.

However, in order to compare with other computations, it may be useful to give
some additional intermediate results.
We therefore also integrated the contribution of the interference term
separately, yielding $A^{\rm int}_{3{\rm s}\rightarrow 1{\rm s}}=-10.810
(-10.810)\,{\rm s}^{-1}$ and $A^{\rm int}_{3{\rm d}\rightarrow 1{\rm
    s}}=-30.019(-30.019)\,{\rm s}^{-1}$.
This shows that, because of interference, the small increase of the decay-rate
due to the non-resonant term (see Table \ref{tab:one}) is completely canceled,
again emphasizing how important the interference term is. In Table
\ref{tab:one} we included a maximal number of summands above the initial
state, which was $n_{\rm sum}=4000$ for $n_i\in\{2, 3\}$ and $n_{\rm
sum}=200$ for $n_i>3$. 

\begin{table}
\caption{The non-resonant contribution to the total two-photon rates for the
transitions $n_i{\rm s}\rightarrow 1{\rm s}$ and $n_i{\rm d}\rightarrow 1{\rm
s}$ up to $n_i=20$.}
\label{tab:one}
\centering
\begin{tabular}{@{}ccc}
\hline
\hline
$n_i$  
& $A^{\rm nr}_{n_i {\rm s}\rightarrow 1{\rm s}}$ & $A^{\rm nr}_{n_i {\rm d}\rightarrow 1{\rm s}}$ \\
\hline
2 & $8.2293\,{\rm s}^{-1}$ & -- 
\\
3 
& $10.556\,{\rm s}^{-1}$
& $7.1474\,{\rm s}^{-1}$
\\
4 
& $11.951\,{\rm s}^{-1}$
& $11.942\,{\rm s}^{-1}$
\\
5 
& $13.741\,{\rm s}^{-1}$
& $15.331\,{\rm s}^{-1}$
\\
6 
& $15.954\,{\rm s}^{-1}$
& $18.004\,{\rm s}^{-1}$
\\
7 
& $18.501\,{\rm s}^{-1}$
& $20.293\,{\rm s}^{-1}$
\\
8 
& $21.301\,{\rm s}^{-1}$
& $22.362\,{\rm s}^{-1}$
\\
9 
& $24.296\,{\rm s}^{-1}$
& $24.296\,{\rm s}^{-1}$
\\
10 
& $27.448\,{\rm s}^{-1}$
& $26.144\,{\rm s}^{-1}$
\\
11 
& $30.722\,{\rm s}^{-1}$
& $27.935\,{\rm s}^{-1}$
\\
\hline
\hline
\end{tabular}
\hspace{2mm}
\begin{tabular}{@{}ccc}
\hline
\hline
$n_i$  
& $A^{\rm nr}_{n_i {\rm s}\rightarrow 1{\rm s}}$ & $A^{\rm nr}_{n_i {\rm d}\rightarrow 1{\rm s}}$ \\
\hline
-- & -- & -- 
\\
12 
& $34.096\,{\rm s}^{-1}$
& $29.687\,{\rm s}^{-1}$
\\
13
& $37.552\,{\rm s}^{-1}$
& $31.410\,{\rm s}^{-1}$
\\
14 
& $41.076\,{\rm s}^{-1}$
& $33.114\,{\rm s}^{-1}$
\\
15
& $44.659\,{\rm s}^{-1}$
& $34.801\,{\rm s}^{-1}$
\\
16 
& $48.290\,{\rm s}^{-1}$
& $36.478\,{\rm s}^{-1}$
\\
17 
& $51.964\,{\rm s}^{-1}$
& $38.146\,{\rm s}^{-1}$
\\
18 
& $55.674\,{\rm s}^{-1}$
& $39.806\,{\rm s}^{-1}$
\\
19 
& $59.416\,{\rm s}^{-1}$
& $41.462\,{\rm s}^{-1}$
\\
20 
& $63.185\,{\rm s}^{-1}$
& $43.113\,{\rm s}^{-1}$
\\
\hline
\hline
\end{tabular}
\end{table}
\subsection{The two-photon rates for the $n\rm s$ and $n\rm d$-states}
For future computations and {more} complete considerations of the
higher order correction to the lifetime of the $n$s and $n$d-states, here we now
give the results for the total contribution of the non-resonant term to the
two-photon decay rate. This contribution does not depend on the treatment of
the poles in the cascade and interference terms.
However, these values should have no direct relation to the total two-photon
correction of the lifetime, but are mainly meant for cross-checks.

\subsubsection{The non-resonant contribution}
\label{sec:nr_large_n}
In Table~\ref{tab:one} we summarize the values for the non-resonant
contribution to the two-photon rates for the $n$s and $n$d-states up to
$n_i=20$.
%
\begin{figure}
\centering 
\includegraphics[width=0.96\columnwidth]
{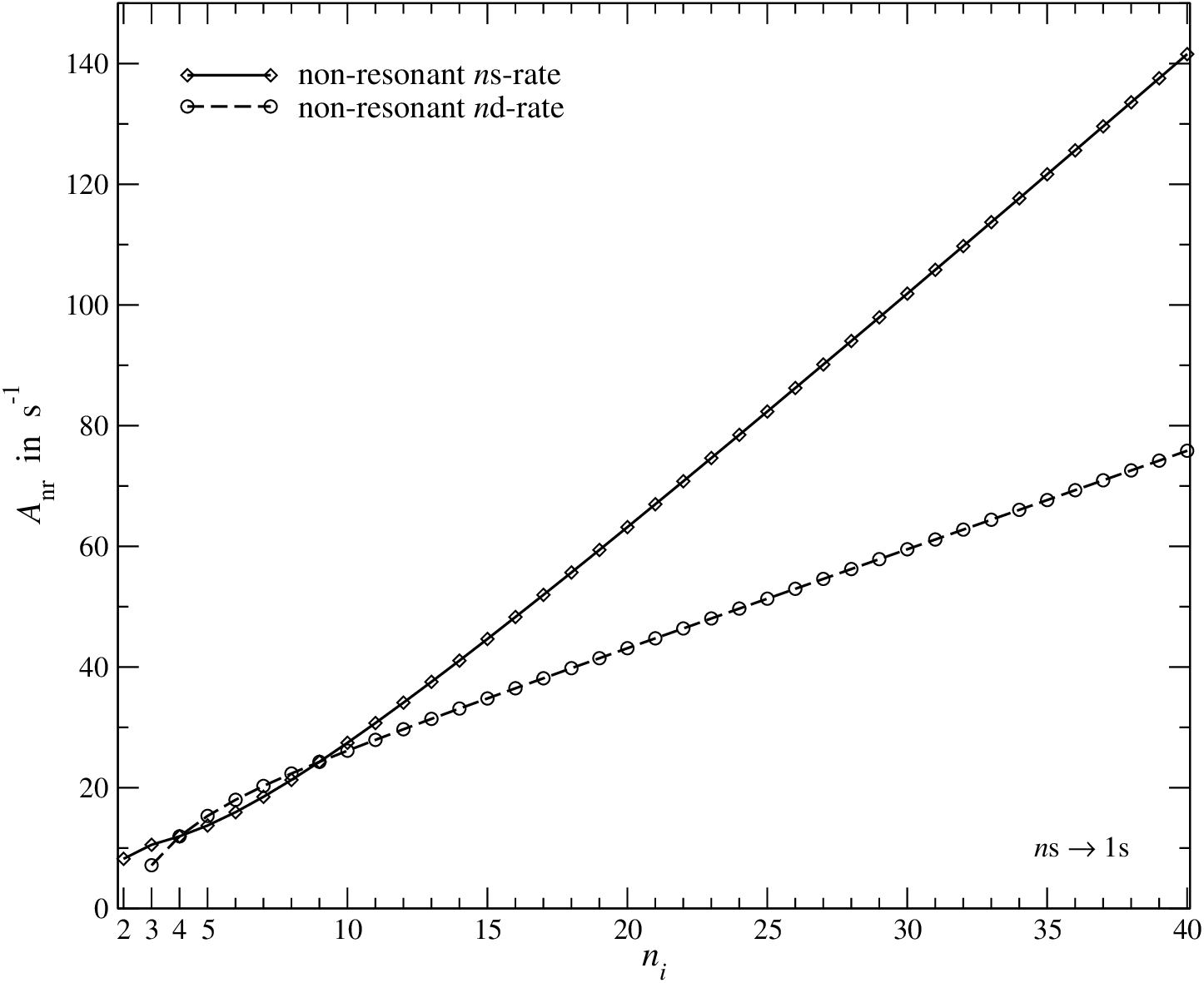}
\caption
{Non-resonant contribution to the total two-photon decay rate in vacuum for
  the $n$s and $n$d-states of the hydrogen atom. The results were computed
  using the first 200 terms above $n_i$.}
\label{fig:one}
\end{figure}
The dependence of the non-resonant contribution to the total two-photon decay
rate on $n_i$ is presented in Fig.~\ref{fig:one}. For large $n_i$ in both
cases, the rates scale roughly linear, increasing towards larger $n_i$.
The slope is slightly steeper for the s-states.
We find that for $n_i\gtrsim 20$ one can use
\bsub
\label{eq:A_lin}
\beal
A^{\rm nr}_{n_i {\rm s}\rightarrow 1{\rm s}}&\approx -15.857\,{\rm s}^{-1}+3.930\,{\rm s}^{-1}\,n_i
\\
A^{\rm nr}_{n_i {\rm d}\rightarrow 1{\rm s}}&\approx 10.432\,{\rm s}^{-1}+1.636\,{\rm s}^{-1}\,n_i
\end{align}
\esub
within percent accuracy up to $n_i\sim 40$. 
\change{Explicitly computing} the values for $n_i=50$ ($n_{\rm sum}=200$), we
  find $A^{\rm nr}_{50{\rm s}\rightarrow 1{\rm s}}=181.74\,{\rm s}^{-1}$ and
  $A^{\rm nr}_{50{\rm d}\rightarrow 1{\rm s}}=92.19\,{\rm s}^{-1}$, using our
  full numerical treatment, and $180.64\,{\rm s}^{-1}$ and $92.23\,{\rm
  s}^{-1}$ with the approximations \eqref{eq:A_lin}.
We did not check up to which value of $n_i$ the formulae \eqref{eq:A_lin} are
applicable. Also, one should bear in mind that, above some value of $n_i\gg
1$, the \change{usual} dipole approximation for the transition matrix elements
breaks down \citep{Dubrovich2005, SwitzerHirata2007II} and other methods
should be used.

The linear scaling of the non-resonant contribution to the two-photon decay
rate for $n_i\gg 1$ was expected \citep{Dubrovich1987, Dubrovich2005}, but
here we have included all virtual intermediate states in the sum. However,
one should keep in mind that, due to the interference term, it is difficult to
interpret this contribution separately.


\section{Astrophysical application}
\label{sec:astro_app}

\subsection{Two-photon process in the single photon picture}
\label{sec:1and1_gamma}
As {described} in the introduction, the standard procedure for treating the atomic
transitions of electrons involving more than one photon is to break them down
into independent, single-step, one-photon processes. This approximation should
be especially good in the presence of many perturbing particles (free
electrons and ions), such as in {stellar atmospheres}, which destroy
the {\it coherence} of processes involving more than one transition.
Here we now explain how the two-photon process can be formulated in the
simplified $1+1$-single photon picture.

\subsubsection{Distribution of the high frequency photon}
\label{sec:low_phot}
As an example, we consider the decay of the 3s-level {in vacuum}.
If there are no perturbing particles, two photons will be released and the
emission profile (see Fig.~\ref{fig:3s3d_spec}) is described by the two-photon
formulae discussed in the previous sections.
In the $1+1$-single photon picture, with very high probability the electron
after a short time ($\sim \pot{1.6}{-7}\,{\rm s}$) decays to the 2p-state,
emitting a photon close to the Balmer-$\alpha$ frequency. Then it
independently releases a second photon, for which the frequency distribution,
in the rest frame of the atom, is given by the natural line profile. Therefore
the number of photons appearing per second in the frequency interval $\nu$ and
$\nu+\id \nu$ in the vicinity of the Lyman-$\alpha$ resonance due to the
transition from the 3s-state is given by\footnote{\change{The factor of
$1/\pi$ is required due to the normalization of $L(a, b)$.}}
\beal 
\label{eq:phi_3s2p1s}
\phi^{3\rm s, 1+1\gamma}_{\rm 2p\rightarrow 1s}(\nu)\,\id\nu
=\frac{A^{1\gamma}_{\rm 3s\rightarrow
2p}}{\pi}\,L\left(\frac{\Gamma^{1\gamma}_{\rm 2p\rightarrow 1s}}{4\pi},
\nu-\nu_{\alpha}\right)\,\id\nu,
\end{align}
where $L(a, b)$ is defined in Eq. \eqref{eq:L} and $\nu_{\alpha}$ is the
Lyman-$\alpha$ central frequency. 
Integrating over $\nu$ yields $\int\phi^{3\rm s, 1+1\gamma}_{\rm 2p\rightarrow
  1s}(\nu)\,\id\nu=A^{1\gamma}_{\rm 3s\rightarrow 2p}$, i.e. the total rate at
which electrons are added to the 2p-state after the transition from the
initial 3s-level.
\change{Note that here we have assumed $\Gamma_{\rm
    2p}\approx\Gamma^{1\gamma}_{\rm 2p\rightarrow 1s}$.}

For \eqref{eq:phi_3s2p1s} one assumes that there is no coherence or
correlation between the first and second photon, and consequently the
Lyman-$\alpha$ line-profile in the $1+1$-photon picture is a pure Lorentzian
up to very large distances from the resonance.
\change{This is also equivalent to assuming that the transition from the
  3s-state leads to a `natural' distribution of electrons within the 2p-state
  \citep{Mihalas1978}.}
One can also obtain this result using the interpretation of
\citet{Weisskopf1930} for the line width.

Looking at other initial s or d-states, the same argument as above can be
carried out. In the more general case, one simply has to replace
$A^{1\gamma}_{\rm 3s\rightarrow 2p}$ with the corresponding partial
spontaneous decay rate $A^{1\gamma}_{n\rm s/d\rightarrow 2p}$ to the 2p-state.
This shows that no matter what is the initial level, the shape of the
$1+1$-emission profile in the vicinity of the Lyman-$\alpha$ resonance is
always a Lorentzian. Within the $1+1$-single photon picture, the same is true
for the other possible intermediate resonances (e.g Lyman-$\beta$, $\gamma$,
etc) in the two-photon cascades from high initial s or d-states. However,
there in addition the partial width of the 2p-state due to the transition to
the ground level appearing in Eq. \eqref{eq:phi_3s2p1s} has to be replaced by
the corresponding \change{total (one-photon) width of the intermediate
p-state. Also one has to take into account the branching ratio for transitions
leading directly to the ground state.}

With these comments one then can write
\beal 
\label{eq:phi_ndnp1s}
\phi^{n_i\rm s/d, 1+1\gamma}_{n\rm p\rightarrow 1s}(\nu)\,\id\nu =\frac{A^{1\gamma}_{n_i\rm
    s/d\rightarrow{\it n} p}}{\pi}
\,\frac{\Gamma^{1\gamma}_{n\rm p\rightarrow 1s}}{\Gamma^{1\gamma}_{n\rm p}}\,
L\left(\frac{\Gamma^{1\gamma}_{n\rm p}}{4\pi}, \nu-\nu_{n\rm 1s}\right)\,\id\nu,
\end{align}
where $\nu_{n\rm 1s}$ is the central frequency of the corresponding
\change{Lyman-series transition}.

\subsubsection{Distribution of the low-frequency photon}
In Sect. \ref{sec:low_phot} we have focused on the high-frequency photons
released in the two-photon cascade. If we now consider the low-frequency
photons, then the profiles of these will be given by
\beal 
\label{eq:phi_ndntp}
\phi^{1+1\gamma}_{n_i\rm s/d \rightarrow  {\it n}\rm p}(\nu)\,\id\nu =\frac{A^{1\gamma}_{n_i\rm
    s/d\rightarrow{\it n} p}}{\pi}\,\frac{\Gamma^{1\gamma}_{n\rm p\rightarrow
    1s}}{\Gamma^{1\gamma}_{n\rm p}}\,L\left(\frac{\Gamma^{1\gamma}_{n\rm p}}{4\pi}, \nu-\nu_{n_i n \rm p}\right)\,\id\nu,
\end{align}
where $\nu_{n_i n \rm p}=\nu_{n_i \rm 1s}-\nu_{n\rm 1s}$ is the transition
frequency from the initial $n_i$s or $n_i$d-state to the $n$p-state.

Here one may ask why the width of the line is determined by the width of the
intermediate $n$p-state only and not by $\Gamma^{1\gamma}_{n\rm
p}+\Gamma^{1\gamma}_{n_i\rm s/d}$ as usual.
We simply wanted to be consistent with the approximate treatment of the
cascade and interference terms in the full two-photon formulation (see
Sect.~\ref{sec:cas_int}), for which the width of the initial state was
neglected.
\change{As mentioned above,} physically this is motivated by the idea that,
within the formulation of \citet{Weisskopf1930}, we consider one particular
initial `energy sub-level' and do not specify the process that populated it.
Therefore the final profile is independent of the shape of the distribution of
energy-sub-levels around the mean energy of the initial state.
One can also consider this as equivalent to neglecting any possible
reshuffling of the electron by perturbing particles while it is in the
initial state.
Furthermore, in general $\Gamma^{1\gamma}_{n\rm p}>\Gamma^{1\gamma}_{n_i\rm
  s/d}$ such that $\Gamma^{1\gamma}_{n_i\rm s/d}$ \change{would} not
  contribute much to the total width of the line. \change{But most
  important}, in our computations we do not approach the resonances so
  close that these differences would play any role.

\begin{figure*}
\centering 
\includegraphics[width=0.93\columnwidth]
{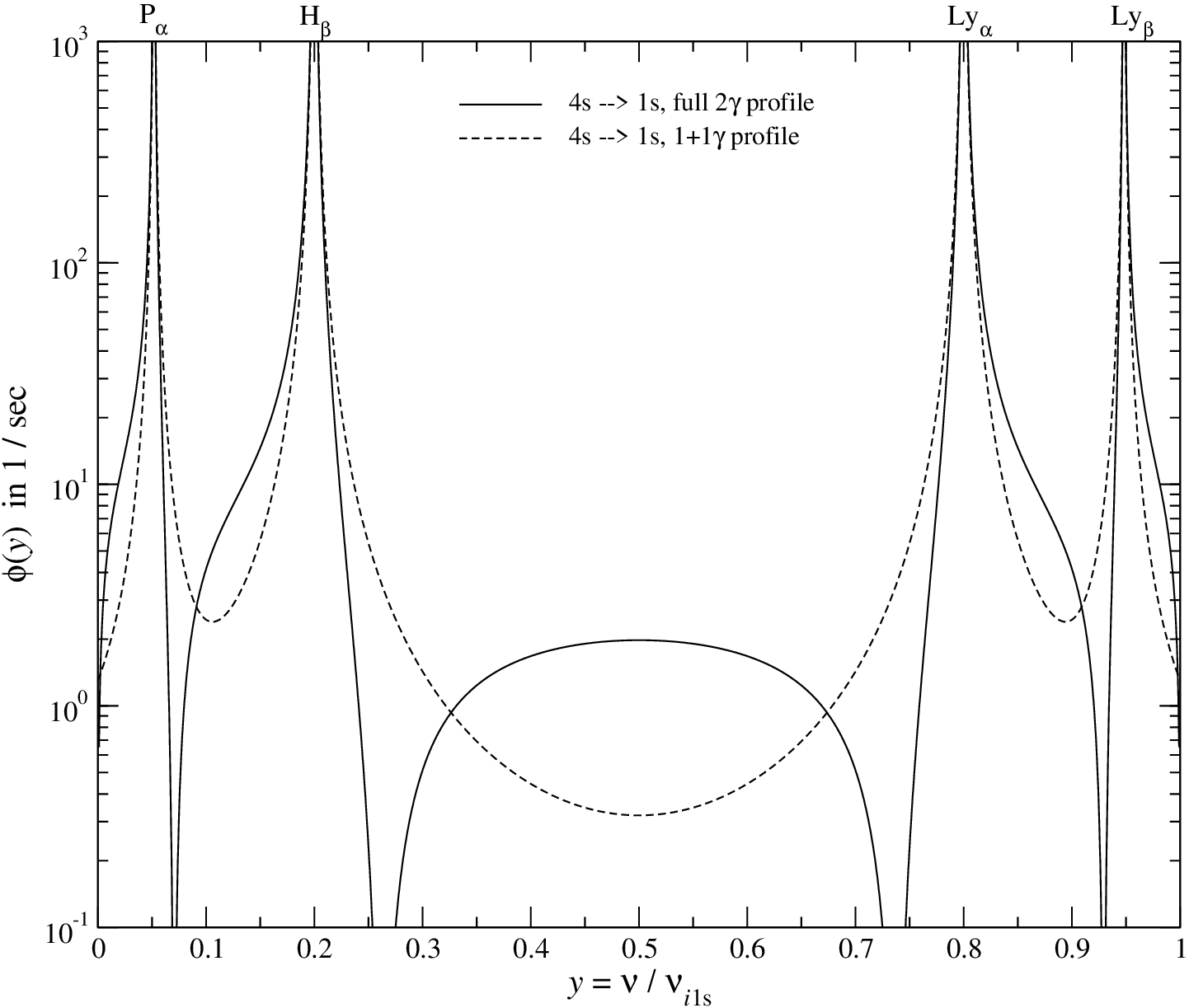}
\hspace{1mm}
\includegraphics[width=0.93\columnwidth]
{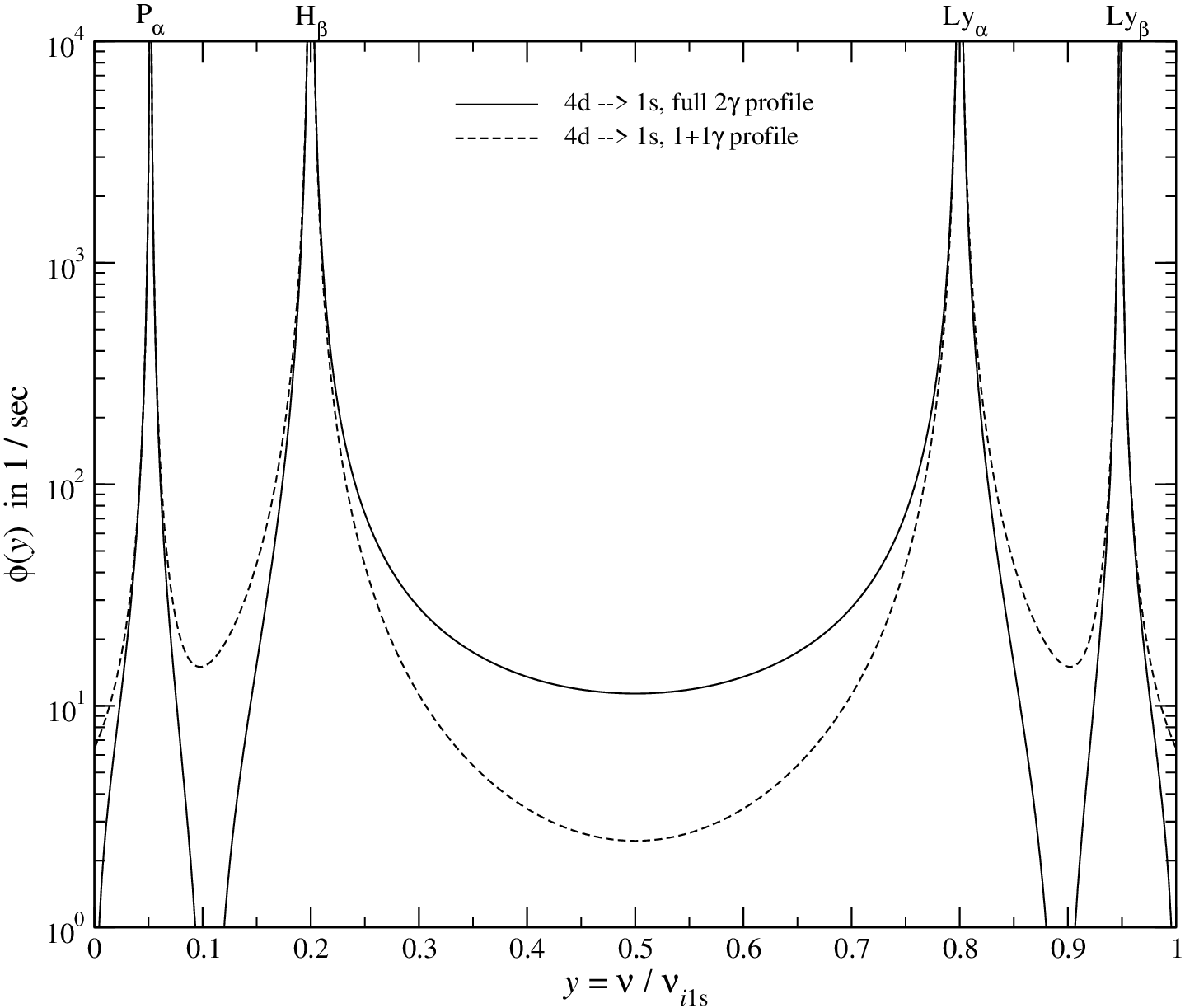}
\caption
{\change{Comparison of the two-photon emission profiles for the 4s and 4d
    states. We show, $\phi^{2\gamma}_{4\rm s/d\rightarrow 1s}$, following from
    the full $2\gamma$ treatment according to Eq.~\eqref{eq:phi_2g}, and,
    $\phi^{1+1\gamma}_{4\rm s/d \rightarrow 1s}$, using the $1+1$-single
    photon description, as given by Eq.~\eqref{eq:phi_11g}.} The first 200
  terms above $n=4$ were included in the sum for $\phi^{2\gamma}_{4\rm
    s/d\rightarrow 1s}$. }
\label{fig:4s4d_1p1}
\end{figure*}
\subsubsection{Total two-photon profile in the single photon picture}
With Eqn. \eqref{eq:phi_ndnp1s} and \eqref{eq:phi_ndntp} it is now possible to
write the total distribution of \change{photons} emitted in the two-photon
decay of an isolated hydrogen atom in some given initial s or d-level within
the $1+1$-single photon formulation:
\beal 
\label{eq:phi_11g}
\phi^{1+1\gamma}_{n_i\rm s/d \rightarrow 1s}(\nu)
&=
\sum_{n=2}^{n_i-1}
\left[
\phi^{1+1\gamma}_{n_i\rm s/d \rightarrow  {\it n}\rm p}(\nu)
+
\phi^{n_i\rm s/d, 1+1\gamma}_{n\rm p\rightarrow 1s}(\nu)
\right].
\end{align}
\change{Integrating $\phi^{1+1\gamma}_{n_i\rm s/d \rightarrow 1s}(\nu)$ over
  all possible frequencies yields $\frac{1}{2}\int \phi^{1+1\gamma}_{n_i\rm
  s/d \rightarrow 1s}(\nu)\id \nu\equiv\sum_{n=2}^{n_i-1} A^{1\gamma}_{n_i\rm
  s/d\rightarrow{\it n} p}\,\frac{\Gamma^{1\gamma}_{n\rm p\rightarrow
  1s}}{\Gamma^{1\gamma}_{n\rm p}}$, which is the total contribution to the
  width of the initial level due to spontaneous transitions that lead to the ground
  state and involve exactly two photons (cf.  Eq.~\eqref{eq:Gamma_n}).}

What are the main differences \change{of the $1+1$-photon profile} with
respect to the full two-photon profile, as defined by Eq. \eqref{eq:phi_2g}?
\begin{itemize}
  
\item[(i)] There is no non-resonant contribution, resulting from virtual
  transitions via intermediate states with energies higher than {or
    equal to} the initial state.
  
\item[(ii)] As a consequence of (i) there is no
  \change{resonance/non-resonance} interference term.
  
\item[(iii)] In contrast to Eq. \eqref{eq:M2_cas_inf_a}, there is no
  interference among the resonances.
  
\item[(iv)] {As a consequence of (ii) and (iii) each resonance has the shape
  of a Lorentzian up to very large distances from their line centers.}
  
\item[(v)] Usually one does not restrict the range of integration to the
  interval $0\leq \nu\leq \nu_{n\rm 1s}$.

\end{itemize}
It is also important that interpreting each resonance appearing in
$\phi^{1+1\gamma}_{n_i\rm s/d \rightarrow 1s}(\nu)$ separately, it is possible
to uniquely define the rates at which electrons flow in and out of a
particular intermediate p-state.
In astrophysical computations this is the usual procedure for solving the
radiative transfer problem for each \change{transition} separately.
Within the full two-photon formulation this is only possible very close to the
centers of the resonances \change{(where the contribution from the other terms
is negligible)}, but in the wings, photons from different intermediate
transitions contribute non-trivially and make this separation difficult.
\change{Also this overlap of the resonances is taken into account in
  Eq.~\eqref{eq:phi_11g}, but it is usually neglected in astrophysical
  computations.}

\change{
  As an example, we illustrate the differences in the two-photon emission
  profiles for the initial 4s and 4d states in Fig. \ref{fig:4s4d_1p1}. }
One can see that in the distant wings of all the resonances the differences of
the profiles are rather big.
This is mainly due to the non-resonant term and its interference with the
cascade contribution, but also the \change{resonance/resonance} interference plays some
role.
Below we now focus our analysis on the deviations of the two-photon
profile from the pure Lorentzian close to the Lyman-$\alpha$ resonance.  
These differences are the main reason for the corrections to the hydrogen
recombination history.

\begin{figure*}
\centering 
\includegraphics[width=0.93\columnwidth]
{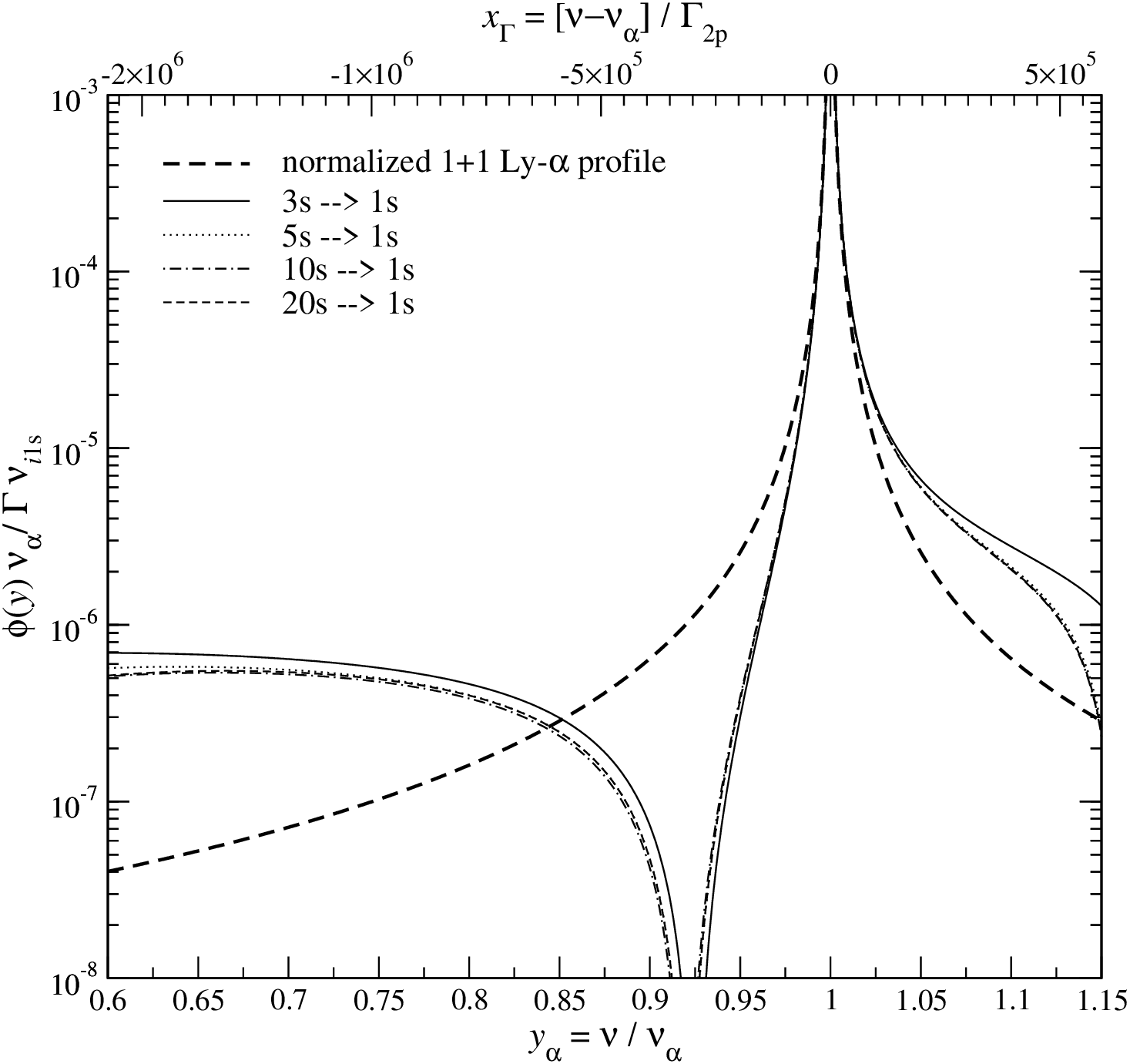}
\hspace{1mm}
\includegraphics[width=0.93\columnwidth]
{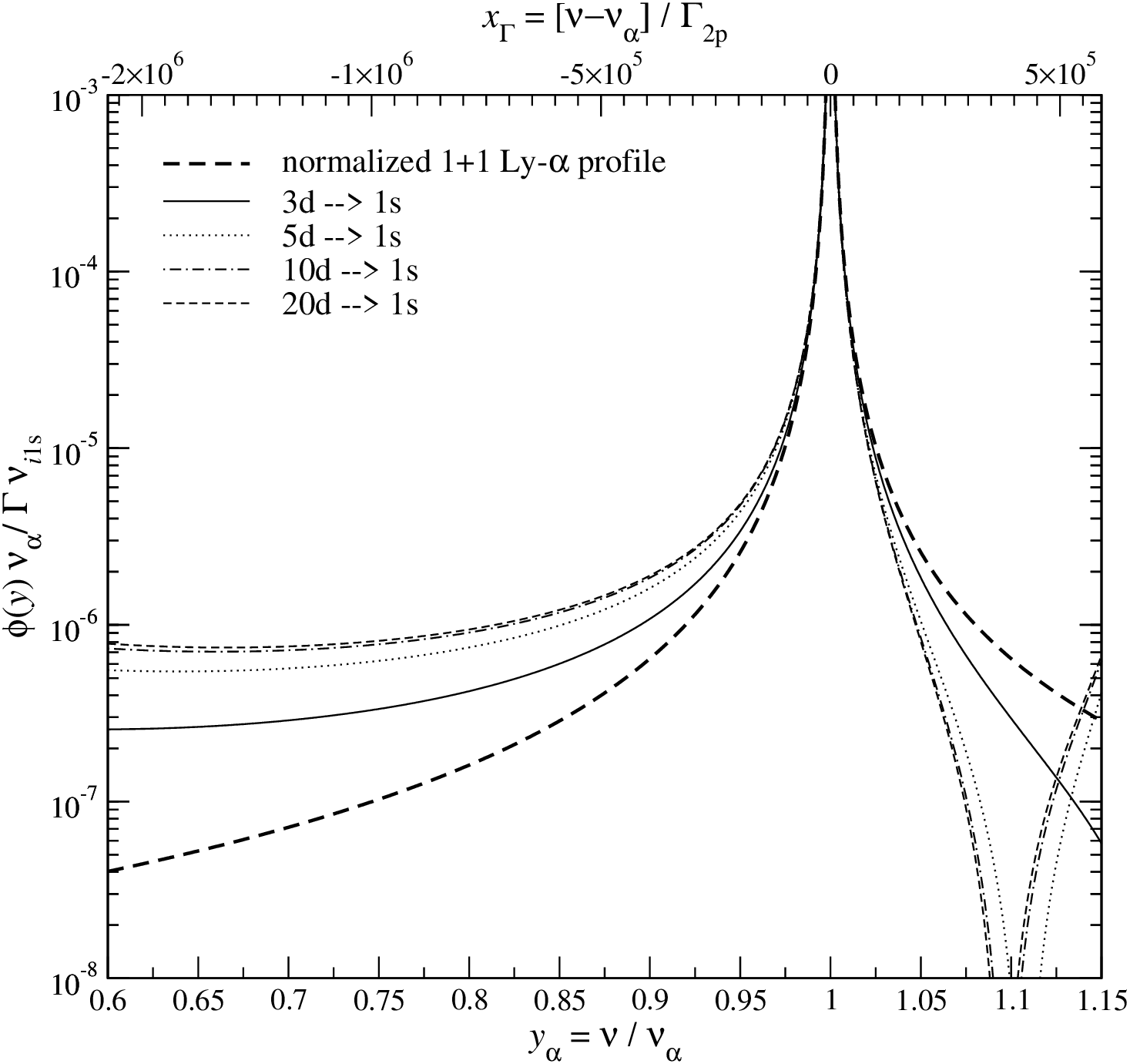}
\caption
{{Normalized $1+1$-two-photon profile $\phi^{n\rm s/d, 1+1\gamma}_{2\rm
      p\rightarrow 1s}$, see Eq. \eqref{eq:phi_ndnp1s}, close to the
    Lyman-$\alpha$ frequency in comparison with the full two-photon-profiles,
    $\phi^{2\gamma}_{n\rm s/d\rightarrow 1s}$ according to Eq.
    \eqref{eq:phi_2g}, for several initial s and d-states. We divided the
    two-photon spectra by their partial one-photon transition rate to the
    2p-state ($\Gamma_{n\rm s/d \rightarrow 2p}$) and transformed to the
    variable $y_\alpha=\nu/\nu_{\alpha}$. For the given curves we included the
    first 500 terms above $n_i$.
\change{The dashed curve corresponds to the usual Lorentzian profile of the
Lyman-$\alpha$ transition.}
}}
\label{fig:Lorentz_nsnd}
\end{figure*}
\begin{figure*}
\centering 
\includegraphics[width=0.93\columnwidth]
{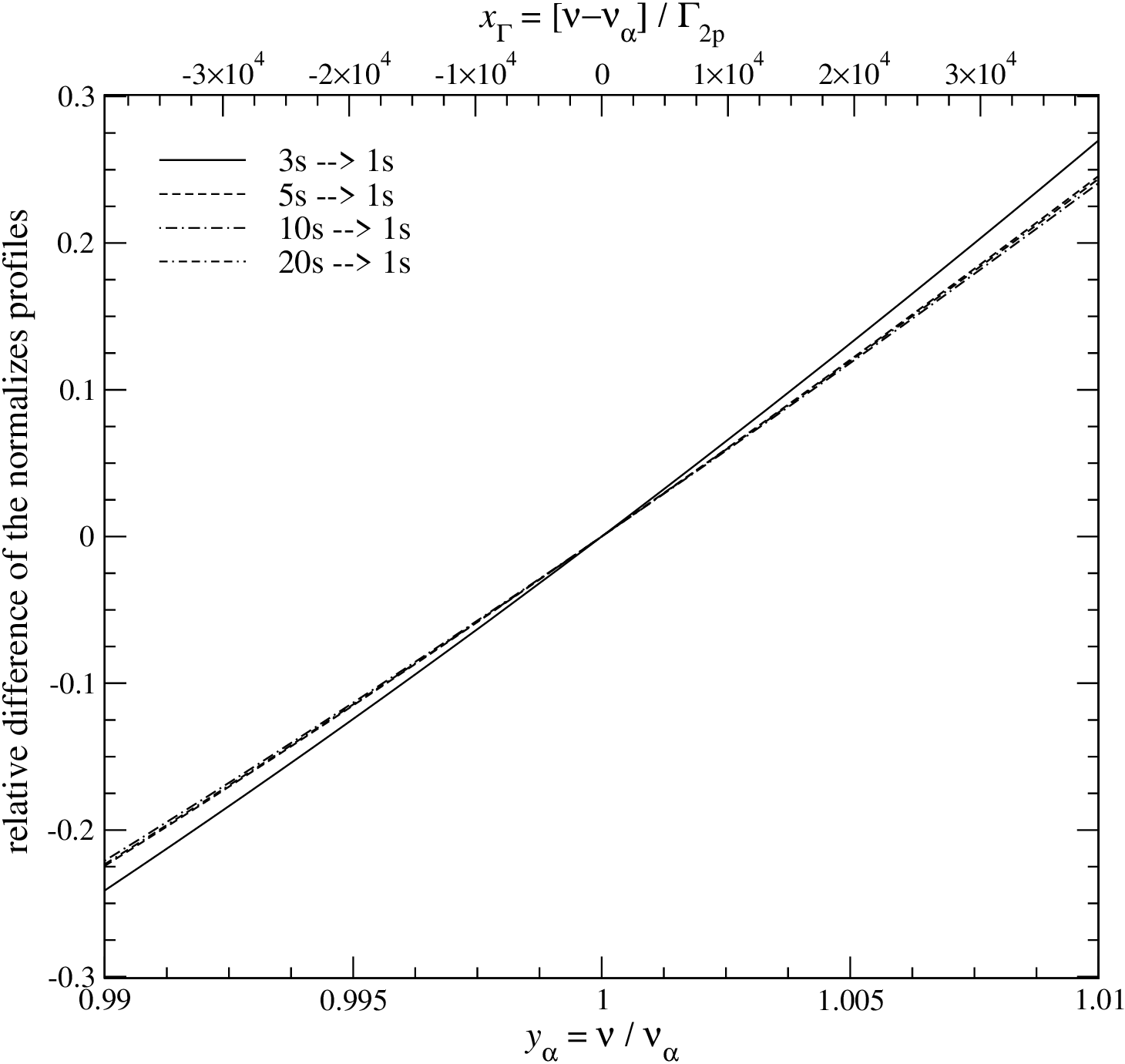}
\hspace{1mm}
\includegraphics[width=0.93\columnwidth]
{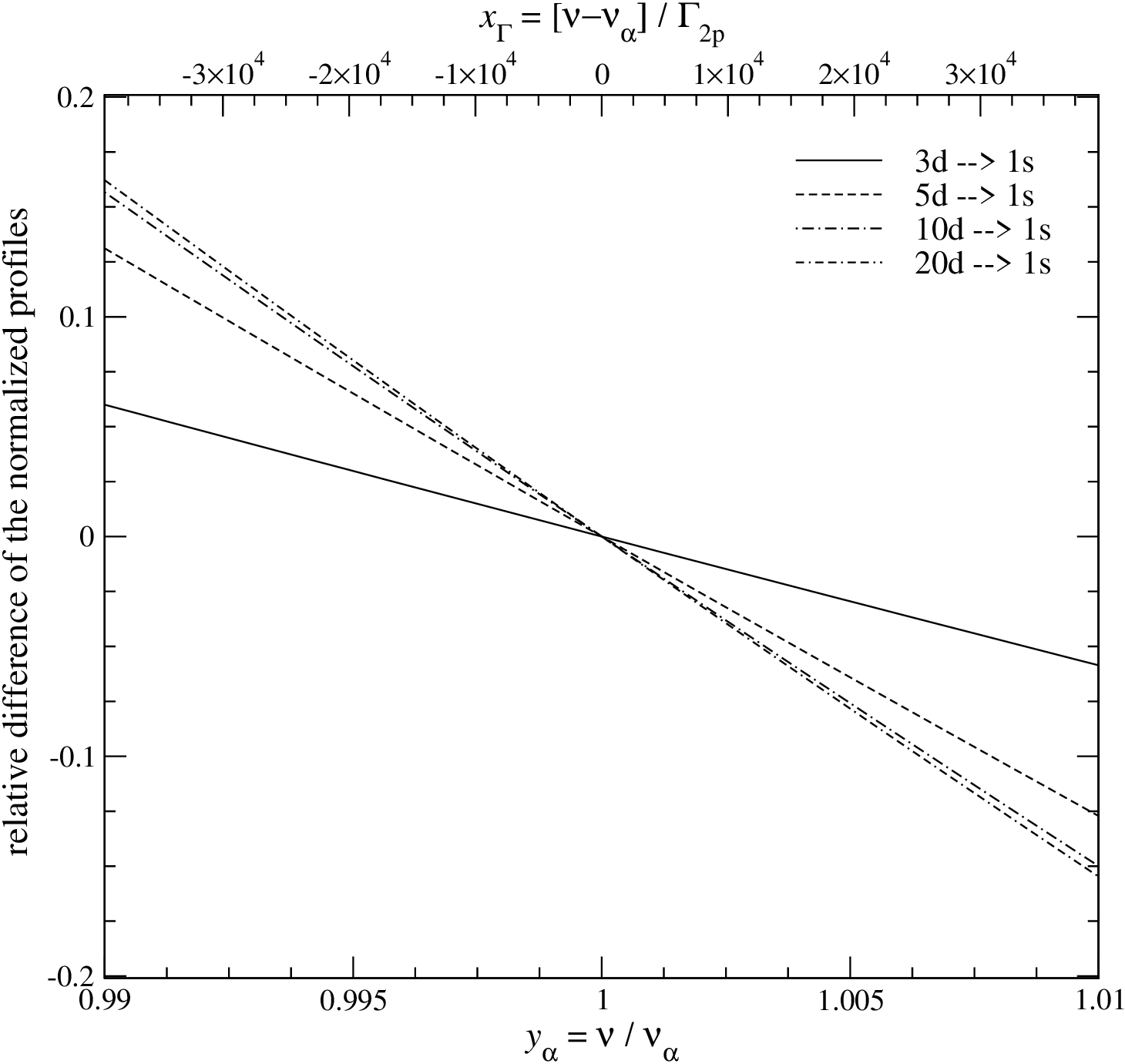}
\caption
{Relative difference, $[\nu_\alpha\,\phi^{2\gamma}_{n\rm s/d \rightarrow 1{\rm
s}}(y_\alpha)/\nu_{n\rm 1s}-\phi^{n\rm s/d, 1+1\gamma}_{\rm 2p\rightarrow 1s}(y_\alpha)]/\phi^{n\rm s/d,
1+1\gamma}_{\rm 2p\rightarrow 1s}(y_\alpha)$, of the curves given in
Fig. \ref{fig:Lorentz_nsnd} with respect to normalized $1+1$-two-photon
profile. Very close to $y_\alpha\sim 1$, the curves should be considered as
extrapolated estimates.}
\label{fig:DLorentz_nsnd}
\end{figure*}
\subsection{Two-photon profiles close to the Lyman-$\alpha$ resonance}
\label{sec:dev_wing}
In low-density plasmas like the expanding Universe during cosmological
hydrogen recombination, hardly any perturbing particle is within the
Weisskopf-radius \citep{Weisskopf1932, Sobel1995}, so that the coherence in
two-photon and possibly multi-photon transitions is maintained at least for the
lower shells.  \change {
  In astrophysical computations the frequency distribution of photons released
  in the Lyman-$\alpha$ transition due to electrons reaching the 2p-state from
  higher levels is usually described by a pure Lorentzian. Within the
  interpretation of \citet{Weisskopf1930}, this means that the electron is
  completely reshuffled among all the possible 2p energy sub-levels.

  In calculations of the cosmological hydrogen recombination problem, we are
  now interested in the deviations of the full two-photon profile from
  the normal Lorentzian shape.
} \change{Here one should mention that in general the deviations of the
  $2\gamma$-emission profile, using the full two-photon treatment as described
  in Sect.~\ref{sec:theory}, from the one in the $1+1$-single photon
  description (Eq.~\eqref{eq:phi_11g}) should also be considered. However, in
  the red wing of the Lyman-$\alpha$ resonance, one can write
\beal 
\label{eq:phi_11g_wing}
\phi^{1+1\gamma}_{n_i\rm s/d \rightarrow 1s}
&\approx
\sum_{n=2}^{n_i-1}
\left[
\frac{A^{1\gamma}_{n_i\rm s/d\rightarrow{\it n} p}}{4\pi^2}\,\frac{\Gamma^{1\gamma}_{n\rm p\rightarrow
    1s}}{\Big[\nu-\nu_{n_i n \rm p}\Big]^2}
+
\frac{A^{1\gamma}_{n_i\rm s/d\rightarrow{\it n} p}}{4\pi^2}\,\frac{\Gamma^{1\gamma}_{n\rm p\rightarrow
    1s}}{\Big[\nu-\nu_{n \rm 1s}\Big]^2}
\right]
\nonumber\\
&\approx\frac{A^{1\gamma}_{n_i\rm s/d\rightarrow 2p}\,\Gamma^{1\gamma}_{\rm
    2p\rightarrow 1s}}{4\pi^2\,\Big[\nu-\nu_{\rm 2p1s}\Big]^2},
%
\end{align}
where the last step is possible, since the distant wings of all the other
resonances do not contribute significantly when one is close enough to the
center of the Lyman-$\alpha$ transition. The more one approaches the
Lyman-$\alpha$ resonance, the better this approximation becomes. For the
estimates carried out below, this approximation is satisfactory.}

To understand the deviations of the two-photon emission profiles close to
the Lyman-$\alpha$ resonance, we now directly compare $\phi^{2\gamma}_{n\rm
  s/d\rightarrow 1s}$ according to Eq.~\eqref{eq:phi_2g} with $\phi^{n_i\rm
  s/d, 1+1\gamma}_{2\rm p\rightarrow 1s}$ as given by
Eq.~\eqref{eq:phi_ndnp1s}.
For convenience we choose $y_\alpha=\nu/\nu_\alpha$ as the common frequency
variable. Then the full two-photon profile in this new coordinate is given by
$\phi^{2\gamma}_{\alpha}(y_\alpha)=\nu_\alpha \phi^{2\gamma}_{n\rm
  s/d\rightarrow 1s}(y)/\nu_{i\rm 1s}$.
The axis of symmetry is then at $y_\alpha=\frac{2}{3}[n^2-1]n^{-2}$ instead of
$y=1/2$.
Since in the vicinity of any particular resonance all the two-photon profiles
scale like $\Gamma_{n\rm s/d\rightarrow {\it n'}p}$, focusing on the
Lyman-$\alpha$ transition, we also re-normalized by $\Gamma_{n\rm
  s/d\rightarrow 2p}$.
In Fig. \ref{fig:Lorentz_nsnd} we give the normalized $1+1$-two-photon profile
in the vicinity of the Lyman-$\alpha$ transition in comparison with the
re-normalized two-photon-profiles for several initial s and d-states.
One can see that at large distances the two-photon profiles \change{in the
  full $2\gamma$-treatment} deviate a lot from the Lorentzian shape.
For both the initial s and d-states, the very distant red wing is several times
above the Lorentzian.
Within the frequency range $0.9\lesssim y_\alpha \lesssim 1.1$ for the
s-states, the red wing lies below, the blue wing above the Lorentzian, whereas
the opposite is true for the d-states.
In particular for the d-states, the red wing is always above the Lorentzian,
and unlike the s-states \change{in the considered frequency range} there is no
additional zero below the Lyman-$\alpha$ resonance.
\change{In Fig.~\ref{fig:Lorentz_nsnd} one can also see that for the chosen
  set of coordinates,} the variation in the profiles is rather small in the
  case of initial s-states, and the modifications become \change{negligible}
  even for initial d-states above $n\sim 10$.

In Fig. \ref{fig:DLorentz_nsnd} we show the relative difference of the curves
given in Fig. \ref{fig:Lorentz_nsnd} with respect to the Lorentzian \change{of
the Lyman-$\alpha$ resonance}. The wing redward of the Lyman-$\alpha$
frequency lies below the Lorentzian for initial s-states, exceeding the level
of $\sim 10$\% at more than $\sim \pot{1.6}{4}$ natural width from the center.
For initial d-states, in all shown cases the wing redward of the Lyman-$\alpha$
frequency lies above the Lorentzian.
The relative correction to the Lorentzian scales roughly linearly with
$\Delta\nu=\nu-\nu_\alpha$ in this regime. Therefore the net change in the
rate of photon production in the red wing of the Lyman-$\alpha$ transition at
frequencies in the range $\nu_1\lesssim \nu\lesssim \nu_2$ depends
logarithmically on the ratio of $\nu_1$ and $\nu_2$: $\Delta
N_\gamma\propto\int_{\nu_1}^{\nu_2}
\frac{1}{[\nu-\nu_\alpha]^2}\times\Delta\nu\id\nu\propto\log[\nu_2/\nu_1]$.
Here we used the wing approximation of the Lorentzian $L\propto
1/[\nu-\nu_\alpha]^2$. This estimate shows that the value of the effective
two-photon decay rate does not depend very strongly on $\nu_2$ (see
Sect.~\ref{sec:recombination}).

\begin{figure*}
\centering 
\includegraphics[width=0.93\columnwidth]
{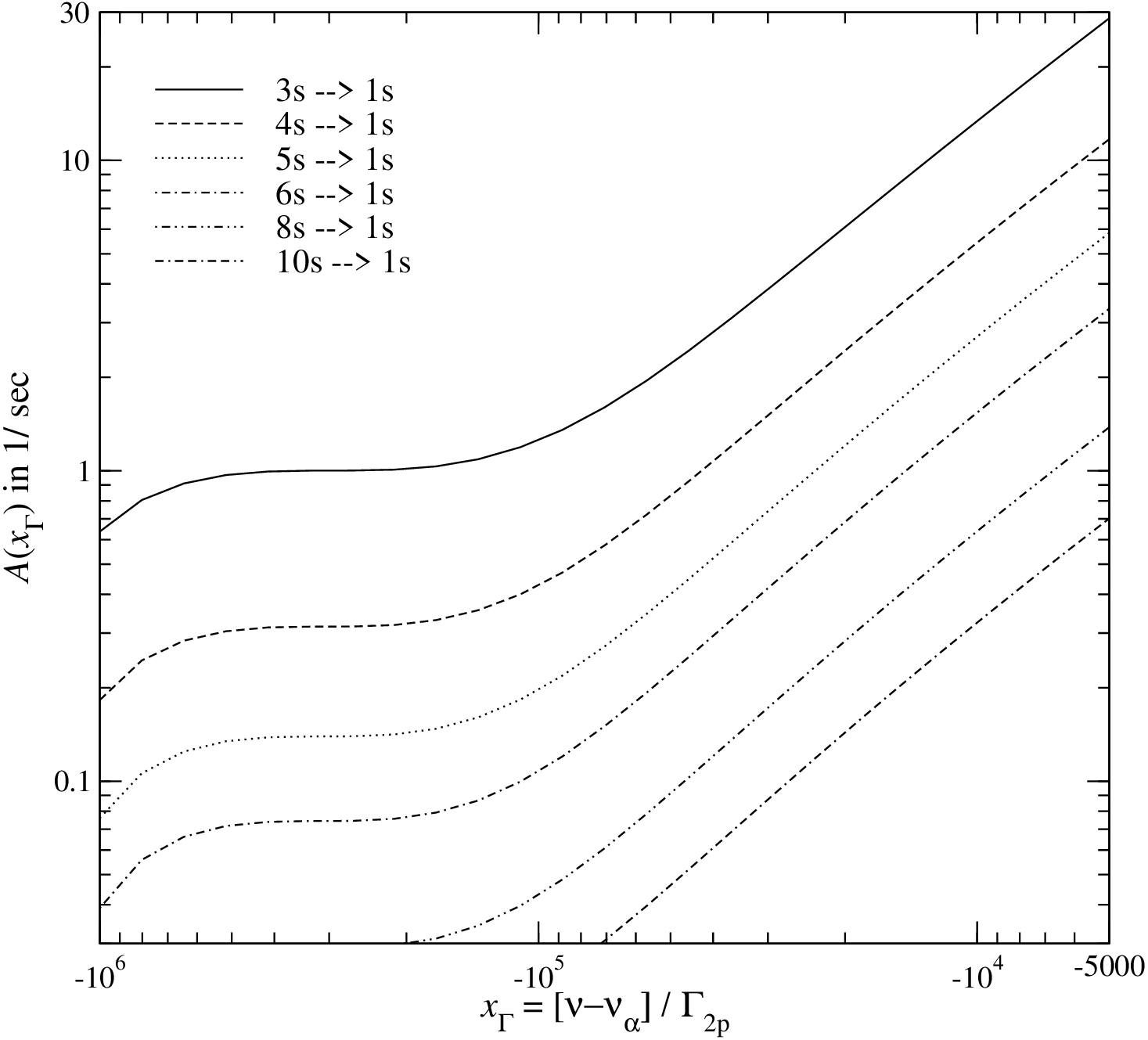}
\hspace{1mm}
\includegraphics[width=0.93\columnwidth]
{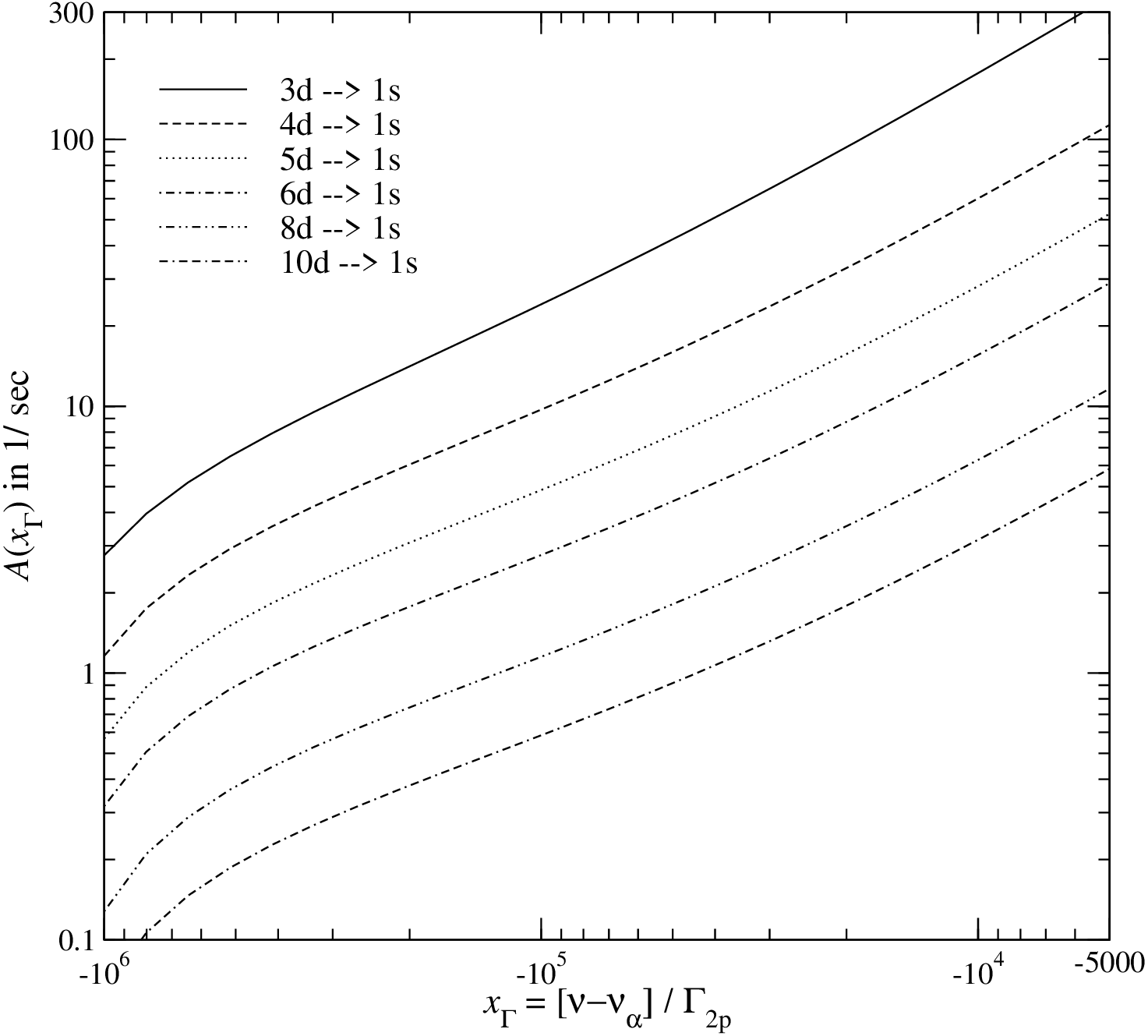}
\caption
{Rate of photon production at frequencies below $x_\Gamma$ according to Eq.
  \eqref{eq:A_nu_c} for several initial s and d-states. {To convert
    to the variable $[\nu-\nu_{\alpha}]/\nu_{\alpha}$, one should multiply
    $x_\Gamma$ by $\sim\pot{2.54}{-7}$.}}
\label{fig:Rate_nsnd}
\end{figure*}
\begin{figure*}
\centering 
\includegraphics[width=0.93\columnwidth]
{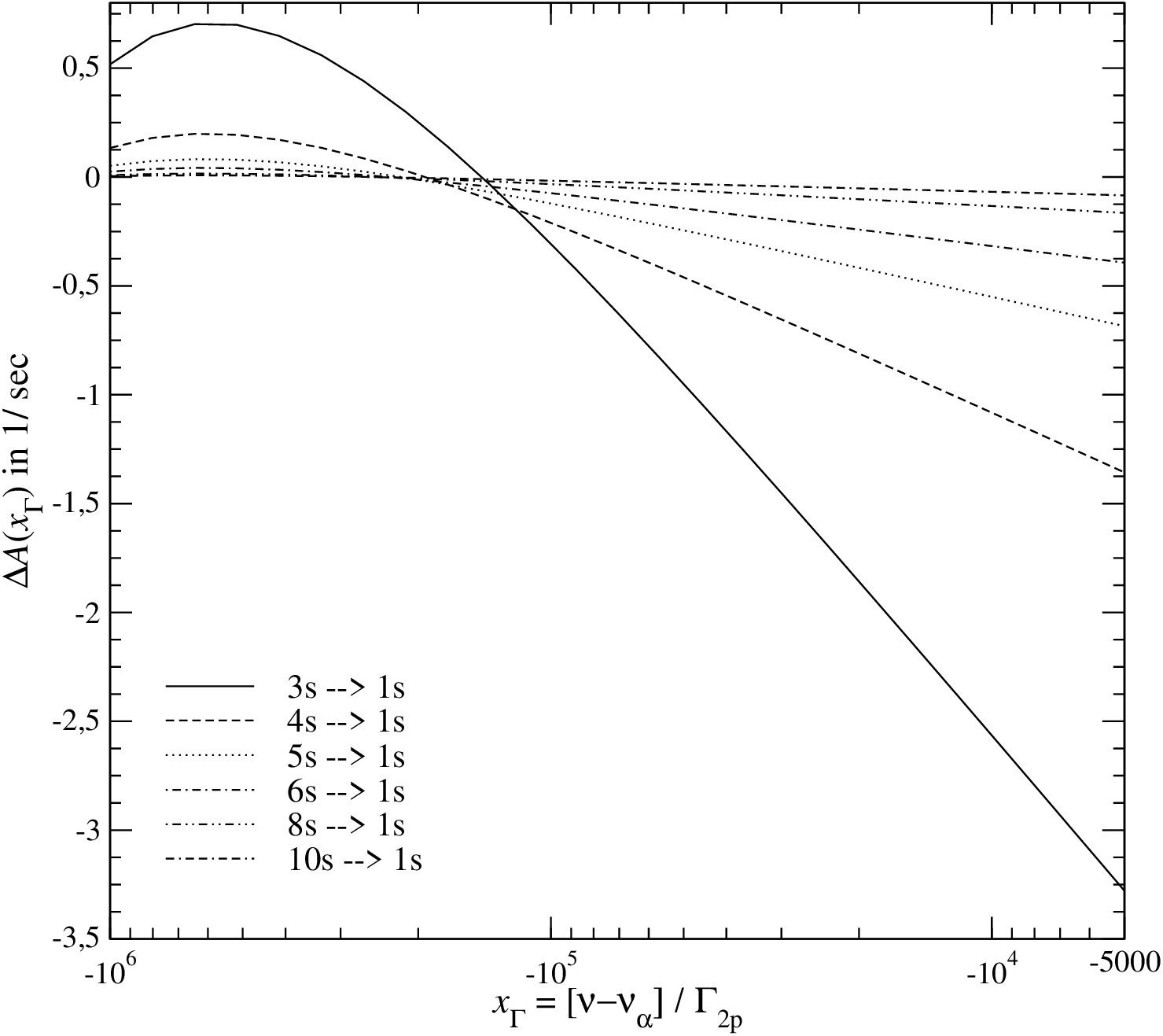}
\hspace{1mm}
\includegraphics[width=0.93\columnwidth]
{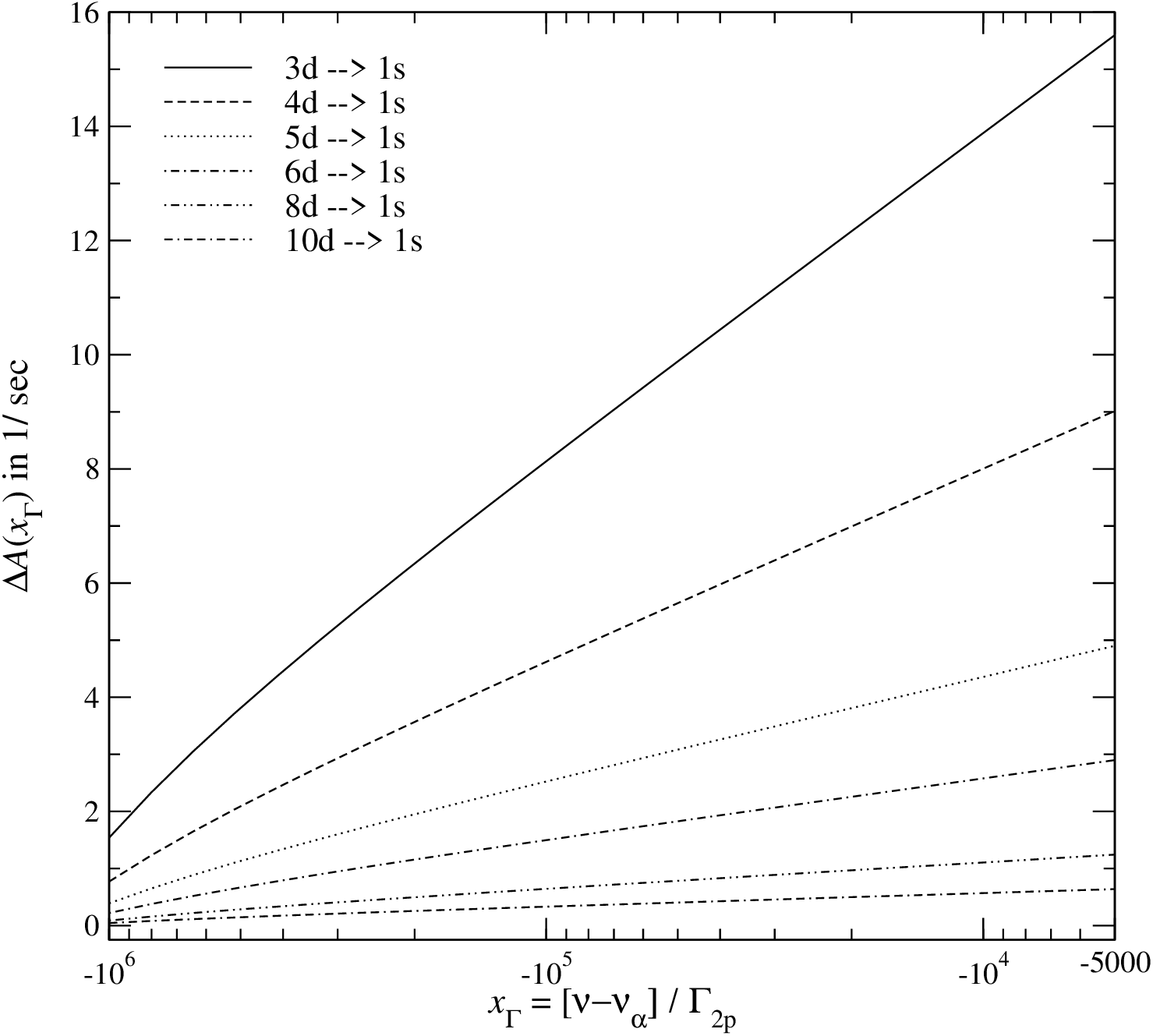}
\caption
{Effective change in the rate of photon production (real profile minus
  Lorentzian) at frequencies below $x_\Gamma$, according to
  Eq. \eqref{eq:DA}. {To convert to the variable
  $[\nu-\nu_{\alpha}]/\nu_{\alpha}$, one should multiply $x_\Gamma$ by
  $\sim\pot{2.54}{-7}$.}}
\label{fig:DRate_nsnd}
\end{figure*}
\subsection{Cosmological hydrogen recombination}
\label{sec:recombination}

\subsubsection{Escape of photons in the red wing of the Lyman-$\alpha$ resonance}
In the context of cosmological hydrogen recombination, the escape of photons
in the red wing of the Lyman-$\alpha$ resonance, which is one of the major
channels to reach the ground state of hydrogen, plays a key role in
controlling the dynamics of recombination \citep{Varshalovich1968,
Grachev1991, Rybicki1994, Chluba2007c}.
At large distances, say at frequencies below $\nu_{\rm c}$ redward of the
Lyman-$\alpha$ central frequency, $\nu_\alpha$, the probability of absorbing a
photon to the continuum, thereby creating a free electron, becomes very
low.
Photons released below $\nu_{\rm c}$ directly escape further interaction with
the neutral hydrogen atoms and lead to the settling of an electron in the
1s-state.
On the other hand, all photons emitted at frequencies $\nu\gtrsim\nu_{\rm c}$
will have a very high probability of being absorbed in the continuum {or
undergoing transitions to higher levels}, possibly after many interactions
with neutral hydrogen atoms or when redshifting into the domain of the
Lyman-$\alpha$ resonance from frequencies $\nu>\nu_\alpha$.

Determining the exact value of $\nu_{\rm c}$ during the epoch of cosmological
hydrogen recombination requires a full treatment of the radiative transfer in
the Lyman-$\alpha$ resonance. Our computations show \citep{Chluba2007c} that
$\nu_{\rm c}$ depends on redshift and should typically lie within
100 to 1000 Doppler width below the Lyman-$\alpha$ frequency. At redshift $z$,
one Doppler width \change{corresponds to}
\beal 
\Delta\nu_{\rm D} 
&\approx
58.0\,\left[\frac{(1+z)}{1100}\right]^{1/2}\!{\rm GHz}
\approx \pot{2.35}{-5}\left[\frac{(1+z)}{1100}\right]^{1/2}\!\nu_\alpha
\nonumber\\
&\approx 92.5\,\left[\frac{(1+z)}{1100}\right]^{1/2}\!\Gamma^{1\gamma}_{\rm 2p\rightarrow 1s};
\end{align}
hence $|\nu_{\rm c}-\nu_\alpha|/\nu_\alpha \sim 10^{-3}-10^{-2}$, \change{or in
terms of the 2p-line width $|\nu_{\rm c}-\nu_\alpha|/\Gamma^{1\gamma}_{\rm
2p\rightarrow 1s}\sim 10^{4}-10^{5}$}.

In computations of the hydrogen recombination history, it is therefore
important to know how many photons reach the very distant red wing of
the Lyman-$\alpha$ resonance directly.
{If we want to estimate this effect, we need to compute the difference in the
number of photons, that are directly escaping in the distant wing by comparing
the emission profiles in the full treatment of two-photon processes with the
one in $1+1$-single photon picture. This will show the relevance of this
process.

If we consider those photons emitted in the red wing of the Lyman-$\alpha$
resonance because of two-photon transitions from upper s or d-states, then
when introducing the dimensionless frequency variable
$x_\Gamma(\nu)=[\nu-\nu_\alpha]/\Gamma_{\rm 2p}$, the results discussed in
Sect. \ref{sec:dev_wing} suggest the following: }
\begin{itemize}
\item[(i)] Because of two-photon processes, more photons will escape in the
very distant red wing of the Lyman-$\alpha$ resonance (below
$x_\Gamma\sim-\pot{5}{5}$) than in the $1+1$-single photon treatment for the
direct cascade emission.

\item[(ii)] For initial s-states, fewer photons are emitted in the range
$-\pot{5}{5}\lesssim x_\Gamma \lesssim -10^4$ than in the $1+1$-single photon
picture.

\item[(iii)] For initial d-states, more photons are emitted in the range
$-\pot{5}{5}\lesssim x_\Gamma \lesssim -10^4$ than in the $1+1$-single photon
picture.

\end{itemize}
Because of (i) and (iii) hydrogen recombination should occur slightly faster,
while (ii) may make it a bit slower. Since the statistical weight of d-states
is 5 times higher than for s-levels, one expects that recombination will in
total be slightly faster than in the standard treatment because of two-photon
processes.

{ In addition to the direct escape of photons in the distant red wing of the
Lyman-$\alpha$ transition, also significant differences close to the line
center arise (see Fig.~\ref{fig:DLorentz_nsnd}). Understanding how these
changes affect the effective escape of photon from the line center requires a
more rigorous treatment of the radiative transfer problem in the line. Also
the feedback of photons emitted in the blue wing of the Lyman-$\alpha$
transition and in particular those coming from the other Lyman-series
transitions, should be slightly modified when taking the full two-photon
process into account. Both aspects are beyond the scope of this paper and will
be addressed in a future work.}

{We can now estimate the effect of the changes in the effective escape of
photons in the distant red wing of the Lyman-$\alpha$ transition.  For this
only the photons between the innermost resonances in the two-photon emission
spectrum are contributing (e.g. photons between the Balmer-$\gamma$ and
Lyman-$\alpha$ transition for the 5s and 5d-two-photon decay, see
Fig. \ref{fig:5s5d_spec}). This is because we only want to count photons up to
$\nu\leq\nu_{\rm c}$ and correspondingly $\nu'\geq \nu_{n\rm 1s}-\nu_{\rm c}$.
Because of the symmetry of the full two-photon profile, it is therefore sufficient
to integrate $\phi^{2\gamma}_{n\rm s/d \rightarrow 1s}(y)$ from $y=1/2$ up
to $y_{\rm c}=\nu_{\rm c}/\nu_{i\rm 1s}$:
\beal 
\label{eq:A_nu_c}
A^{2\gamma}_{n\rm s/d \rightarrow 1s}(\nu_{\rm c})
&=
\frac{1}{2}
\!\int_{1-y_{\rm c}}^{y_{\rm c}}\!\phi^{2\gamma}_{n\rm s/d\rightarrow 1s}(y)\,\id y
\equiv
\!\!\int_{1/2}^{y_{\rm c}}\!\phi^{2\gamma}_{n\rm s/d\rightarrow 1s}(y)\,\id y.
\end{align}
This integral yields the total number of photons that directly escape per
second in the distant red wing of the Lyman-$\alpha$ line. 
%
%
It should be compared with the value computed using the standard
$1+1$-single photon profile.  
}
Since we only consider cases very far in the red wing of the Lyman-$\alpha$
transition, the integral over the Lorentzian resulting in the $1+1$ approach
can be written as
\beal 
\label{eq:phi_int}
A^{n\rm s/d, 1+1\gamma}_{\rm 2p \rightarrow 1s}(\nu_{\rm c})
&=\!\int_0^{\nu_{\rm c}}\!\phi^{n\rm s/d, 1+1\gamma}_{\rm 2p\rightarrow 1s}(\nu)\,\id\nu
\approx
\frac{A_{n\rm s/d\rightarrow 2p}}{4\pi^2}\,\frac{1+x_{\Gamma,\rm
    c}\epsilon}{-x_{\Gamma,\rm c}}
\end{align}
where we introduced $\epsilon=\Gamma_{\rm 2p\rightarrow 1s}/\nu_\alpha\approx
\pot{2.540}{-7}$ and used the variable
$x_\Gamma(\nu)=[\nu-\nu_\alpha]/\Gamma_{\rm 2p}$.
The effective difference in the photon production rate, or equivalently the
photon escape rate in the distant wings, is then given by
\beal 
\label{eq:DA}
\Delta A_{n\rm s/d \rightarrow 1s}(\nu_{\rm c})
=A^{2\gamma}_{n\rm s/d \rightarrow 1s}(\nu_{\rm c})-A^{n\rm s/d, 1+1\gamma}_{\rm 2p \rightarrow 1s}(\nu_{\rm c})
\end{align}
for a fixed frequency $\nu_{\rm c}$. Although in general $\nu_{\rm c}$ is a
function of time, below we assume that it is constant. A more rigorous
treatment will be presented in some future work.

\subsubsection{Approximate inclusion into the multi-level code}
In our formulation, $\Delta A_{n\rm s/d \rightarrow 1s}(\nu_{\rm c})$ plays
the role of the {\it pure} two-photon rate coefficients used in
\citet{Dubrovich2005} and \citet{Wong2006b}.
If we want to estimate the possible impact of our results on the hydrogen
recombination history, we have to take the additional net escape of
photons into account. This can be accomplished by adding
\beal 
\label{eq:DR}
\Delta R^{2\gamma}_{n\rm s/d\rightarrow 1s}
\approx
\Delta A_{n\rm s/d \rightarrow 1s}(\nu_{\rm c})\left[N_{n\rm s/d} - \frac{g_{n\rm
    s/d}}{g_{\rm 1s}}\,N_{\rm 1s}\,e^{-h\nu_{n\rm 1s}/\kB\Tg}\right]
\end{align}
to the rate equation of the 1s-state and subtracting it from the
corresponding rate equation of the $n$s and $n$d-levels. Here $N_{\rm 1s}$,
$N_{n\rm s}$, $N_{n\rm d}$ are the number density of hydrogen atom in the 1s,
$n$s, and $n$d-states, respectively.  Furthermore, $\Tg=T_0 (1+z)$, is the
temperature of the ambient blackbody radiation field, with $T_0=2.725\,$K
\citep{Fixsen2002}. The factors $g_{n\rm s}\equiv g_{\rm 1s}$ and $g_{n\rm
d}\equiv 5\,g_{\rm 1s}$ are due to the statistical weights of the s and d-states.
In \eqref{eq:DR} we have neglected any possible deviation in the radiation
field from a blackbody and also omitted stimulated two-photon
emission. 
{Both processes should only lead to higher order corrections.}
Moreover we have added an inverse term, assuming detailed balance. This term
is not important during the main epoch of hydrogen recombination ($z\lesssim
1600$) and was only included to maintain full thermodynamic equilibrium at
high redshifts.
A self-consistent derivation is beyond the scope of this paper.

\subsubsection{Results for the photon production in the distant wings}
In Fig. \ref{fig:Rate_nsnd} we give the rate of photon production at
frequencies below $x_\Gamma$ within the full two-photon treatment, i.e.
according to Eq. \eqref{eq:A_nu_c}, for several initial s and d-states.
For the d-states the photon production is $\sim 10$ times faster than for the
corresponding s-state.
In the case of initial s-states, the plateau of $A^{2\gamma}_{n\rm s\rightarrow
1s}(\nu_{\rm c})$ close to $x_\Gamma\sim -\pot{4}{5}$ is caused by the zero in
the central region of the two-photon emission spectra (e.g. see
Fig.~\ref{fig:Lorentz_nsnd}). As mentioned in Sect.  \ref{sec:two_em}, this
zero is absent in the two-photon spectra of initial d-states, and consequently
no such plateau appears for $A^{2\gamma}_{n\rm d \rightarrow 1s}(\nu_{\rm
c})$.
In both cases the rate of photon production decreases when increasing $n$.
Looking at Fig. \ref{fig:one}, just from the non-resonant term one would
expect the opposite behavior. However, due to destructive interference this
does not happen.

In Fig. \ref{fig:DRate_nsnd} the net change in the rate of photon production
at frequencies below $x_\Gamma$ is shown. {The photon production due to
the two-photon decay of initial s-states}, at relevant distances from the
Lyman-$\alpha$ center ($x_\Gamma>-10^5$), is actually slower than in the
$1+1$-single photon picture. This suggests that due to the full treatment of
the two-photon process for the s-states alone, cosmological hydrogen
recombination is expected to be slower than in the standard computations.
This contrasts to the work of \citet{Wong2006b}, where both the s and
d-state two-photon process leads to an increase in the rate of recombination.

On the other hand, for the d-states the effective photon escape rate is higher
than in the $1+1$-single photon picture, hence one expects an increase in
the rate of recombination. Since the statistical weights of the d-states are 5
times larger than the s-states, and also the effective increase in the wing
photon production rate is roughly additional 5 times higher (cf.
Fig.~\ref{fig:DRate_nsnd}), one still expects that, even when including the
combined effect of the s and d-state, two-photon process, cosmological hydrogen
recombination in total will proceed faster than in the standard treatment.

We would like to mention that using the analytic approximations given in
Appendix~\ref{app:fits} for the non-resonant term in connection with the
formulae in Sect. \ref{sec:cas_int} we were able to reproduce the rates
presented in this section.

\subsubsection{Comparing with earlier works}
In the works of \citet{Dubrovich2005} and \citet{Wong2006b}, only the combined
effect of the two-photon process for the $n$s and $n$d-states on the hydrogen
recombination history was discussed.
To compare our results for the effective photon production rates with their
values, we also write the combined \change{effective} decay rate
\beal 
\label{eq:DAnsnd}
6\Delta A_{n\rm s+\it n\rm d \rightarrow 1s}(\nu_{\rm c}) 
=\Delta A_{n\rm s\rightarrow 1s}(\nu_{\rm c})
+5\,\Delta A_{n\rm d \rightarrow 1s}(\nu_{\rm c}),
\end{align}
where we implicitly assumed that the $n$s and $n$d-states are in full
statistical equilibrium with each other ($N_{n\rm d}=5\,N_{n\rm s}$).
At the relevant redshift, the deviations from full statistical
equilibrium are rather small \citep{Jose2006, Chluba2007}, so that this
approximation is possible (see Sect. \ref{sec:DXe}).

\begin{figure}
\centering 
\includegraphics[width=0.93\columnwidth]
{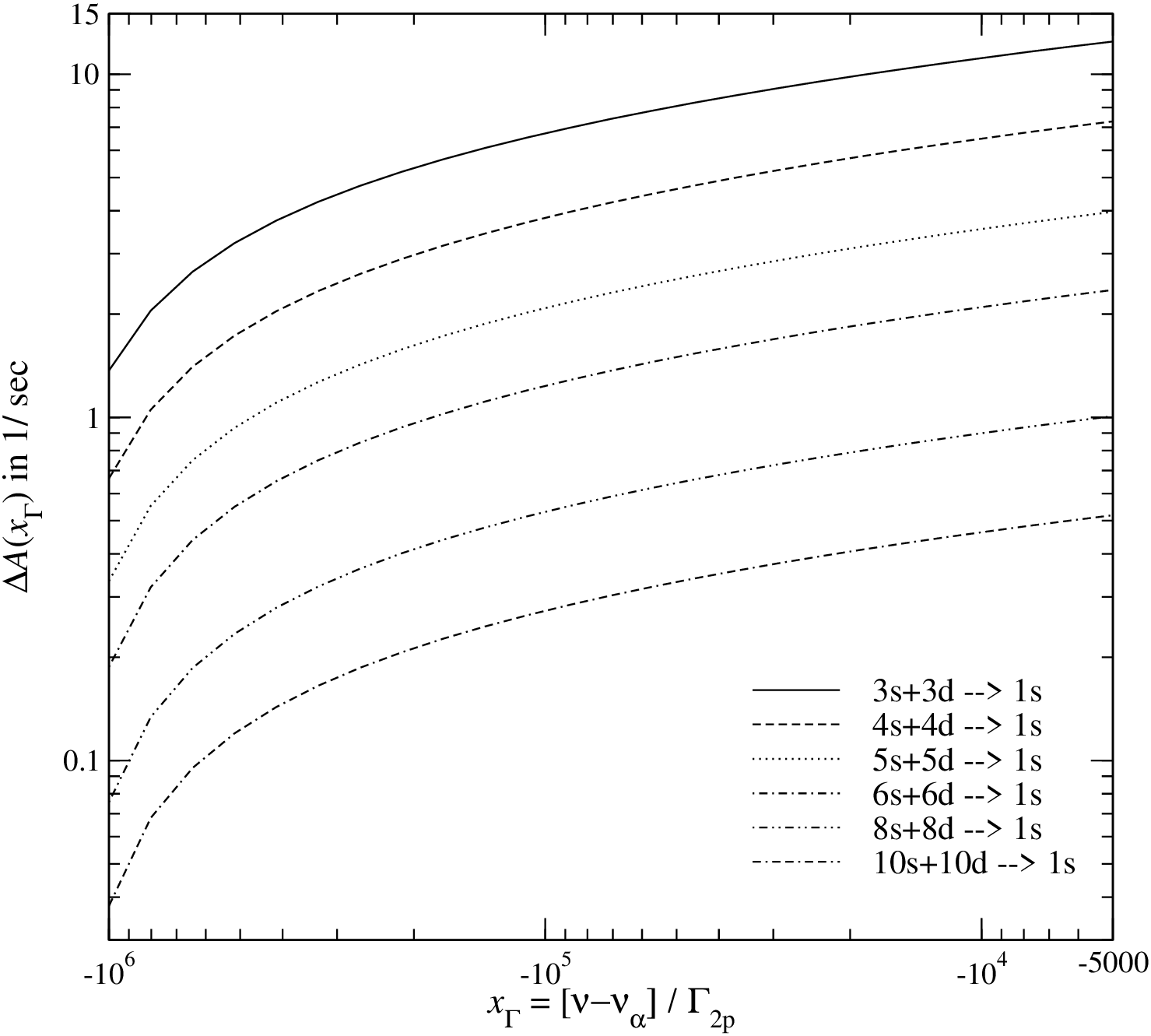}
\caption
{Combined effective two-photon photon production rate, $\Delta A_{n\rm s+\it
    n\rm d \rightarrow 1s}(\nu_{\rm c})$, computed according to Eq.
  \eqref{eq:DAnsnd}.  {To convert to the variable
    $[\nu-\nu_{\alpha}]/\nu_{\alpha}$, one should multiply $x_\Gamma$ by
    $\sim\pot{2.54}{-7}$.}}
\label{fig:DRate_nspnd}
\end{figure}
In Fig. \ref{fig:DRate_nspnd} we \change{present} the results for $\Delta
  A_{n\rm s+\it n\rm d \rightarrow 1s}(\nu_{\rm c})$ for several shells.
If we consider the effective rate for the 3s and 3d-levels then, even for very
conservative values of $\nu_{\rm c}$, say 1000 Doppler width or $\sim 10^5$
natural width below the Lyman-$\alpha$ resonance, we obtain $\Delta A_{\rm
  3s+3d \rightarrow 1s}\sim 6.7\,\rm s^{-1}$, whereas from the formulae in
\citet{Dubrovich2005} and \citet{Wong2006b} one can find $\Delta A^{\rm
  DG}_{\rm 3s+3d\rightarrow 1s}\sim 22\,\rm s^{-1}$ and $\Delta A^{\rm
  WS}_{\rm 3s+3d\rightarrow 1s}\sim 1.5\,\rm s^{-1}$, respectively.
Our value is only $\sim 3.3$ times lower than the one of \citet{Dubrovich2005}
but $\sim 4.5$ times higher than in \citet{Wong2006b}.
We would argue that for the third shell even values up to $10\,\rm
s^{-1}$ still are reasonable, in particular at very low ($z \lesssim 1000$)
and high ($z \gtrsim 1300$) redshifts, where the probability of absorption
decreases.
In Table \ref{tab:DR_Gamma} we give a few values of $\Delta A_{n\rm s/d
\rightarrow 1s}(\nu_{\rm c})$ for different frequencies $\nu_{\rm c}$. Given
are the values of $\Delta A_{n\rm s/d \rightarrow 1s}$ in $\rm 1/sec$ for
different frequencies $x_{\Gamma,\rm c}=[\nu_{\rm c}-\nu_\alpha]/\Gamma_{\rm
2p\rightarrow 1s}$. In each column the first value is for the s-levels, the
second for the d-states.

Figure \ref{fig:DRate_nspnd} also shows that, in contrast to the works of
\citet{Dubrovich2005} and \citet{Wong2006b}, the net photon escape rate due to
the combined effect of the s and d-state, two-photon process decreases with
increasing $n$. This implies that the relevance of the two-photon emission
from higher shells is significantly less than in their computations,
because the sharp drop in \change{the populations of levels} with $n$ will no
longer be partially canceled by the assumed linear increase in the effective
two-photon-decay rate.

\begin{table}
\caption{Effective difference in the photon production rate in the distant
wings using Eq. \eqref{eq:DA}. }
\label{tab:DR_Gamma}
\centering
{
\begin{tabular}{@{}cccc}
\hline
\hline
$n$  & $x_{\Gamma,\rm c}=-10^5$ & $x_{\Gamma,\rm c}=-5\times10^4$ & $x_{\Gamma,\rm c}=-10^4$ \\
\hline
3 & -0.307 / 8.133 &-0.951 / 9.882 &-2.565 / 13.883
\\
4 & -0.210 / 4.619 &-0.461 / 5.648 &-1.082 / 8.004
\\
5 & -0.122 / 2.523 &-0.245 / 3.082 &-0.551 / 4.356
\\
6 & -0.074 / 1.497 &-0.144 / 1.8272  &-0.317 / 2.578
\\
7 & -0.048 / 0.954 &-0.091 / 1.164 &-0.199 / 1.641
\\
8 & -0.033 / 0.643 &-0.061 / 0.785 &-0.133 / 1.106
\\
9 & -0.023 / 0.454 &-0.043 / 0.553 &-0.093 / 0.779
\\
10 & -0.017 / 0.331 &-0.032 / 0.404 &-0.068 / 0.569
\\
\hline
\hline
\end{tabular}
}
\end{table}
\subsubsection{Differences in the free electron fraction}
\label{sec:DXe}
{ We modified our multi-level hydrogen code \citep[for more details
  see][]{Jose2006, Chluba2007} to take into account the additional escape of
  photons in the distant wings of the Lyman-$\alpha$ resonance due to the
  two-photon process using Eq.~\eqref{eq:DR}.
  \change{For the hydrogen atom} we typically included the first 30 shells
  in our computations, following the evolution of the populations for each
  angular-momentum substate separately. We also performed computations with
  more shells, but this did not alter the results significantly with our approach.

  The additional two-photon process was included for s and d-states with
  $n\leq n_{2\gamma}$, where the parameter $n_{2\gamma}$ gives the highest
  shell for which the additional two-photon decay was taken into account.
  We only used $n_{2\gamma}\leq 10$, but because of the strong decrease of
  $|\Delta A_{n\rm s/d \rightarrow 1s}(\nu_{\rm c})|$ with $n$ (see
  Fig.~\ref{fig:DRate_nsnd}) and the drop in the populations of higher shells,
  we do not expect any significant differences when going beyond this.
  For simplicity we also assumed that the value of $\nu_{\rm c}$ is constant
  with time. This makes our estimates more conservative, since both at very
  low and very high redshifts, $\nu_{\rm c}$ should be closer to $\nu_\alpha$
  and therefore may increase the impact of the two-photon process on the
  recombination history.
  We performed computations with three different values of $\nu_{\rm c}$.  The
  \change{effective rates for these cases} are summarized in
  Table~\ref{tab:DR_Gamma}. We consider the case with $x_{\Gamma,\rm c}=-10^5$
  as pessimistic, whereas the case $x_{\Gamma,\rm c}=-10^4$ may be optimistic.
}

  We also ran computations using the formulae according to
  \citet{Dubrovich2005} and \citet{Wong2006b}. In the paper of
  \citet{Dubrovich2005}, the s and d-rates were not given separately, but
  assuming $\Delta A^{2\gamma}_{n\rm s\rightarrow 1s}\equiv\Delta
  A^{2\gamma}_{n\rm d\rightarrow 1s}$ for simplicity, one finds
\beal
\label{eq:A_DG}
\Delta A^{2\gamma,\rm DG}_{n\rm s\rightarrow 1s}
&\equiv
\Delta A^{2\gamma,\rm DG}_{n\rm d\rightarrow 1s}
=8.2293\,{\rm s^{-1}}\times S_{\rm DG}(n),
\end{align}
where $S_{\rm
  DG}(n)=9\,\left[\frac{n-1}{n+1}\right]^{2n}\frac{11\,n^2-41}{n}$. Since the
deviations from full statistical equilibrium are rather small \citep{Jose2006,
  Chluba2007}, this assumption should not be very critical and, in any case, is
only meant for comparison.

\citet{Wong2006b} explicitly give the rates for the 3s and 3d-states and then
assume the same $n$-scaling as \citet{Dubrovich2005}. This yields
\bsub
\label{eq:A_WS}
\beal 
\Delta A^{2\gamma,\rm WS}_{n\rm s\rightarrow 1s}
&=8.2197\,{\rm s^{-1}}\times S_{\rm WS}(n)
\\
\Delta A^{2\gamma,\rm WS}_{n\rm d\rightarrow 1s}
&=0.13171\,{\rm s^{-1}}\times S_{\rm WS}(n),
\end{align}
\esub
with $S_{\rm WS}(n)=S_{\rm DG}(n)/S_{\rm
  DG}(3)=\frac{96}{29}\,\left[\frac{n-1}{n+1}\right]^{2n}\frac{11\,n^2-41}{n}$.
  Comparing with Eq. \eqref{eq:A_DG}, one can see that the difference in the
  approach of \citet{Dubrovich2005} and \citet{Wong2006b} is mainly because
  they used a much lower rate for the d-states \change{(a factor $\sim
  170$!). The assumed rate for the s-states is only $\sim 2.7$ times lower
  than in the computations of \citet{Dubrovich2005}.}

\begin{figure}
\centering 
\includegraphics[width=0.93\columnwidth]
{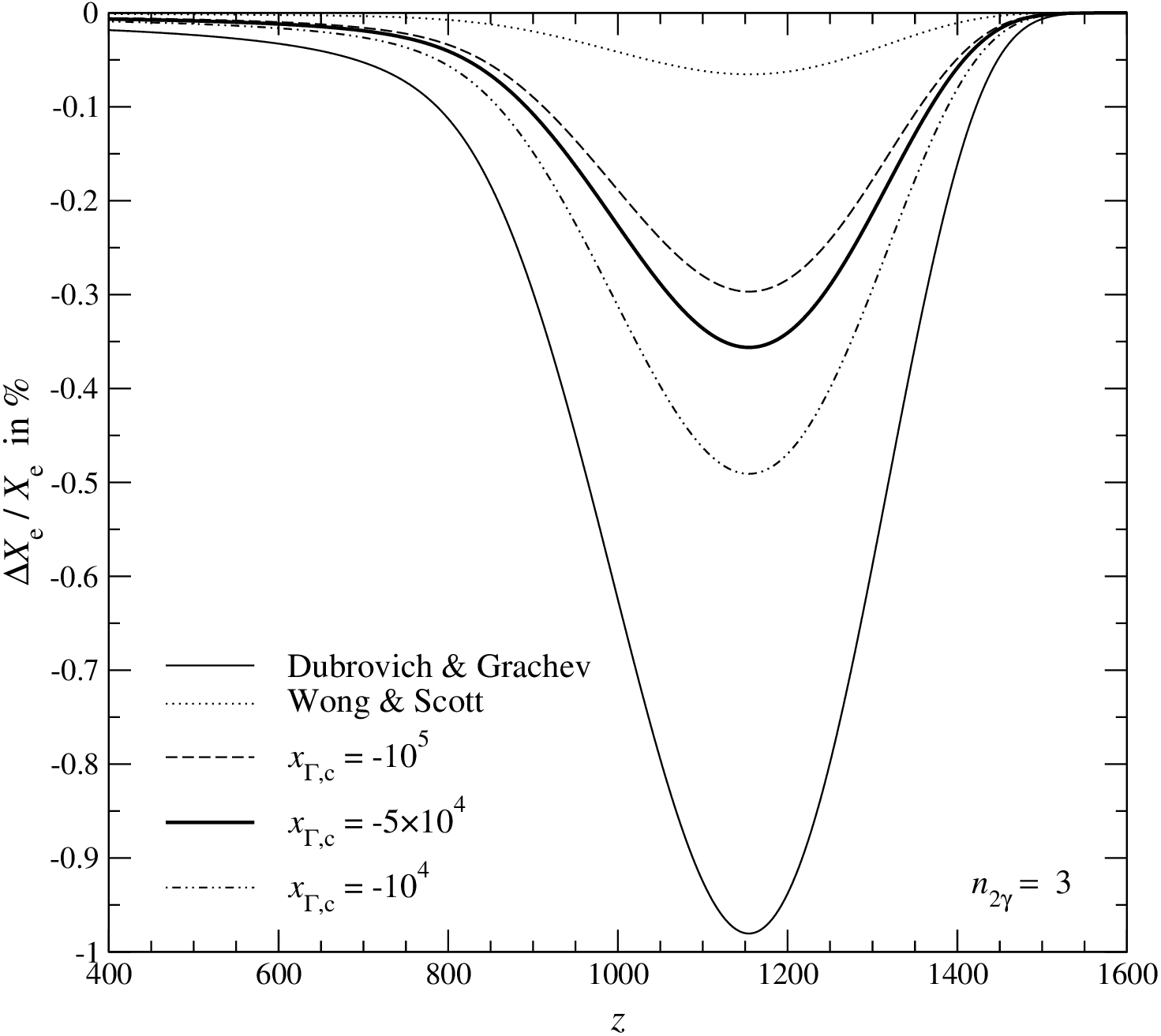}
\caption
{Relative change in the free electron fraction. Here we only included the
  additional two-photon process for the 3s and 3d-states. The computations
  were performed for a 30-shell hydrogen atom. The effective two-photon rates
  for three different values of $\nu_{\rm c}$ according to
  Table~\ref{tab:DR_Gamma} were used. For comparison, we show the results that
  were obtained using the 3s and 3d two-photon decay rates of
  \citet{Dubrovich2005} and \citet{Wong2006b}.}
\label{fig:DX_3s3d}
\end{figure}
In Fig. \ref{fig:DX_3s3d} we present the relative change in the free electron
fraction when only including the additional two-photon process for 3s and
3d-states. For comparison we show the results obtained using the decay rates of
\citet{Dubrovich2005} and \citet{Wong2006b}.
{One can clearly see that the dependence on the adopted value of
  $\nu_{\rm c}$ is not very strong.}
For our optimistic value of $\nu_{\rm c}$, close to the maximum the effect is
roughly 2 times smaller than for the values of \citet{Dubrovich2005}, and even
in our pessimistic model, it is still more than $\sim 4$ times greater than
within the framework of \citet{Wong2006b}.
Comparing the curves, which we obtained within the approach of
\citet{Dubrovich2005} and \citet{Wong2006b}, with those in Fig. 3 of
\citet{Wong2006b} one can see that our results for the changes in the electron
fraction are slightly smaller. We checked that this is not due to our detailed
treatment of the angular-momentum substates. This is expected since the
deviations from full statistical equilibrium at the relevant redshifts are too
small to have any effect here \citep{Jose2006, Chluba2007}.
Also we computed the same correction using 50 shells, but found no significant
increase.  

\begin{figure}
\centering \includegraphics[width=0.93\columnwidth] {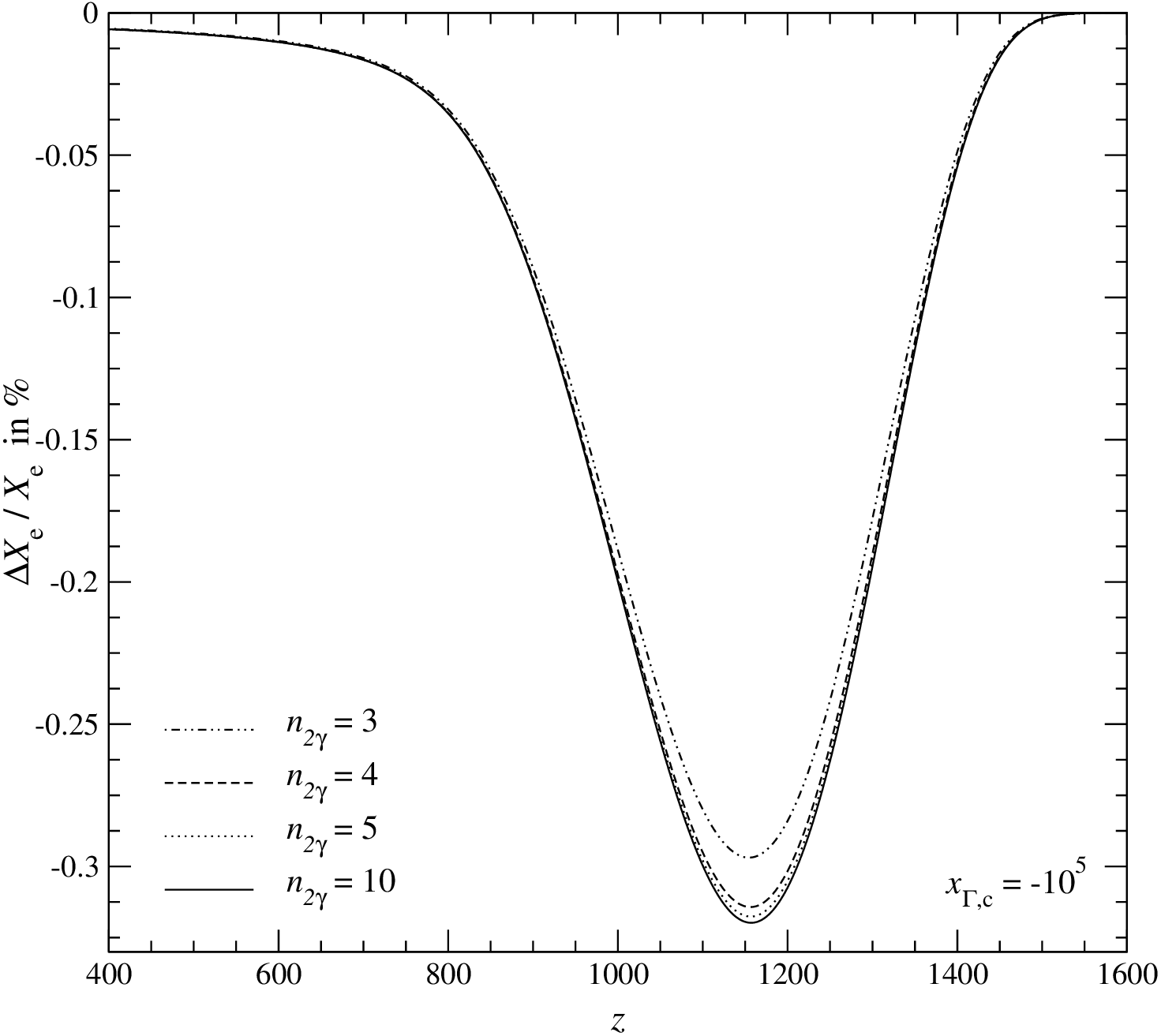}
\caption
{Relative change in the free electron fraction for different values of
  $n_{2\gamma}$. The computations were performed including 30 shells.
}
\label{fig:DX_nsnd}
\end{figure}
In Fig. \ref{fig:DX_nsnd} we illustrate the impact of the two-photon process
from higher shells. With our estimates of the effective two-photon decay
rates, like in the studies of \citet{Dubrovich2005} and \citet{Wong2006b}, the
effect increases with $n_{2\gamma}$.
However, the strong decrease in the effective rates within our computations
(see Table \ref{tab:DR_Gamma}) implies that the result practically does not
change when including the additional two-photon effect for more than 5 shells.
This strongly contrasts the works of \citet{Dubrovich2005} and
\citet{Wong2006b}, where the total change in the free electron fraction
radically depends on the chosen value of $n_{2\gamma}$ (even up to
$n_{2\gamma}=40$ was considered).
As mentioned above, in these computations the increase in the two-photon decay
rates with $n$ (cf.  Eqn.  \eqref{eq:A_DG} and \eqref{eq:A_WS}) partially
cancels the decrease in the population of the higher levels, and therefore
enhances the impact of their contribution as compared to the lower shells.
\change{For example, at $z\sim 1200$ \change{(i.e close to the maximum of the
changes in $N_{\rm e}$)} the populations of the excited states are still
nearly in Saha-equilibrium with the continuum \citep{Chluba2007}. Therefore
the population of the fourth shell is roughly a factor of
$e^{-h\nu_{43}/kT_\gamma}\sim 0.19$ smaller than in the third shell. Also the
effective $2\gamma$-rate decreases by $\sim 1.8$, whereas in the picture of
\citet{Dubrovich2005} and \citet{Wong2006b} it would have increased $\sim 1.9$
times. }
From Fig.~\ref{fig:DX_nsnd} it is also clear that the strongest effect for our
estimates of the effective decay rates comes from the 3s and 3d-levels alone.
This again is in strong opposition to the computations of
\citet{Dubrovich2005} and \citet{Wong2006b} where more than $\sim 75\%$ of the
correction is due to the combined effect of higher shells.
%

\begin{figure}
\centering 
\includegraphics[width=0.93\columnwidth]
{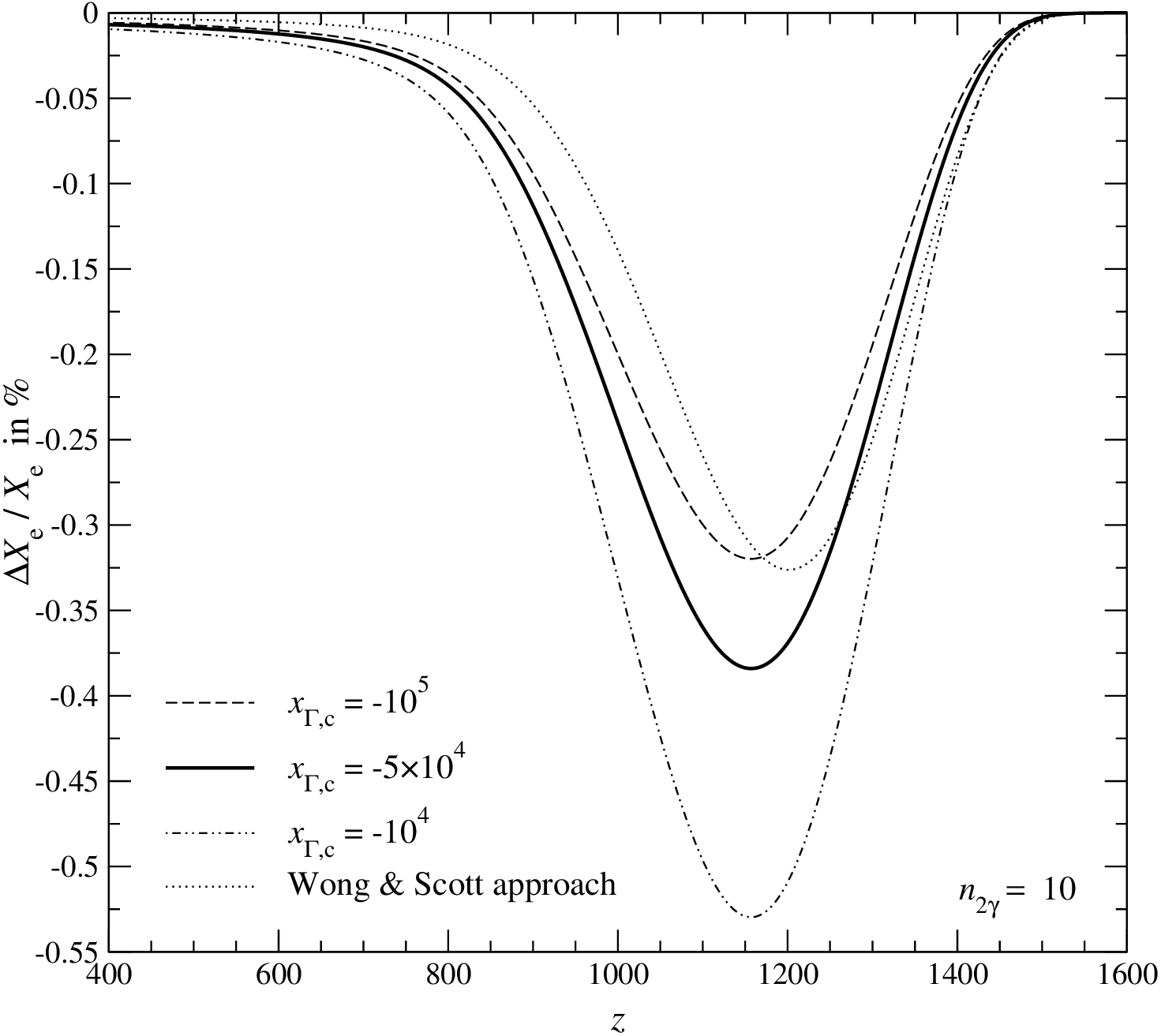}
\caption
{Relative change in the free electron fraction when taking the additional
  two-photon emission for up to 10 shells into account. The computations were
  performed including 30 shells, for three different values of $\nu_{\rm c}$.
  {For comparison, the result that was obtained within the approach of
  \citet{Wong2006b} is shown, but for $n_{2\gamma}=40$ and using a 50-shell
  model for the hydrogen atom.}}
\label{fig:DX_10s10d}
\end{figure}
In Fig. \ref{fig:DX_10s10d} we give our final estimates for the possible
changes in the recombination history. In our optimistic model the change is
$\Delta X_{\rm e}/X_{\rm e}\sim -0.53\%$ at redshift $z\sim 1150$, and it drops
to $\Delta X_{\rm e}/X_{\rm e}\sim -0.32\%$ for the pessimistic case.
Including more shells in the model for the hydrogen atom did not change
these results.
For comparison we also computed the changes in the ionization history by
applying the formulae of \citet{Wong2006b}, but using $n_{2\gamma}=40$ and 50
shells for the model of the hydrogen atom.
Although our discussion has shown that the values computed by
\citet{Cresser1986} for 3s and 3d-states are likely not related to the
cosmological hydrogen recombination problem and that extrapolating those
values to higher shells is rather rough, our final results are
numerically compatible with those obtained using the approach of
\citet{Wong2006b}.
However, within our approach the changes in the ionization history close to
the maximum of the Thomson visibility function \citep{Sunyaev1970} are larger
than in the computations of \citet{Wong2006b}. Therefore the changes in the
cosmic microwave background temperature and polarization power spectra are
also expected to be a bit larger.
Indeed it seems that the corrections to the ionization history due to the
two-photon decay from higher shells does not reach the percent level, and that
the impact of this process was overestimated by \citet{Dubrovich2005}.

\subsubsection{Additional remarks}
{ 
  Here we have only investigated the bound-bound two-photon transitions
  directly leading to the ground state. 
  Equations \eqref{eq:general_nsnd} and \eqref{eq:A_nsnd} are also applicable
  when the final state is any s-level. We also checked the rate for the
  two-photon transition $3{\rm s}\rightarrow 2{\rm s}$ and $3{\rm
  d}\rightarrow 2{\rm s}$ and, as expected, found very low values ($\sim
  0.0885\,{\rm s}^{-1}$ for the 3s and $0.0278\,{\rm s}^{-1}$ in the case of
  3d).  In addition, because all dipole transitions to the second shell are
  optically thin in the recombination problem, these corrections should never
  be important within this context.
  Similarly, the $2{\rm p}\rightarrow 1{\rm s}$ two-photon transition due to
  its low probability \citep[$\sim {\rm few}\times 10^{-6}\,{\rm s}^{-1}$,
  see][]{Labzowsky2005} can be completely ignored.

  One may in addition consider the problem of two-photon transitions starting
  from the continuum, e.g. the recombination of electrons to the 2p-state and
  subsequent release of a Lyman-$\alpha$ photon. Here deviations of the line
  profile from the normal Lorentzian shape can also be expected and may lead to an
  increase in the {effective} Lyman-$\alpha$ escape rate.
  However, since the supply of photons to the 2p-state by transitions from
  higher shells is several times faster, the total impact of this effect is
  very likely less than \change{the one from the $2\gamma$-transitions already
  discussed here}.
  
  As mentioned above, under physical conditions like those in our Universe
  during the epoch of hydrogen recombination, the coherence of two and
  possibly multi-photon processes is maintained. Consequently one should
  investigate how strong the deviations of the corresponding emission profiles
  from a Lorentzian are when more than two photons are involved.  \change{This
  requires a QED multi-photon treatment, which is beyond the scope of this
  paper.}
But as we have seen above, adding the two-photon process for the \change{4s-
and 4d-states} (due to the drop in the population \change{of these levels} and
decrease in \change{their effective two-photon decay rate) has affected the
recombination history at a level of $\sim 0.02\%$ in addition to the 3s and 3d
(see Fig.  \ref{fig:DX_nsnd}).
\change{For the $3\gamma$-decay of the 4f-level, one expects that the {\it
relative correction} will be close to the one from to the 3d two-photon
decay. This is because the largest term in this $3\gamma$-description should
involve at least one nearly resonant transition, since the other contributions
should be suppressed in addition. For photons appearing close to the
Lyman-$\alpha$ line, a nearly resonant $4{\rm f}\rightarrow 3{\rm d}$
transition, followed by a quasi $2\gamma$-decay of the 3d-state, is most
likely. Therefore one has
\beal
\label{eq:DA_4f}
\Delta A^{3\gamma}_{4\rm f\rightarrow 1s}
\sim A^{3\gamma}_{4\rm f\rightarrow 3d}
\times
\frac{\Delta A_{3\rm d\rightarrow 1s}}{A^{2\gamma}_{3\rm d\rightarrow 1s}}
\sim
\frac{A^{1\gamma}_{4\rm f\rightarrow 3d}}{A^{1\gamma}_{3\rm d\rightarrow 2p}}\times\Delta A_{3\rm d\rightarrow 1s},
\end{align}
where we have used\footnote{In both cases the branching ratios are very close
to unity.} $A^{3\gamma}_{4\rm f\rightarrow 3d}\sim A^{1\gamma}_{4\rm
f\rightarrow 3d}$ and $A^{2\gamma}_{3\rm d\rightarrow 2p}\sim
A^{1\gamma}_{3\rm d\rightarrow 2p}$. With the value in
Table~\ref{tab:DR_Gamma}, one finds $\Delta A^{3\gamma}_{4\rm f\rightarrow
1s}\sim 2.1\,{\rm s^{-1}}$. Due to the statistical weight of the 4f-state, one
then should obtain another $\sim 7\times 2.1\,{\rm s^{-1}}\sim 14.7\,{\rm
s^{-1}}$ addition to the overall decay rate of the fourth level. This is
roughly a $\sim 50\%$ correction to the total two-photon correction of the
fourth shell.}
Therefore, one could expect another $\sim 0.01\%$ correction to the free
electron faction when including the three-photon decay from the 4f.
  Furthermore, for some high level \change{(likely above $n\sim 20-50$),} even
  the residual disturbances from the small amount of perturbers that are
  present in the Universe will destroy the coherence of the {possible
  multi-photon emission coming from very high initial levels}.

  One could also consider the three-photon decay of the 2p-state. Here, just
  as in the 2s-two-photon decay, no intermediate resonances are involved; and
  due to momentum conservation, this process is allowed. However, simple
  estimates show that this process has a rate lower than $\sim 10^{-2}\,{\rm
  s}^{-1}$ and hence negligible at the \change{$0.1\%$-level}.}



\section{Conclusions}
{We have studied in detail the emission of photons due to two-photon
  transitions from high s and d-states to the ground level.  
  Up to $n=20$ we found simple analytic fitting formulae to represent the full
  two-photon emission profile with very high accuracy. 
  We have discussed the deviations in the two-photon emission profiles from
  the natural Lorentzian shape and investigated the importance of the
  non-resonant, cascade, and interference term separately.

  Applying our results to the cosmological hydrogen recombination shows that
  the corrections to the ionization history due to the additional two-photon
  process from higher shells likely do not reach the percent level. For
  conservative assumptions we find a correction $\Delta X_{\rm e}/X_{\rm
  e}\sim-0.4\%$ at redshift $z\sim 1160$.
  This is numerically similar to the result of \citet{Wong2006b};
  however, the physics leading to this conclusion is rather different.
  In particular we find that the two-photon process for initial s-states
  actually slows the recombination process down. In addition, the effective
  two-photon rates connecting the high s and d-level directly to the 1s-level
  decrease with principle quantum number $n$.
  Both aspects contrast to the rate estimates used in the studies by
  \citet{Dubrovich2005} and \citet{Wong2006b}.
Here it is very important that the destructive interference between the
  cascade and non-resonant term cancels a large part of the additional
  non-resonant emission in the distant red wings of the Lyman-$\alpha$
  transition.
Furthermore, in our computations the main correction to the ionization history
  stems from the 3s and 3d-states, while in the computations of
  \citet{Dubrovich2005} and \citet{Wong2006b} more than $\sim 75\%$ of the
  correction is due to the combined effect of higher shells.

\acknowledgements{The authors are glad to thank S.G.~Karshenboim for many
  useful discussions and consultations about details of the two-photon
  processes and for pointing us towards several useful references.
  They are also grateful to L.N.~Labzowsky for his advice and detailed
  discussions of the two-photon emission. In particular J.C. thanks
  L.N.~Labzowsky for hospitality during his visit in Dresden, December 2006.
  We also wish to thank S.G.~Karshenboim and V.G.~Ivanov for the possibility to
  compare our results with their computations on the 3s and 3d rates prior to
  publication.
  { It was a pleasure to discuss the detailed physics of recombination with
    C.~Hirata during the visit to the IAS, September 2006.  }
\change{Furthermore the authors thank E.E.~Kholupenko for his detailed
comments on the paper.}
}

\bibliographystyle{aa}
\bibliography{Lit}

\begin{appendix}

\section{Radial integrals}
\label{app:A}
\subsection{Bound-bound radial integrals}
For the the required bound-bound radial integrals up to $n_i=5$, one has
\bsub
\beal
\label{eq:bbmat}
\left<R_{1{\rm s}}|\,r\,| R_{n{\rm p}}\right>
&=2^4 \,n^{7/2}\,\frac{(n-1)^{n-5/2}}{(n+1)^{n+5/2}}
\\
\left<R_{2{\rm s}}|\,r\,| R_{n{\rm p}}\right>
&=2^8\sqrt{2} \,n^{7/2}(n^2-1)^{1/2}\,\frac{(n-2)^{n-3}}{(n+2)^{n+3}}
\\
\left<R_{3{\rm s}}|\,r\,| R_{n{\rm p}}\right>
&=2^4 3^3\sqrt{3}\, n^{7/2}(n^2-1)^{1/2}[7n^2-27]\,\frac{(n-3)^{n-4}}{(n+3)^{n+4}}
\\
\left<R_{4{\rm s}}|\,r\,| R_{n{\rm p}}\right>
&=\frac{2^{13}}{3}\, n^{7/2}(n^2-1)^{1/2}[23n^4-288n^2+768]
\,\frac{(n-4)^{n-5}}{(n+4)^{n+5}}
\\
\left<R_{5{\rm s}}|\,r\,| R_{n{\rm p}}\right>
&=\frac{2^{4}5^{4}\sqrt{5}}{3}\,
n^{7/2}(n^2-1)^{1/2}
\nonumber
\\
&\qquad
\times[91n^6-2545n^4 +20625n^2-46875]\,\frac{(n-5)^{n-6}}{(n+5)^{n+6}}
\\
\left<R_{3{\rm d}}|\,r\,| R_{n{\rm p}}\right>
&=\frac{2^5 3^3\sqrt{2}\sqrt{3}}{\sqrt{5}}\, n^{11/2}(n^2-1)^{1/2}
\,\frac{(n-3)^{n-4}}{(n+3)^{n+4}}
\\
\left<R_{4{\rm d}}|\,r\,| R_{n{\rm p}}\right>
&=\frac{2^{14}}{3\sqrt{5}}
\,n^{11/2}(n^2-1)^{1/2}[7n^2-48]
\,\frac{(n-4)^{n-5}}{(n+4)^{n+5}}
\\
\left<R_{5{\rm d}}|\,r\,| R_{n{\rm p}}\right>
&=\frac{2^5 5^4\sqrt{2}\sqrt{5}}{3\sqrt{7}}
\,n^{11/2}(n^2-1)^{1/2}
\nonumber
\\
&\qquad\qquad\qquad
\times[29n^4-590n^2+2625]
\,\frac{(n-5)^{n-6}}{(n+5)^{n+6}}.
\end{align}
\esub
In addition one needs $\left<R_{n, l-1}|\,r\,| R_{nl}\right>=-\frac{3}{2} n\sqrt{n^2-l^2}$.

\subsection{Bound-free radial integrals}
For the necessary bound-free radial integrals up to $n_i=5$, using the
definition of the radial functions for the continuum states \citep[e.g. see \S
36][]{LandauQM}, one obtains
\bsub
\beal
\label{eq:bcmat}
\left<R_{1{\rm s}}|\,r\,| R_{x 1}\right>
&=2^4\,\frac{ x^{1/2}}{(1+x^2)^{5/2}}\,\frac{e^{-2\arctan(x)/x}}{\sqrt{1-e^{-2\pi/x}}}
\\
\left<R_{2{\rm s}}|\,r\,| R_{x 1}\right>&=2^8 \sqrt{2}
\,\frac{ x^{1/2}(1+x^2)^{1/2}}{(1+4x^2)^3}\,
\frac{e^{-2\arctan(2x)/x}}{\sqrt{1-e^{-2\pi/x}}}
\\
\left<R_{3{\rm s}}|\,r\,| R_{x 1}\right>&=2^4 3^3\sqrt{3}
\,\frac{ x^{1/2}(1+x^2)^{1/2}[7+27x^2]}{(1+9x^2)^4}
\,\frac{e^{-2\arctan(3x)/x}}{\sqrt{1-e^{-2\pi/x}}}
\\
\left<R_{4{\rm s}}|\,r\,| R_{x 1}\right>&=\frac{2^{13}}{3}
\,\frac{ x^{1/2}(1+x^2)^{1/2}[23+96x^2(3+8x^2)]}{(1+16x^2)^5}
\,\frac{e^{-2\arctan(4x)/x}}{\sqrt{1-e^{-2\pi/x}}}
\\
\left<R_{5{\rm s}}|\,r\,| R_{x 1}\right>&=\frac{2^{4}5^4\sqrt{5}}{3}
\,\frac{ x^{1/2}(1+x^2)^{1/2}[91+5x^2(509+4125x^2+9375x^4)]}{(1+25x^2)^6}
\nonumber
\\
&\qquad\qquad\qquad\qquad
\times
\frac{e^{-2\arctan(5x)/x}}{\sqrt{1-e^{-2\pi/x}}}
\\
\left<R_{3{\rm d}}|\,r\,| R_{x 1}\right>&=\frac{2^{5}3^3\sqrt{2}\sqrt{3}}{\sqrt{5}}
\,\frac{ x^{1/2}(1+x^2)^{1/2}}{(1+9x^2)^4}
\,\frac{e^{-2\arctan(3x)/x}}{\sqrt{1-e^{-2\pi/x}}}
\\
\left<R_{4{\rm d}}|\,r\,| R_{x 1}\right>&=\frac{2^{14}}{3\sqrt{5}}
\,\frac{ x^{1/2}(1+x^2)^{1/2}[7+48x^2]}{(1+16x^2)^5}
\,\frac{e^{-2\arctan(4x)/x}}{\sqrt{1-e^{-2\pi/x}}}
\\
\left<R_{5{\rm d}}|\,r\,| R_{x 1}\right>&=\frac{2^{5}5^4\sqrt{2}\sqrt{5}}{3\sqrt{7}}
\,\frac{ x^{1/2}(1+x^2)^{1/2}[29+590x^2+2625x^4]}{(1+25x^2)^6}
\nonumber
\\
&\qquad\qquad\qquad\qquad
\times
\frac{e^{-2\arctan(5x)/x}}{\sqrt{1-e^{-2\pi/x}}}.
\end{align}
\esub
The value of $x$ ranges from $0$ to $\infty$. 

\section{Non-linear fitting functions non-resonant emission spectra}
\label{app:fits}
The Tables \ref{tab:two} and \ref{tab:three} contain the non-linear fitting
coefficients for the non-resonant emission spectra. The non-resonant emission
spectra are then given by $\phi^{\rm nr}_{n_i l_i\rightarrow 1{\rm
s}}(y)=\sigma(y)^2$, with
$\sigma(y)=w^{1/2}[a_0+b_0\,w^{\beta}(1+b_1\,w^{1/8}+b_2\,w^{1/7}+b_3\,w^{1/6}+b_4\,w^{1/5}+b_5\,w^{1/4}+b_6\,w^{1/3}+b_7\,w^{1/2}+b_8\,w^{1}+b_{9}\,w^2)]$
and $w=y(1-y)$. In this definition one has $M_{\rm nr}=\sigma(y)/\sqrt{G_{n_i
l_i}}\,w^{3/2}$.
The first 200 terms in the infinite sum were taken into account. Within the
  assumptions, the accuracy of these approximations should be better than
  $\sim 0.1\%$. \change{Note that $a_0$ and $b_0$ have dimension $\rm
  sec^{-1/2}$, and $\phi^{\rm nr}(1/2)$ has dimension $\rm sec^{-1}$}.

\onecolumn
\begin{table*}
\centering
\caption{Non-linear fitting coefficients for the non-resonant $n{\rm
s}\rightarrow 1{\rm s}$ emission spectra within the frequency range
$10^{-3}\leq y\leq 0.999$ for $n\leq 20$.}
\label{tab:two}
\begin{tabular*}{\columnwidth}{lrrrrr}
\hline
\hline
 \\[-2.8mm] 
 & 2s  & 3s  & 4s  & 5s  & 6s 
 \\ 
 \hline 
 \\[-2.8mm] 
 $a_0$ & $\pot{-1.3974426528}{1}$ & $\pot{-2.0848318929}{1}$ & $\pot{-2.5412664413}{1}$ & $\pot{-2.8856705303}{1}$ & $\pot{-3.1567507558}{1}$ 
 \\ 
 $\beta$ & $\pot{2.5591291935}{-1}$ & $\pot{5.5198239644}{-1}$ & $\pot{8.0397418838}{-1}$ & $\pot{4.8110524091}{-1}$ & $\pot{2.5266888083}{-1}$ 
 \\ 
 $b_0$ & $\pot{6.6487585307}{0}$ & $\pot{4.6179953215}{2}$ & $\pot{7.6795669026}{2}$ & $\pot{8.3598144011}{2}$ & $\pot{5.6902488765}{2}$ 
 \\ 
 $b_1$ & $\pot{-5.1211954644}{1}$ & $\pot{-9.8891531378}{1}$ & $\pot{-4.2588083735}{2}$ & $\pot{5.1404483119}{1}$ & $\pot{4.5519082179}{1}$ 
 \\ 
 $b_2$ & $\pot{3.4686628383}{1}$ & $\pot{1.0769127960}{2}$ & $\pot{7.0510361599}{2}$ & $\pot{-7.2853184204}{1}$ & $\pot{-6.7551149725}{1}$ 
 \\ 
 $b_3$ & $\pot{3.8515432953}{1}$ & $\pot{6.1301937755}{1}$ & $\pot{-3.6120946105}{1}$ & $\pot{-4.9378611706}{1}$ & $\pot{-3.8327617310}{1}$ 
 \\ 
 $b_4$ & $\pot{6.7503966084}{0}$ & $\pot{-4.1466090643}{1}$ & $\pot{-4.4049919721}{2}$ & $\pot{4.7383284338}{1}$ & $\pot{4.4401360958}{1}$ 
 \\ 
 $b_5$ & $\pot{-3.1431622488}{1}$ & $\pot{-8.4016250015}{1}$ & $\pot{2.4657329140}{2}$ & $\pot{9.9191282175}{1}$ & $\pot{7.4195208626}{1}$ 
 \\ 
 $b_6$ & $\pot{-1.3697636757}{1}$ & $\pot{7.1191755771}{1}$ & $\pot{-6.5900037717}{1}$ & $\pot{-1.0379836854}{2}$ & $\pot{-7.9312304169}{1}$ 
 \\ 
 $b_7$ & $\pot{2.4016413274}{1}$ & $\pot{-1.8695504961}{1}$ & $\pot{1.8226732411}{1}$ & $\pot{2.9602921680}{1}$ & $\pot{2.1866188936}{1}$ 
 \\ 
 $b_8$ & $\pot{-9.5149995633}{0}$ & $\pot{2.2794449683}{0}$ & $\pot{-2.8692508508}{0}$ & $\pot{-2.8636914451}{0}$ & $\pot{-1.9806370670}{0}$ 
 \\ 
 $b_9$ & $\pot{2.6555412799}{0}$ & $\pot{-4.3875761781}{-1}$ & $\pot{4.4042663362}{-1}$ & $\pot{3.8579054257}{-1}$ & $\pot{2.6143991878}{-1}$ 
 \\ 
 $\phi^{\rm nr}(1/2)$ & $\pot{2.1303295046}{1}$ & $\pot{2.5955835234}{1}$ & $\pot{3.0008781647}{1}$ & $\pot{3.5729374807}{1}$ & $\pot{4.2778194752}{1}$ 
\\[1mm]
\hline
\end{tabular*} 
\begin{tabular*}{\columnwidth}{lrrrrr}
\hline
 \\[-2.8mm] 
 & 7s  & 8s  & 9s  & 10s  & 11s 
 \\ 
 \hline 
 \\[-2.8mm] 
 $a_0$ & $\pot{-3.3720076517}{1}$ & $\pot{-3.5420358274}{1}$ & $\pot{-3.6749872667}{1}$ & $\pot{-3.7779791307}{1}$ & $\pot{-3.8573952826}{1}$ 
 \\ 
 $\beta$ & $\pot{8.5151589950}{-2}$ & $\pot{-4.2719713121}{-2}$ & $\pot{-1.3118006971}{-1}$ & $\pot{-2.1932912263}{-1}$ & $\pot{-2.9226372396}{-1}$ 
 \\ 
 $b_0$ & $\pot{3.7656038819}{2}$ & $\pot{2.7408956380}{2}$ & $\pot{2.8136510570}{2}$ & $\pot{1.5264948382}{2}$ & $\pot{8.6437136666}{1}$ 
 \\ 
 $b_1$ & $\pot{2.7980511952}{1}$ & $\pot{7.4703207553}{0}$ & $\pot{-1.8981579892}{1}$ & $\pot{-1.9235840320}{1}$ & $\pot{-1.9681933480}{1}$ 
 \\ 
 $b_2$ & $\pot{-4.9359948916}{1}$ & $\pot{-2.6259765669}{1}$ & $\pot{6.6356770418}{0}$ & $\pot{7.1592304894}{0}$ & $\pot{7.8432338862}{0}$ 
 \\ 
 $b_3$ & $\pot{-2.0507457345}{1}$ & $\pot{-3.3913231244}{0}$ & $\pot{1.3837014244}{1}$ & $\pot{1.3979846789}{1}$ & $\pot{1.4343371719}{1}$ 
 \\ 
 $b_4$ & $\pot{3.6291414100}{1}$ & $\pot{2.5464049259}{1}$ & $\pot{7.8618686503}{0}$ & $\pot{7.3730033289}{0}$ & $\pot{6.8679571684}{0}$ 
 \\ 
 $b_5$ & $\pot{4.3713955958}{1}$ & $\pot{1.6166653183}{1}$ & $\pot{-8.0358407914}{0}$ & $\pot{-8.6532015972}{0}$ & $\pot{-9.6695240617}{0}$ 
 \\ 
 $b_6$ & $\pot{-5.2721158916}{1}$ & $\pot{-2.8335792233}{1}$ & $\pot{-4.8006902023}{0}$ & $\pot{-3.6779369765}{0}$ & $\pot{-2.2948815920}{0}$ 
 \\ 
 $b_7$ & $\pot{1.4831730999}{1}$ & $\pot{8.6649789093}{0}$ & $\pot{2.8424870765}{0}$ & $\pot{2.4362978935}{0}$ & $\pot{2.0390764869}{0}$ 
 \\ 
 $b_8$ & $\pot{-1.3230829528}{0}$ & $\pot{-7.9156543417}{-1}$ & $\pot{-3.2207517000}{-1}$ & $\pot{-2.7645228408}{-1}$ & $\pot{-2.3641043114}{-1}$ 
 \\ 
 $b_9$ & $\pot{1.7474071394}{-1}$ & $\pot{1.0596166967}{-1}$ & $\pot{4.6227764509}{-2}$ & $\pot{3.9732841793}{-2}$ & $\pot{3.4157538873}{-2}$ 
 \\ 
 $\phi^{\rm nr}(1/2)$ & $\pot{5.0793294231}{1}$ & $\pot{5.9525977879}{1}$ & $\pot{6.8805689403}{1}$ & $\pot{7.8512293219}{1}$ & $\pot{8.8558644431}{1}$ 
\\[1mm]
\hline
\end{tabular*} 
\begin{tabular*}{\columnwidth}{lrrrrr}
\hline
 \\[-2.8mm] 
 & 12s  & 13s  & 14s  & 15s  & 16s 
 \\ 
 \hline 
 \\[-2.8mm] 
 $a_0$ & $\pot{-3.9188326323}{1}$ & $\pot{-3.9670102478}{1}$ & $\pot{-4.0057506143}{1}$ & $\pot{-4.0380409686}{1}$ & $\pot{-4.0661454371}{1}$ 
 \\ 
 $\beta$ & $\pot{-3.5484078761}{-1}$ & $\pot{-4.0844617686}{-1}$ & $\pot{-4.5428570381}{-1}$ & $\pot{-4.6335483910}{-1}$ & $\pot{-4.9050805652}{-1}$ 
 \\ 
 $b_0$ & $\pot{4.9431541043}{1}$ & $\pot{2.9197464029}{1}$ & $\pot{1.7971514946}{1}$ & $\pot{1.5982980641}{1}$ & $\pot{1.1724743879}{1}$ 
 \\ 
 $b_1$ & $\pot{-1.9750664049}{1}$ & $\pot{-1.9723076238}{1}$ & $\pot{-1.9651279272}{1}$ & $\pot{-2.0606406024}{1}$ & $\pot{-2.1081283167}{1}$ 
 \\ 
 $b_2$ & $\pot{8.1838767684}{0}$ & $\pot{8.3673624537}{0}$ & $\pot{8.4788048970}{0}$ & $\pot{9.3338234414}{0}$ & $\pot{9.8388495127}{0}$ 
 \\ 
 $b_3$ & $\pot{1.4326519992}{1}$ & $\pot{1.4291867041}{1}$ & $\pot{1.4206704409}{1}$ & $\pot{1.4985604253}{1}$ & $\pot{1.5395638329}{1}$ 
 \\ 
 $b_4$ & $\pot{6.4270125576}{0}$ & $\pot{6.0967371906}{0}$ & $\pot{5.8366404588}{0}$ & $\pot{5.7371156374}{0}$ & $\pot{5.5633379574}{0}$ 
 \\ 
 $b_5$ & $\pot{-1.0032296167}{1}$ & $\pot{-1.0363858125}{1}$ & $\pot{-1.0518387038}{1}$ & $\pot{-1.1546724399}{1}$ & $\pot{-1.2242342508}{1}$ 
 \\ 
 $b_6$ & $\pot{-1.4087194335}{0}$ & $\pot{-6.8184121010}{-1}$ & $\pot{-2.9722754662}{-1}$ & $\pot{1.1575092351}{-1}$ & $\pot{4.8942663430}{-1}$ 
 \\ 
 $b_7$ & $\pot{1.8482968593}{0}$ & $\pot{1.8529967851}{0}$ & $\pot{2.1519310722}{0}$ & $\pot{2.2891230659}{0}$ & $\pot{2.6782065008}{0}$ 
 \\ 
 $b_8$ & $\pot{-2.1185600113}{-1}$ & $\pot{-1.9498101086}{-1}$ & $\pot{-1.8067549402}{-1}$ & $\pot{-2.0119354736}{-1}$ & $\pot{-1.8223251218}{-1}$ 
 \\ 
 $b_9$ & $\pot{3.0555243512}{-2}$ & $\pot{2.7915669075}{-2}$ & $\pot{2.5650331467}{-2}$ & $\pot{3.2820939449}{-2}$ & $\pot{3.0707648484}{-2}$ 
 \\ 
 $\phi^{\rm nr}(1/2)$ & $\pot{9.8879969305}{1}$ & $\pot{1.0942707459}{2}$ & $\pot{1.2016192323}{2}$ & $\pot{1.3105436054}{2}$ & $\pot{1.4208118814}{2}$ 
\\[1mm]
\hline
\end{tabular*} 
\begin{tabular*}{\columnwidth}{lrrrrr}
\hline
\\[-2.8mm] 
 & 17s  & 18s  & 19s  & 20s 
 \\ 
 \hline 
 \\[-2.8mm] 
 $a_0$ & $\pot{-4.0917355907}{1}$ & $\pot{-4.1160166944}{1}$ & $\pot{-4.1398378313}{1}$ & $\pot{-4.1637819958}{1}$ 
 \\ 
 $\beta$ & $\pot{-4.7538949275}{-1}$ & $\pot{-4.4937224788}{-1}$ & $\pot{-4.4703263317}{-1}$ & $\pot{-4.8244405309}{-1}$ 
 \\ 
 $b_0$ & $\pot{1.2547183851}{1}$ & $\pot{1.4317441971}{1}$ & $\pot{1.3339226887}{1}$ & $\pot{9.1516183240}{0}$ 
 \\ 
 $b_1$ & $\pot{-2.1972421315}{1}$ & $\pot{-2.2875778234}{1}$ & $\pot{-2.3839004652}{1}$ & $\pot{-2.3934619536}{1}$ 
 \\ 
 $b_2$ & $\pot{1.0636513367}{1}$ & $\pot{1.1277442747}{1}$ & $\pot{1.2219984677}{1}$ & $\pot{1.2479746639}{1}$ 
 \\ 
 $b_3$ & $\pot{1.5943304131}{1}$ & $\pot{1.6640858983}{1}$ & $\pot{1.7332324934}{1}$ & $\pot{1.7384969031}{1}$ 
 \\ 
 $b_4$ & $\pot{5.4874376556}{0}$ & $\pot{5.6104613925}{0}$ & $\pot{5.3841616235}{0}$ & $\pot{5.1298737644}{0}$ 
 \\ 
 $b_5$ & $\pot{-1.2655779187}{1}$ & $\pot{-1.3260181912}{1}$ & $\pot{-1.4146776145}{1}$ & $\pot{-1.4462821999}{1}$ 
 \\ 
 $b_6$ & $\pot{4.9579530934}{-1}$ & $\pot{5.2627269447}{-1}$ & $\pot{1.0461078949}{0}$ & $\pot{1.3716208664}{0}$ 
 \\ 
 $b_7$ & $\pot{2.5916598871}{0}$ & $\pot{2.4342837775}{0}$ & $\pot{2.3937323753}{0}$ & $\pot{2.8470359883}{0}$ 
 \\ 
 $b_8$ & $\pot{-2.2938505101}{-1}$ & $\pot{-2.7807376757}{-1}$ & $\pot{-2.8519500457}{-1}$ & $\pot{-2.4816096711}{-1}$ 
 \\ 
 $b_9$ & $\pot{4.2155577449}{-2}$ & $\pot{5.4040208750}{-2}$ & $\pot{5.6731652666}{-2}$ & $\pot{4.9205565009}{-2}$ 
 \\ 
 $\phi^{\rm nr}(1/2)$ & $\pot{1.5322255378}{2}$ & $\pot{1.6446312975}{2}$ & $\pot{1.7579037822}{2}$ & $\pot{1.8719377215}{2}$ 
\\[1mm]
\hline
\hline
\end{tabular*} 
\end{table*}

\begin{table*}
\centering
\caption{Non-linear fitting coefficients for the non-resonant $n{\rm
d}\rightarrow 1{\rm s}$ emission spectra within the frequency range
$10^{-3}\leq y\leq 0.999$ for $n\leq 20$.}
\label{tab:three}
{
\begin{tabular*}{\columnwidth}{lrrrrr}
\hline
\hline
 \\[-2.8mm] 
 & 3d  & 4d  & 5d  & 6d  & 7d 
 \\ 
 \hline 
 \\[-2.8mm] 
 $a_0$ & $\pot{-1.0484392971}{1}$ & $\pot{-1.4527047719}{1}$ & $\pot{-1.7350074013}{1}$ & $\pot{-1.9527421713}{1}$ & $\pot{-2.1271595034}{1}$ 
 \\ 
 $\beta$ & $\pot{2.6115596268}{-1}$ & $\pot{8.1050542323}{-1}$ & $\pot{4.6206291704}{-1}$ & $\pot{2.3419462081}{-1}$ & $\pot{6.7999906938}{-2}$ 
 \\ 
 $b_0$ & $\pot{5.3546036998}{1}$ & $\pot{1.7065087104}{2}$ & $\pot{2.0902445563}{2}$ & $\pot{1.4535416639}{2}$ & $\pot{1.1921878150}{2}$ 
 \\ 
 $b_1$ & $\pot{-9.7319568527}{1}$ & $\pot{-4.2936500531}{2}$ & $\pot{5.8562543507}{1}$ & $\pot{5.9612688683}{1}$ & $\pot{3.0991642200}{1}$ 
 \\ 
 $b_2$ & $\pot{1.0624929126}{2}$ & $\pot{7.1072286522}{2}$ & $\pot{-7.8501054392}{1}$ & $\pot{-8.0969461685}{1}$ & $\pot{-5.2470184998}{1}$ 
 \\ 
 $b_3$ & $\pot{6.0569007312}{1}$ & $\pot{-3.6472867939}{1}$ & $\pot{-5.8200368181}{1}$ & $\pot{-5.3437128442}{1}$ & $\pot{-2.3362093646}{1}$ 
 \\ 
 $b_4$ & $\pot{-4.1091307521}{1}$ & $\pot{-4.4275350080}{2}$ & $\pot{4.8846690629}{1}$ & $\pot{4.9735078265}{1}$ & $\pot{3.7636220223}{1}$ 
 \\ 
 $b_5$ & $\pot{-8.3346792794}{1}$ & $\pot{2.4703670467}{2}$ & $\pot{1.1281858951}{2}$ & $\pot{9.8247844915}{1}$ & $\pot{4.8276486304}{1}$ 
 \\ 
 $b_6$ & $\pot{6.9608223453}{1}$ & $\pot{-6.6433743910}{1}$ & $\pot{-1.1364879196}{2}$ & $\pot{-9.8596391243}{1}$ & $\pot{-5.6611227459}{1}$ 
 \\ 
 $b_7$ & $\pot{-1.7046798510}{1}$ & $\pot{1.8959198585}{1}$ & $\pot{3.1841053645}{1}$ & $\pot{2.6560347162}{1}$ & $\pot{1.5852011862}{1}$ 
 \\ 
 $b_8$ & $\pot{1.6052249571}{0}$ & $\pot{-3.1192847294}{0}$ & $\pot{-3.0563352329}{0}$ & $\pot{-2.3898992832}{0}$ & $\pot{-1.4245686402}{0}$ 
 \\ 
 $b_9$ & $\pot{-2.2422235582}{-1}$ & $\pot{5.0786391006}{-1}$ & $\pot{4.1302480839}{-1}$ & $\pot{3.1693775497}{-1}$ & $\pot{1.8991428606}{-1}$ 
 \\ 
 $\phi^{\rm nr}(1/2)$ & $\pot{2.0915177571}{1}$ & $\pot{3.4902680010}{1}$ & $\pot{4.4901877444}{1}$ & $\pot{5.2879363207}{1}$ & $\pot{5.9763238933}{1}$ 
\\[1mm]
\hline
\end{tabular*} 
\begin{tabular*}{\columnwidth}{lrrrrr}
\hline
 \\[-2.8mm] 
 & 8d  & 9d  & 10d  & 11d  & 12d 
 \\ 
 \hline 
 \\[-2.8mm] 
 $a_0$ & $\pot{-2.2687581599}{1}$ & $\pot{-2.3840999089}{1}$ & $\pot{-2.4781138994}{1}$ & $\pot{-2.5548956191}{1}$ & $\pot{-2.6179506753}{1}$ 
 \\ 
 $\beta$ & $\pot{-4.0416246997}{-2}$ & $\pot{-1.4739834639}{-1}$ & $\pot{-2.3472013799}{-1}$ & $\pot{-3.0648469505}{-1}$ & $\pot{-3.6629270128}{-1}$ 
 \\ 
 $b_0$ & $\pot{1.8805941117}{2}$ & $\pot{1.0632962733}{2}$ & $\pot{6.0922960536}{1}$ & $\pot{3.6160368635}{1}$ & $\pot{2.2144091289}{1}$ 
 \\ 
 $b_1$ & $\pot{-1.8483714555}{1}$ & $\pot{-1.8910499138}{1}$ & $\pot{-1.9083232938}{1}$ & $\pot{-1.9357145527}{1}$ & $\pot{-1.9508239494}{1}$ 
 \\ 
 $b_2$ & $\pot{5.8156915967}{0}$ & $\pot{6.5026790075}{0}$ & $\pot{6.8739045830}{0}$ & $\pot{7.3456517995}{0}$ & $\pot{7.6357892226}{0}$ 
 \\ 
 $b_3$ & $\pot{1.3504738304}{1}$ & $\pot{1.3795221365}{1}$ & $\pot{1.3924013399}{1}$ & $\pot{1.4140788681}{1}$ & $\pot{1.4314213871}{1}$ 
 \\ 
 $b_4$ & $\pot{8.5177417832}{0}$ & $\pot{8.0023959097}{0}$ & $\pot{7.6277097156}{0}$ & $\pot{7.2310576094}{0}$ & $\pot{6.9389854602}{0}$ 
 \\ 
 $b_5$ & $\pot{-7.0817594819}{0}$ & $\pot{-7.9116375657}{0}$ & $\pot{-8.4667321100}{0}$ & $\pot{-9.1736549163}{0}$ & $\pot{-9.8020559467}{0}$ 
 \\ 
 $b_6$ & $\pot{-6.3232881999}{0}$ & $\pot{-5.0886630492}{0}$ & $\pot{-4.1097110628}{0}$ & $\pot{-3.0436236228}{0}$ & $\pot{-2.1203514774}{0}$ 
 \\ 
 $b_7$ & $\pot{3.4340616701}{0}$ & $\pot{2.9768149183}{0}$ & $\pot{2.6171546155}{0}$ & $\pot{2.2949189350}{0}$ & $\pot{2.0887633671}{0}$ 
 \\ 
 $b_8$ & $\pot{-3.9664209528}{-1}$ & $\pot{-3.4089403622}{-1}$ & $\pot{-2.9766246686}{-1}$ & $\pot{-2.6063687833}{-1}$ & $\pot{-2.3211600328}{-1}$ 
 \\ 
 $b_9$ & $\pot{5.7497854559}{-2}$ & $\pot{4.9213789949}{-2}$ & $\pot{4.2727779792}{-2}$ & $\pot{3.7171055385}{-2}$ & $\pot{3.2750870685}{-2}$ 
 \\ 
 $\phi^{\rm nr}(1/2)$ & $\pot{6.6011443629}{1}$ & $\pot{7.1866675229}{1}$ & $\pot{7.7466332091}{1}$ & $\pot{8.2892729834}{1}$ & $\pot{8.8197557986}{1}$ 
\\[1mm]
\hline
\end{tabular*} 
\begin{tabular*}{\columnwidth}{lrrrrr}
\hline
 \\[-2.8mm] 
 & 13d  & 14d  & 15d  & 16d  & 17d 
 \\ 
 \hline 
 \\[-2.8mm] 
 $a_0$ & $\pot{-2.6702542838}{1}$ & $\pot{-2.7142749615}{1}$ & $\pot{-2.7520108750}{1}$ & $\pot{-2.7850443691}{1}$ & $\pot{-2.8146062687}{1}$ 
 \\ 
 $\beta$ & $\pot{-4.1655350067}{-1}$ & $\pot{-4.5817894101}{-1}$ & $\pot{-4.9230994423}{-1}$ & $\pot{-4.8799319840}{-1}$ & $\pot{-5.1046103153}{-1}$ 
 \\ 
 $b_0$ & $\pot{1.4031825050}{1}$ & $\pot{9.3536034158}{0}$ & $\pot{6.5800387184}{0}$ & $\pot{6.5003185806}{0}$ & $\pot{5.0269776925}{0}$ 
 \\ 
 $b_1$ & $\pot{-1.9675880116}{1}$ & $\pot{-1.9851518345}{1}$ & $\pot{-2.0183574589}{1}$ & $\pot{-2.0542531107}{1}$ & $\pot{-2.0953818322}{1}$ 
 \\ 
 $b_2$ & $\pot{8.0418685007}{0}$ & $\pot{8.3266612119}{0}$ & $\pot{8.8233502592}{0}$ & $\pot{8.8786234533}{0}$ & $\pot{9.3483218939}{0}$ 
 \\ 
 $b_3$ & $\pot{1.4344987353}{1}$ & $\pot{1.4507266322}{1}$ & $\pot{1.4666068986}{1}$ & $\pot{1.5088704084}{1}$ & $\pot{1.5370738183}{1}$ 
 \\ 
 $b_4$ & $\pot{6.5426767345}{0}$ & $\pot{6.3117201027}{0}$ & $\pot{6.0075843919}{0}$ & $\pot{6.3138200610}{0}$ & $\pot{6.1350437791}{0}$ 
 \\ 
 $b_5$ & $\pot{-1.0108197620}{1}$ & $\pot{-1.0587100158}{1}$ & $\pot{-1.0920565892}{1}$ & $\pot{-1.1277066540}{1}$ & $\pot{-1.1714497606}{1}$ 
 \\ 
 $b_6$ & $\pot{-1.4725216928}{0}$ & $\pot{-9.1174802515}{-1}$ & $\pot{-6.0110879243}{-1}$ & $\pot{-8.1971758897}{-1}$ & $\pot{-6.0125221865}{-1}$ 
 \\ 
 $b_7$ & $\pot{2.0501069105}{0}$ & $\pot{2.1660824963}{0}$ & $\pot{2.4592074550}{0}$ & $\pot{2.6330961336}{0}$ & $\pot{2.9489602434}{0}$ 
 \\ 
 $b_8$ & $\pot{-2.0980025116}{-1}$ & $\pot{-1.8682379725}{-1}$ & $\pot{-1.5933095829}{-1}$ & $\pot{-2.0505355745}{-1}$ & $\pot{-1.7710483631}{-1}$ 
 \\ 
 $b_9$ & $\pot{2.9129276734}{-2}$ & $\pot{2.5589737997}{-2}$ & $\pot{2.1812553698}{-2}$ & $\pot{3.3518563369}{-2}$ & $\pot{2.9819941280}{-2}$ 
 \\ 
 $\phi^{\rm nr}(1/2)$ & $\pot{9.3414593519}{1}$ & $\pot{9.8566665797}{1}$ & $\pot{1.0366966228}{2}$ & $\pot{1.0873478648}{2}$ & $\pot{1.1377060842}{2}$ 
\\[1mm]
\hline
\end{tabular*} 
\begin{tabular*}{\columnwidth}{lrrrrr}
\hline
 \\[-2.8mm] 
 & 18d  & 19d  & 20d 
 \\ 
 \hline 
 \\[-2.8mm] 
 $a_0$ & $\pot{-2.8416405767}{1}$ & $\pot{-2.8668633507}{1}$ & $\pot{-2.8908128993}{1}$ 
 \\ 
 $\beta$ & $\pot{-5.2741247590}{-1}$ & $\pot{-4.8873340484}{-1}$ & $\pot{-4.7794020321}{-1}$ 
 \\ 
 $b_0$ & $\pot{4.1457341691}{0}$ & $\pot{5.2856466775}{0}$ & $\pot{5.2735742805}{0}$ 
 \\ 
 $b_1$ & $\pot{-2.1898431473}{1}$ & $\pot{-2.2963670035}{1}$ & $\pot{-2.3833638246}{1}$ 
 \\ 
 $b_2$ & $\pot{1.0391335216}{1}$ & $\pot{1.1284085887}{1}$ & $\pot{1.2121030366}{1}$ 
 \\ 
 $b_3$ & $\pot{1.6037472635}{1}$ & $\pot{1.6719978965}{1}$ & $\pot{1.7273259137}{1}$ 
 \\ 
 $b_4$ & $\pot{5.7762172996}{0}$ & $\pot{5.7383329948}{0}$ & $\pot{5.5675791948}{0}$ 
 \\ 
 $b_5$ & $\pot{-1.2705538272}{1}$ & $\pot{-1.3194456779}{1}$ & $\pot{-1.3752468820}{1}$ 
 \\ 
 $b_6$ & $\pot{-1.8363323426}{-2}$ & $\pot{7.2380465060}{-2}$ & $\pot{3.0584919862}{-1}$ 
 \\ 
 $b_7$ & $\pot{3.1624501247}{0}$ & $\pot{2.7857591072}{0}$ & $\pot{2.7454079863}{0}$ 
 \\ 
 $b_8$ & $\pot{-1.4050292467}{-1}$ & $\pot{-2.4398422587}{-1}$ & $\pot{-2.7060311659}{-1}$ 
 \\ 
 $b_9$ & $\pot{2.5244597061}{-2}$ & $\pot{4.6094366553}{-2}$ & $\pot{5.2504979483}{-2}$ 
 \\ 
 $\phi^{\rm nr}(1/2)$ & $\pot{1.1878311140}{2}$ & $\pot{1.2377675855}{2}$ & $\pot{1.2875544820}{2}$ 
\\[1mm]
\hline
\hline
\end{tabular*} 
}
\end{table*}
\end{appendix}

\end{document}

%% file: Befehle.tex
\newcommand{\id}{{\,\rm d}}

\newcommand{\beq}{\begin{equation}}   %

\newcommand{\eeq}{\end{equation}}   %

\newcommand{\beqa}{\begin{eqnarray}}   %

\newcommand{\eeqa}{\end{eqnarray}}   %

\newcommand{\beal}{\begin{align}}

\newcommand{\enal}{\end{align}}

\newcommand{\bspl}{\begin{split}}

\newcommand{\espl}{\end{split}}

\newcommand{\bsub}{\begin{subequations}}

\newcommand{\esub}{\end{subequations}}

\newcommand{\bmulti}{\begin{multline}}   %

\newcommand{\beqm}{\begin{mathletters}}   %

\newcommand{\eeqm}{\end{mathletters}}   %

\newcommand{\kB}{k_{\rm B}}

\newcommand{\Tg}{T_{\gamma}}

\newcommand{\pot}[2]{#1 \times 10^{#2}}

